\documentclass[11pt,english]{article}
\usepackage[T1]{fontenc}
\usepackage[latin9]{inputenc}
\usepackage{geometry}
\usepackage{color}
\usepackage{babel}
\usepackage{longtable}
\usepackage{booktabs}
\usepackage{multirow}
\usepackage{amsmath}
\usepackage{amsthm}
\usepackage{amssymb}
\usepackage{graphicx}
\usepackage{setspace}
\usepackage[authoryear]{natbib}
\usepackage[unicode=true]
 {hyperref}

\makeatletter

\providecommand{\tabularnewline}{\\}

\theoremstyle{plain}
\newtheorem{thm}{\protect\theoremname}
\theoremstyle{definition}
\newtheorem{example}[thm]{\protect\examplename}

\@ifundefined{date}{}{\date{}}
\usepackage{babel}
\bibpunct{(}{)}{,}{autheryear}{,}{,}

\allowdisplaybreaks

\usepackage{tikz}
\usetikzlibrary{fit,positioning}

\allowdisplaybreaks

\newtheorem{assumption}{Assumption}\newtheorem{theorem}{Theorem}\newtheorem{corollary}{Corollary}\newtheorem{proposition}{Proposition}\newtheorem{remark}{Remark}

\newcommand*{\indep}{%
  \mathbin{%
    \mathpalette{\@indep}{}%
  }%
}
\newcommand*{\nindep}{%
  \mathbin{
    \mathpalette{\@indep}{\not}
  }%
}
\newcommand*{\@indep}[2]{%
  \sbox0{$#1\perp\m@th$}
  \sbox2{$#1=$}
  \sbox4{$#1\vcenter{}$}
  \rlap{\copy0}
  \dimen@=\dimexpr\ht2-\ht4-.2pt\relax
  \kern\dimen@
  {#2}%
  \kern\dimen@
  \copy0 
} 

\newcommand{\cZ}{\mathcal{Z}} 
\newcommand{\cD}{\mathcal{D}} 
\newcommand{\cB}{\mathcal{B}} 
\newcommand{\pr}{\mathbb{P}}

\newcommand{\R}{\mathbb{R}}
\newcommand{\de}{\mathrm{d}}
\newcommand{\T}{\mathrm{\scriptscriptstyle T}}
\newcommand{\logit}{\text{logit}}

\newcommand{\aipw}{\mathrm{aipw}}

\newcommand{\eff}{\mathrm{eff}}

\newcommand{\N}{\mathcal{N}}

\newcommand{\rmN}{\mathrm{Normal }}

\newcommand{\E}{\mathbb{E}}
\newcommand{\V}{\mathbb{V}}
\newcommand{\cC}{\mathbb{C}}

\newcommand{\rt}{\mathrm{rt}}
\newcommand{\os}{\mathrm{rw}}
\newcommand{\rtrw}{\mathrm{eff}}
\newcommand{\rw}{\mathrm{rw}}

\newcommand{\pp}{\mathbb{P}}

\newcommand{\elas}{\mathrm{elas}}

\newcommand{\rmt}{\mathrm{t}}
\newcommand{\rteff}{\textnormal{rt-eff}}

\newcommand{\cI}{\mathcal{I}}
\newcommand{\bone}{\mathbf{1}}

\newcommand{\bias}{\textnormal{bias}}
\newcommand{\mse}{\textnormal{mse}}

\providecommand{\examplename}{Example}
\providecommand{\theoremname}{Theorem}

\makeatother

\providecommand{\examplename}{Example}
\providecommand{\theoremname}{Theorem}

\begin{document}
\title{Elastic Integrative Analysis of Randomized Trial and Real-World Data
for Treatment Heterogeneity Estimation}
\author{Shu Yang\thanks{Department of Statistics, North Carolina State University, North Carolina
27695, U.S.A. Email: syang24@ncsu.edu.}, Chenyin Gao$^{*}$, Donglin Zeng\thanks{Department of Biostatistics, University of North Carolina at Chapel
Hill}, and Xiaofei Wang\thanks{Department of Biostatistics and Bioinformatics, Duke University}
\thanks{This research is supported by the NSF grant DMS 1811245, NIA grant
1R01AG066883, and NIEHS 1R01ES031651. The authors would also like
to thank the Associate editor and anonymous reviewers for their valuable
comments and suggestions to improve the quality of the paper.}}
\maketitle
\begin{abstract}
We propose a test-based elastic integrative analysis of the randomized
trial and real-world data to estimate treatment effect heterogeneity
with a vector of known effect modifiers. When the real-world data
are not subject to bias, our approach combines the trial and real-world
data for efficient estimation. Utilizing the trial design, we construct
a test to decide whether or not to use real-world data. We characterize
the asymptotic distribution of the test-based estimator under local
alternatives. We provide a data-adaptive procedure to select the test
threshold that promises the smallest mean square error and an elastic
confidence interval with a good finite-sample coverage property.

\end{abstract}

\section{Introduction\label{sec:Introduction}}

\noindent \textcolor{black}{Precision medicine \citep{hamburg2010path},
which aims at customizing medical treatments to individual patient
characteristics, has recently received lots of attention. A critical
step toward precision medicine is to characterize the heterogeneity
of treatment effect (HTE; \citealp{rothwell2005subgroup,rothwell2005subgroups})
entailing how patient characteristics are related to treatment effect.
}Randomized trials (RTs) are the gold-standard method for treatment
effect evaluation because randomization of treatment ensures that
treatment groups are comparable and biases are minimized to the extent
possible. However, due to high costs and eligibility criteria for
recruiting patients, the trial sample is often small and limited in
the patient diversity, which renders the trial underpowered to estimate
the HTE and unable to estimate the HTE for specific patient characteristics.
On the other hand, extensive real-world (RW) data are increasingly
available for research purposes, such as electronic health records,
claims databases, and disease registries, with much larger sample
sizes and broader demographic and diversity than RT cohorts. Several
national organizations \citep{norris2010selecting} and regulatory
agencies \citep{sherman2016real} have recently advocated using RW
data to have a faster and less costly drug discovery process. Indeed,
big data provide unprecedented opportunities for new scientific discovery;
however, \textcolor{black}{they also present challenges with possible
incomparability with RT data due to selection bias, unmeasured confounding,
lack of concurrency, data quality, outcome validity, etc \citep{us2019rareD}. }

The motivating application is to evaluate adjuvant chemotherapy for
resected non-small cell lung cancer (NSCLC) at early-stage disease.
Adjuvant chemotherapy for resected NSCLC was shown to be effective
in late-stage II and IIIA disease based on RTs \citep{le2003results}.
However, the benefit of adjuvant chemotherapy in stage IB NSCLC disease
is unclear. Cancer and Leukemia Group B (CALGB) 9633 is the only RT
designed specifically for stage IB NSCLC \citep{strauss2008adjuvant};
however, it comprises about $300$ patients, which was undersized
to detect clinically meaningful improvements for adjuvant chemotherapy
\citep{katz2009calgb}. \textit{``Who can benefit from adjuvant chemotherapy
with stage IB NSCLC?''} remains an important clinical question. An
exploratory analysis of CALGB 9633 showed that patients with tumor
size $\geq4.0$ cm might benefit from adjuvant chemotherapy \citep{strauss2008adjuvant}.
On the other hand, the National Cancer Database (NCDB) is a clinical
oncology registry database that captures the information from approximately
$75\%$ of all newly diagnosed cancer patients in the US. Our goal
is to integrate the CALGB 9633 trial with a cohort selected under
the same trial eligibility criteria from the NCDB. We expect that
an integrated analysis of the CALGB 9633 and NCDB data can considerably
improve the efficiency of the HTE estimation on adjuvant chemotherapy
regarding tumor size over the RT-only analysis. Although such population-based
disease registries provide rich information citing the real-world
usage of adjuvant chemotherapy, the concern is the potential bias
associated with RW data.

Many authors have proposed methods for generalizing treatment effects
from RTs to the target population, whose covariate distribution can
be characterized by the RW data \citep{buchanan2018generalizing,zhao2019robustify,colnet2020causal,lee2021improving,lee2022generalizable}.
When both RT and RW data provide covariate, treatment, and outcome
information, there are two main approaches for integrative analysis:
meta-analyses of summary statistics \citep[e.g.,][]{verde2015combining}
and pooled patient data \citep{sobel2017causal}. The major drawback
of meta-analyses of the first kind is that they use only aggregated
information and do not distinguish the roles of the RT and RW data,
both having unique strengths and weaknesses. Meta-analyses of the
second kind include all patients, but pooling the data from two sources
breaks the randomization of treatments and relies on causal inference
methods to adjust for confounding bias \citep[e.g.,][]{prentice2005combined}.
More importantly, one cannot rule out possible unmeasured confounding
in the RW data. In addition, most existing integrative methods focused
on average treatment effects (ATEs) but not on HTEs, which lies at
the heart of precision medicine.

To acknowledge the advantages of the RT and RW data, we propose an
elastic algorithm for combining the RT and RW data for accurate and
robust estimation of the HTE function with a vector of known effect
modifiers. The primary identification assumptions underpinning our
method are \textcolor{black}{(i) the transportability of the HTE from
the RT data to the target population and (ii) the strong ignorability
of treatment assignment in the RT data. Transportability is a common
assumption in the trial generalizability literature, which holds if
the HTE function captures all the treatment effect modifiers, or the
study sample is a random sample from the target population. The well-controlled
trial design can also ensure the strong ignorability of treatment
assignment. If the RW sample satisfies the parallel assumptions (i)
and (ii), it is comparable to the RT sample in estimating the HTE.
In this case, integrating the RW sample would increase the efficiency
of HTE estimation. Toward this end, we use the semiparametric efficiency
theory \citep{bickel1993efficient,robins1994correcting} to derive
the semiparametrically efficient integrative estimator of the HTE.}
\textcolor{black}{However, due to many practical limitations, the
RW sample may violate the desirable comparability assumption (i) or
(ii). In this case, integrating the RW sample would lead to bias in
HTE estimation. }Utilizing the design advantage of RTs, we derive
a preliminary test statistic to gauge the \textcolor{black}{comparability
and} reliability of the RW data and decide whether or not to use the
RW data in an integrative analysis. Therefore, our test-based elastic
integrative estimator uses the efficient combination strategy for
estimation if the violation test is insignificant and retains only
the RT data if the violation test is significant.

The proposed estimator belongs to pre-test estimation by construction
\citep{giles1993pre} and is non-regular. \textcolor{black}{We consider
null, local, and fixed alternative hypotheses for the pre-testing,
representing three scenarios when the comparability assumption required
for the RW data is zero, weakly, and strongly violated, respectively.
Notably, the fixed alternative formulates the bias of the RW score
of the HTE parameter to be fixed, under which the pre-test statistic
goes to infinity as the sample size increases. Thus, the inference
under the fixed alternative can not capture the finite-sample behavior
of the test and estimator well and lacks uniform validity. A common
strategy to obtain uniform inference validity for non-regular estimators
is considering the local alternative, which formulates the bias of
the RW score to be in the $n^{-1/2}$ neighborhood of zero. The inference
under the local alternative provides a better approximation of the
finite-sample behavior of the proposed estimator. Such strategies
have been considered in designing trials for sample size/power calculation
and in the weak instrument, partial identification, and classification
literature \citep{staiger1994instrumental,cheng2008robust,laber2011adaptive}.
Under the local alternative, when the testing distribution is non-degenerate},
exact inference for pre-test estimation is complex because the estimator
depends on the randomness of the test procedure. This issue cannot
be solved by splitting the sample into two parts for testing and estimation
separately \citep{toyoda1979pre}. The reason is that sample splitting
cannot bypass the issue of the additional randomness due to pre-testing,
and therefore the impact of pre-testing remains. Also, our test statistic
and estimator are constructed based on the whole sample data. To consider
the effect of pre-testing, we decompose the test-based elastic integrative
estimator into orthogonal components; one is affected by the pre-testing,
and the other is not. This step reveals the asymptotic distributions
of the proposed estimator to be mixture distributions involving a
truncated normal component with ellipsoid truncation and a normal
component. Under this framework, we provide a data-adaptive procedure
to select the threshold of the test statistic that promises the smallest
mean square error (MSE) of the proposed estimator. Lastly, we propose
an elastic procedure to construct confidence intervals (CIs), which
are adaptive to the local and fixed alternative and have good finite-sample
coverage properties.

This article is organized as follows. Section \ref{sec:Basic-setup}
introduces the basic setup, HTE, identification assumptions, and semiparametric
efficient estimation. Section \ref{subsec: robust and elastic integ}
establishes a test statistic for gauging the comparability of the
RW data with the RT data, a test-based elastic integrative estimator,
the asymptotic properties, and an elastic inference procedure. Section
\ref{sec:Simulation} presents a simulation study to evaluate the
performance of the proposed estimator in terms of robustness and efficiency.
Section \ref{sec:Application} applies the proposed method to combined
CALGB 9633 (RT) and NCDB (RW) data to characterize the HTE of adjuvant
chemotherapy in patients with stage IB non-small cell lung cancer.
We relegate technical details and all proofs to the supplementary
material.

\section{Basic setup\label{sec:Basic-setup}}

\subsection{Notation, the HTE, and two data sources}

Let $A\in\{0,1\}$ be the binary treatment, $Z$ a vector of pre-treatment
covariates of interest with the first component being $1$, $X$ a
vector of auxiliary variables including $Z$, and $Y$ the outcome
of interest. We consider $Y$ to be continuous or binary to fix ideas,
although our framework can be extended to general-type outcomes, including
the survival outcome. To define causal effects, we follow the potential
outcomes framework \citep{splawa1990application,rubin1974estimating}.
Under the Stable Unit of Treatment Value assumption, let $Y(a)$ be
the potential outcome had the subject been given treatment $a$, for
$a=0,1$. And, by the causal consistency assumption, the observed
outcome is $Y=Y(A)=AY(1)+(1-A)Y(0)$.

Based on the potential outcomes, the individual treatment effect is
$Y(1)-Y(0)$, and $\tau(Z)=\E\{Y(1)-Y(0)\mid Z\}$ characterizes the
HTE. For a binary outcome, $\tau(Z)$ is also called the causal risk
difference. In clinical settings, the parametric family of HTEs is
desirable and has wide applications in precision medicine to discover
optimal treatment regimes tailored to individual characteristics \citep{chakraborty2013statistical}.
We assume the HTE function to be 
\begin{equation}
\tau(Z)=\tau_{\psi_{0}}(Z)=\E\{Y(1)-Y(0)\mid Z;\psi_{0}\},\label{eq:SNMM}
\end{equation}
where $\psi_{0}\in\R^{p}$ is a vector of unknown parameters and $p$
is fixed.

We illustrate the HTE function in the following examples. 
\begin{example}
\label{eg. continuous outcome}(\citealp{lu2014asimplemethod}; \citealp{shi2016RobustLearning})
For a continuous outcome, a linear HTE function is $\tau_{\psi_{0}}(Z)=Z^{\T}\psi_{0}$,
where each component of $\psi_{0}$ quantifies how the treatment effect
varies over each $Z$. 
\end{example}

\begin{example}
\label{eg. binary outcome}(\citealp{lu2014asimplemethod}; \citealp{richardson2017modeling})
For a binary outcome, an HTE function for the causal risk difference
is $\tau_{\psi_{0}}(Z)=\{\exp(Z^{\T}\psi_{0})-1\}/\{\exp(Z^{\T}\psi_{0})+1\}$,
ranging from $-1$ to $1$. 
\end{example}

To evaluate the effect of adjuvant chemotherapy, let $Y$ be the indication
of cancer recurrence within one year of surgery. Consider the HTE
function in Example \ref{eg. binary outcome} with $Z=(1,{\rm age},{\rm tumor}\ {\rm size})^{\T}$
and $\psi_{0}=(\psi_{0,0},\psi_{0,1},\psi_{0,2})^{\T}$. This model
entails that, on average, the treatment would increase or decrease
the risk of cancer recurrence by $|\tau_{\psi_{0}}(Z)|$ had the patient
received adjuvant chemotherapy, and the magnitude of increase depends
on age and tumor size. If $Z^{\T}\psi_{0}<0$, it indicates that the
treatment is beneficial for this patient. Moreover, if $\psi_{0,1}<0$
and $\psi_{0,2}<0$, then older patients with larger tumor sizes would
benefit more from adjuvant chemotherapy.

We consider two independent data sources: one from the RT study and
the other from the RW study. Let $\delta=1$ denote RT participation,
and let $\delta=0$ denote RW study participation. Let $V$ summarize
the entire record of observed variables $(A,X,\delta,Y)$. The RT
data consist of $\{V_{i}:i\in\mathcal{A}\}$ with sample size $m$,
and the RW data consist of $\{V_{i}:i\in\mathcal{B}\}$ with sample
size $n$, where $\mathcal{A}$ and $\mathcal{B}$ are sample index
sets for the two data sources. Our setup requires the RT and RW samples
to contain $Z$'s information but may include different sets of auxiliary
information in $X$. The total sample size is $N=m+n$. Generally,
$n$ is larger than $m$. In our asymptotic framework, we assume both
$m$ and $n$ go to infinity, and $m/n\rightarrow\rho$, where $0<\rho<1$.

For simplicity of exposition, we use the following notations throughout
the paper: $\mathbb{P}_{N}$ denotes the empirical measure over the
combined RT and RW data, $M^{\otimes2}$ denotes $MM^{\T}$ for a
vector or matrix $M$, $\E_{a}(\cdot)$ and $\V_{a}(\cdot)$ are the
asymptotic expectation and variance of a random variable, $A_{n}\indep B_{n}$
denotes $A_{n}$ is independent of $B_{n}$, $A_{n}\sim B_{n}$ denotes
that $A_{n}$ follows the same distribution as $B_{n}$, and $A_{n}\stackrel{\cdot}{\sim}B_{n}$
denotes that $A_{n}$ and $B_{n}$ have the same asymptotic distribution
as $n\rightarrow\infty$. Let $e_{\delta}(X)=\pr(A=1\mid X,\delta)$
be the propensity score.

\subsection{Identification of the HTE from the RT and RW data}

The fundamental problem of causal inference is that $Y(0)$ and $Y(1)$
are not jointly observable. Therefore, the HTE is not identifiable
without additional assumptions.

\textcolor{black}{We view the RT sample as the gold standard for HTE
estimation, satisfying the following assumption. }

\begin{assumption}[RT validity]\label{Asump:rand-rct}(i) $\E\{Y(1)-Y(0)\mid X,\delta=1\}=\tau(Z)$,
and (ii) $Y(a)\indep A\mid(X,\delta=1)$ for $a\in\{0,1\}$ and $0<e_{1}(X)<1$
for all $(X,\delta=1)$.

\end{assumption}

Assumption \ref{Asump:rand-rct}(i) states that the HTE function is
transportable from the RT sample to the target population. This assumption
is a common assumption in the data integration literature. Stronger
versions of Assumption \ref{Asump:rand-rct}(i) have also been considered
in the literature, including the ignorability of study participation,
i.e., $\{Y(0),Y(1)\}\indep\delta\mid X$ \citep{stuart2011use,buchanan2018generalizing},
or the mean exchangeability, i.e., $\E\{Y(a)\mid X,\delta\}=\E\{Y(a)\mid X\}$
for $a=0,1$ \citep{dahabreh2019generalizing}. Assumption \ref{Asump:rand-rct}(i)
holds if $Z$ captures the heterogeneity of effect modifiers or if
the study sample is a random sample from the target population. Under
the structural equation model framework, \citet{pearl2011transportability}
provided graphical conditions for transportability. The graphical
representation can aid the investigator in assessing the plausibility
of Assumption \ref{Asump:rand-rct}(i). Assumption \ref{Asump:rand-rct}(ii)
entails that treatment assignment in the RT study follows a randomization
mechanism based on the pre-treatment variables $X$, and all subjects
have positive probabilities of receiving each treatment. Assumption
\ref{Asump:rand-rct}(ii) holds by the design of complete randomization
of treatment, where the treatment is independent of the potential
outcomes and covariates, i.e., ${\color{black}\{Y(a),X\}}\indep A\mid\delta=1$.
It also holds by the design of stratified block randomization of treatment
based on discrete $X$, where the treatment is independent of the
potential outcomes within each stratum of $X$. The propensity score
$e_{1}(X)$ is known by design.

\textcolor{black}{We consider a parallel assumption for the RW sample,
termed RW comparability. }

\begin{assumption}[RW comparability]\label{Asump:rand-rwd}\textcolor{black}{(i)
$\E\{Y(1)-Y(0)\mid X,\delta=0\}=\tau(Z)$,{} }and (ii) $Y(a)\indep A\mid(X,\delta=0)$
for $a\in\{0,1\}$ and $0<e_{0}(X)<1$ for all $(X,\delta=0)$.

\end{assumption}

\textcolor{black}{Although Assumption \ref{Asump:rand-rwd} appears
similar to Assumption \ref{Asump:rand-rct}, its implications differ
substantively. Assumption \ref{Asump:rand-rwd}(i) states that the
HTE function is transportable from the RW sample to the target population.
To make this assumption more plausible, one can use the same trial
eligibility criteria to select the RW sample to ensure a sufficient
overlap of the RW covariate space with the RT sample. However, this
assumption can be violated in various ways. For example, RT and RW
studies may be conducted in different care settings (large academic
medical centers versus smaller community hospitals), contexts (geography,
policy-related or socio-structural factors), or time frames. Each
of these concerns can violate Assumption \ref{Asump:rand-rwd}(i).
In addition, due to the lack of control of treatment assignment in
RW data, Assumption \ref{Asump:rand-rwd}(ii) implies that the observed
covariates $X$ capture all the confounding variables related to the
treatment and outcome. This assumption may also be restrictive in
practice. For example, in the NCDB cohort, the physicians or patients
decided, based on experiences or preferences, whether patients received
adjuvant chemotherapy after tumor resection. While the database captures
many site-level and patient-level information, there may be unmeasured
confounding variables that associate with the treatment selection
and clinical outcome, e.g., financial status and accessibility to
health care facilities.}

By trial design, we assume Assumption \ref{Asump:rand-rct} for the
RT data holds throughout the paper; however, we regard Assumption
\ref{Asump:rand-rwd} for the RW data as an idealistic assumption,
which may be violated. If Assumption \ref{Asump:rand-rwd} holds,
we will use a semiparametric efficient strategy to combine both data
sources for optimal estimation. However, if Assumption \ref{Asump:rand-rwd}
is violated, our proposed method will automatically detect the violation
and retain only the RT data for estimation. \textcolor{black}{In practice,
it is important to identify a ``similar'' RW sample to be integrated
with the RT sample. \citet{hernan2016using} provided a framework
for using big real-world data to emulate a target trial when a randomized
trial is unavailable. When selecting an RW sample, we can check the
rubrics for the eligibility criteria that defines the target population,
treatment definitions, assignment procedures, follow-up time, outcome,
and effect contrast of interest, to increase the chance of successfully
integrating the RW sample with the RT sample. }

\textcolor{black}{Unlike our focus on testing the comparability of
the RW in HTE estimation, testing transportability alone may be of
more importance in some contexts. Under Assumptions \ref{Asump:rand-rct}(ii)
and \ref{Asump:rand-rwd}(ii), i.e., the treatment ignorability holds,
possible tests can be adopted to test $\E\{Y(1)-Y(0)\mid X,\delta=1\}=\E\{Y(1)-Y(0)\mid X,\delta=0\}$,
e.g., the U-statistics-based test \citep{luedtke2019omnibus}. }

Under Assumptions \ref{Asump:rand-rct} and \ref{Asump:rand-rwd},
the following identification formula holds for the HTE: 
\begin{equation}
\E\left\{ \left.\frac{AY}{e_{\delta}(X)}-\frac{(1-A)Y}{1-e_{\delta}(X)}\right\vert Z,\delta\right\} =\tau(Z).\label{eq:ipw}
\end{equation}
The identification formula motivates regression analysis based on
the modified outcome $A\{e_{\delta}(X)\}^{-1}Y-(1-A)\{1-e_{\delta}(X)\}^{-1}Y$
to estimate the HTE. This approach involves the inverse of the treatment
probability, and thus the resulting estimator may be unstable if some
estimated treatment probabilities are close to zero or one. It calls
for a principled way to construct improved estimators of the HTE.
\citet{rudolph2017robust} derived the semiparametric efficiency score
(SES) and bound for the average treatment effect. In the next subsection,
we derive the SES of the HTE under Assumptions \ref{Asump:rand-rct}
and \ref{Asump:rand-rwd} that motivates improved estimators.

\subsection{Semiparametric efficiency score\label{subsec:The-efficient-score} }

The semiparametric model consists of model (\ref{eq:SNMM}) with the
parameter of interest $\psi_{0}$ and the unspecified distribution.\textcolor{black}{{}
Assumptions \ref{Asump:rand-rct} and \ref{Asump:rand-rwd}} impose
restrictions on $\psi_{0}$. To see this, define 
\begin{equation}
H_{\psi}=Y-\tau_{\psi}(Z)A.\label{eq:def of H(k)}
\end{equation}
Intuitively, $H_{\psi_{0}}$ subtracts from the subject's observed
outcome $Y$ the treatment effect of the subject's observed treatment
$\tau_{\psi_{0}}(Z)A$, which mimics the potential outcome $Y(0)$.
Formally, following \citet{robins1994correcting}, we can show that
$\E(H_{\psi_{0}}\mid A,X,\delta)=\E\{Y(0)\mid A,X,\delta\}$. Therefore,
by Assumptions \ref{Asump:rand-rct} and \ref{Asump:rand-rwd}, $\psi_{0}$
must satisfy the restriction\textit{: 
\begin{equation}
\E(H_{\psi_{0}}\mid A,X,\delta)=\E(H_{\psi_{0}}\mid X,\delta).\label{eq:model-part1}
\end{equation}
}For simplicity of exposition, denote 
\[
\E(H_{\psi_{0}}\mid X,\delta)=\mu_{\delta}(X),\quad\V(H_{\psi_{0}}\mid X,\delta)=\sigma_{\delta}^{2}(X),
\]
where $\mu_{\delta}(X)$ is the outcome mean function and $\sigma_{\delta}^{2}(X)$
is the outcome variance function. By viewing $(X,\delta)$ jointly
as the set of confounders, we invoke the SES of the structural nested
mean model in \citet{robins1994correcting}. We further make a simplifying
assumption that 
\begin{equation}
\E(H_{\psi_{0}}^{2}\mid A,X,\delta)=\E(H_{\psi_{0}}^{2}\mid X,\delta),\label{eq:simplyassumption}
\end{equation}
which is a natural extension of (\ref{eq:model-part1}). This assumption
allows us to derive the SES of $\psi_{0}$ as 
\begin{equation}
S_{\psi_{0}}(V)=q^{*}(X,\delta)\{H_{\psi_{0}}-\mu_{\delta}(X)\}\{A-e_{\delta}(X)\},\ \ q^{*}(X,\delta)=\left\{ \partial\tau_{\psi_{0}}(Z)/\partial\psi\right\} \left\{ \sigma_{\delta}^{2}(X)\right\} ^{-1},\label{eq:eff psi}
\end{equation}
which separates the term with the outcome, i.e., $H_{\psi_{0}}-\mu_{\delta}(X)$,
and the term with the treatment, i.e., $A-e_{\delta}(X)$. This feature
relaxes model assumptions of the nuisance functions while retaining
root-$n$ consistency in the estimation of $\psi_{0}$; see Section
\ref{subsec:From-SES}. Even without the simplifying assumption in
(\ref{eq:simplyassumption}), by the mean independence property in
(\ref{eq:model-part1}), we can verify that 
\[
\E\{S_{\psi_{0}}(V)\}=\E[q^{*}(X,\delta)\E\{H_{\psi_{0}}-\mu_{\delta}(X)\mid X,\delta\}\times\E\{A-e_{\delta}(X)\mid X,\delta\}]=0.
\]
Therefore, if (\ref{eq:simplyassumption}) holds, $S_{\psi_{0}}(V)$
is the SES of $\psi_{0}$; if (\ref{eq:simplyassumption}) does not
hold, $S_{\psi_{0}}(V)$ is unbiased and permits robust estimation.
We provide examples to elucidate the SES below before delving into
robust estimation in the following subsection. 
\begin{example}
\label{example(A3.2)} For a continuous outcome and the HTE function
given in Example \ref{eg. continuous outcome}, the SES of $\psi_{0}$
is 
\[
S_{\psi_{0}}(V)=Z\left\{ \sigma_{\delta}^{2}(X)\right\} ^{-1}\{H_{\psi_{0}}-\mu_{\delta}(X)\}\{A-e_{\delta}(X)\}.
\]
For a binary outcome and the HTE function given in Example \ref{eg. binary outcome},
the SES of $\psi_{0}$ is 
\[
S_{\psi_{0}}(V)=Z\frac{2\exp(Z^{\T}\psi_{0})}{\{\exp(Z^{\T}\psi_{0})+1\}^{2}}[\mu_{\delta}(X)\{1-\mu_{\delta}(X)\}]^{-1}\{H_{\psi_{0}}-\mu_{\delta}(X)\}\{A-e_{\delta}(X)\}.
\]
\end{example}

\begin{remark}[Comparison with\textcolor{black}{{} other doubly
robust}\textcolor{red}{{} }approaches]\label{rmk:aipw}

The identification formula (\ref{eq:ipw}) motivates the inverse probability
weighted (IPW)-adjusted regression. However, IPW is known to be inefficient
and sensitive to model misspecification of the propensity score. \textcolor{black}{Alternatively,
\citet{kennedy2020optimal} proposed a pseudo-outcome regression approach
using augmented IPW (AIPW) pseudo-outcomes that leverages weighting
and outcome mean functions and improves the performance of IPW-adjusted
regression. The doubly robust loss function for the treatment contrast
or blip function in \citet{luedtke2016super} also exploits weighting
and outcome mean functions. }Both IPW and AIPW use weighting to remove
confounding biases; differently, the SES in (\ref{eq:eff psi}) uses
the mean independence of $H_{\psi_{0}}-\mu_{\delta}(X)$ and $A-e_{\delta}(X)$
to construct unbiased estimating equations. The simulation study in
Section \ref{subsec:Comparing-AIPW-and-SES} shows that the SES approach
outperforms the AIPW-adjusted approach when the propensity score can
be close to zero or one.

\end{remark}

\subsection{From SES to robust estimation\label{subsec:From-SES}}

In principle, an efficient estimator for $\psi_{0}$ can be obtained
by solving $\mathbb{P}_{N}S_{\eff,\psi}(V)=0$. However, $S_{\eff,\psi}$
depends on the unknown distribution through $e_{0}(X)$, $\mu_{\delta}(X)$,
and $\sigma_{\delta}^{2}(X)$, and thus solving $\mathbb{P}_{N}S_{\eff,\psi}(V)=0$
is infeasible. Nevertheless, the state-of-art causal inference literature
suggests that estimators constructed based on SES are robust to approximation
errors using machine learning methods, the so-called rate double robustness;
see, e.g., \citet{chernozhukov2018double} and \citet{rotnitzky2019characterization}.

In order to obtain a robust estimator with good efficiency properties,
we consider approximating the unknown functions using non-parametric
or machine learning methods. In summary, our algorithm for the estimation
of $\psi_{0}$ proceeds as follows. 
\begin{description}
\item [{Step$\ 1.$}] Obtain an estimator of $e_{0}(X)$ using non-parametric
or machine learning methods, denoted by $\widehat{e}_{0}(X),$ based
on $\{(A_{i},X_{i},\delta_{i}=0):i\in\mathcal{B}\}$. 
\item [{Step$\ 2.$}] Obtain a preliminary estimator $\widehat{\psi}_{\mathrm{p}}$
by solving $\sum_{i\in\mathcal{A}}[q^{*}(X_{i},\delta_{i})\{A_{i}-e_{1}(X_{i})\}H_{\psi,i}]=0$,
based on $\{(A_{i},X_{i},Y_{i},\delta_{i}=1):i\in\mathcal{A}\}$. 
\item [{Step$\ 3.$}] Obtain the estimators of $\mu_{1}(X)$ and $\mu_{0}(X)$
using non-parametric or machine learning methods, denoted by $\widehat{\mu}_{1}(X)$
and $\widehat{\mu}_{0}(X)$, based on $\{(A_{i},X_{i},H_{\widehat{\psi}_{\mathrm{p}},i},$
$\delta_{i}=1):i\in\mathcal{A}\}$ and $\{(A_{i},X_{i},H_{\widehat{\psi}_{\mathrm{p}},i},\delta_{i}=0):i\in\mathcal{B}\}$,
respectively. 
\item [{Step$\ 4.$}] Let $\widehat{S}_{\eff,\psi}(V)$ be $S_{\eff,\psi}(V)$
with the unknown quantities replaced by the estimated parametric models
in Steps 1 and 3. Obtain the efficient integrative estimator $\widehat{\psi}_{\eff}$
by solving 
\begin{equation}
\mathbb{P}_{N}\widehat{S}_{\psi}(V)=0.\label{eq:optimal ee}
\end{equation}
\end{description}
The estimator $\widehat{\psi}_{\eff}$ depends on the approximation
of nuisance functions. To establish the asymptotic properties of $\widehat{\psi}_{\eff}$,
we provide the regularity conditions.

\begin{assumption}\label{assump:o1} (i) $\Vert\widehat{e}_{0}(X)-e_{0}(X)\Vert=o_{\pp}(1)$
and $\Vert\widehat{\mu}_{\delta}(X)-\mu_{\delta}(X)\Vert=o_{\pp}(1)$;
(ii) $\Vert\widehat{e}_{0}(X)-e_{0}(X)\Vert\times\Vert\widehat{\mu}_{\delta}(X)-\mu_{\delta}(X)\Vert=o_{\pp}(n^{-1/2})$;
and (iii) additional regularity conditions in Assumption \ref{asump:DML}.\end{assumption}

Assumption \ref{assump:o1} is typical regularity conditions for Z-estimation
or M-estimation \citep{Vaart2000}. Assumption \ref{assump:o1}(i)
states that we require the posited models to be consistent for the
two nuisance functions. Assumption \ref{assump:o1}(ii) states that
the \textit{combined rate} of convergence of the posited models is
$o_{\pp}(n^{-1/2})$. Assumption \ref{asump:DML} regularizes the
complexity of the functional space. Importantly, these conditions
ensure $\widehat{\psi}_{\eff}$ retains the parametric-rate consistency,
allowing flexible data-adaptive models and not restricting to stringent
parametric models.

\begin{theorem}\label{Thm:Consistency-DML}

Suppose Assumptions \ref{Asump:rand-rct}--\ref{assump:o1} hold.
Then, $\widehat{\psi}_{\eff}$ is root-$n$ consistent for $\psi_{0}$
and asymptotically normal.

\end{theorem}

Theorem \ref{Thm:Consistency-DML} implies that asymptotically, $\widehat{\psi}_{\eff}$
can be viewed as the solution to $\mathbb{P}_{N}S_{\psi}(V)=0$ when
the nuisance functions are known. Therefore, for consistent variance
estimation of $\widehat{\psi}_{\eff}$, we can use the standard sandwich
formula \citep{stefanski2002calculus} or the perturbation-based resampling
\citep{hu2000estimating}, treating the nuisance functions to be known.

\section{Test-based elastic integrative analysis\label{subsec: robust and elastic integ}}

A major concern for integrating the RT and RW data lies in the possibly
poor quality of the RW data. Then, combining the RT and RW data into
an integrative analysis would lead to a biased HTE estimator. This
section addresses the critical challenge of preventing any biases
present in the RW data from leaking into the proposed estimator.

\subsection{Detection of the RW incompatibility \label{subsec:Detection-of-violation}}

We consider all assumptions in Theorem \ref{Thm:Consistency-DML}
hold except that Assumption \ref{Asump:rand-rwd} may be violated.
We derive a test that detects the violation of this crucial assumption
for using the RW data. For simplicity, we denote the SES based solely
on the RT or RW data as 
\[
S_{\rt,\psi}(V)=\delta S_{\psi}(V),\ S_{\os,\psi}(V)=(1-\delta)S_{\psi}(V),
\]
respectively. Moreover, let $\widehat{S}_{\rt,\psi}(V)$ and $\widehat{S}_{\os,\psi}(V)$
be $S_{\rt,\psi}(V)$ and $S_{\os,\psi}(V)$ with the nuisance functions
replaced by their estimates, and let ${\cI}_{\rt}=\E\{S_{\rt,\psi_{0}}(V)^{\otimes2}\mid\delta=1\}$
and ${\cI}_{\rw}=\E\{S_{\os,\psi_{0}}(V)^{\otimes2}\mid\delta=0\}$
be Fisher information matrices.

We now formulate the null hypothesis ${\rm H}_{0}$ for the case when
Assumption \ref{Asump:rand-rwd} holds and fixed and local alternatives
${\rm H}_{a}$ and ${\rm H}_{a,n}$ for the case when Assumption \ref{Asump:rand-rwd}
is violated: 
\begin{description}
\item [{${\rm H}_{0}$}] (Null) $\E\{S_{\os,\psi_{0}}(V)\}=0$. 
\item [{${\rm H}_{a}$}] (Fixed alternative) $\E\{S_{\os,\psi_{0}}(V)\}=\eta_{{\rm fix}}$
, where $\eta_{{\rm fix}}$ is a $p$-vector of constants with at
least one nonzero component. 
\item [{${\rm H}_{a,n}$}] (Local alternative) $\E\{S_{\os,\psi_{0}}(V)\}=n^{-1/2}\eta$
, where $\eta$ is a $p$-vector of constants with at least one nonzero
component. 
\end{description}
Considering the fixed alternative is common to establish asymptotic
properties of standard estimators and tests; however, the local alternative
is useful to study finite-sample properties and regularity of non-standard
estimators and tests. In finite samples, the violation of Assumption
\ref{Asump:rand-rwd} may be weak; e.g., there exists a hidden confounder
in the RW data, but the association between the hidden confounder
and the outcome or the treatment is small. \textcolor{black}{In such
cases, the test statistic can be small or moderate. The fixed alternative
formulates the bias of the RW score to be fixed, implying that the
test statistic goes to infinity with the sample size. Consequently,
the fixed alternative inference can not capture the finite-sample
behavior well in the cases of weak violation and does not have uniform
validity. That is, there exist scenarios where the finite-sample coverage
probability from standard inference is far from the nominal level
for any sample size. The local alternative asymptotics is a common
approach to obtaining uniform inference validity for non-regular estimators.}
In the local alternative ${\rm H}_{a,n}$, the bias of $S_{\os,\psi_{0}}(V)$
may be small as quantified by $n^{-1/2}\eta$. The values of $\eta$
represent different tracks that the bias of $S_{\os,\psi_{0}}(V)$
follows to \textcolor{black}{converge} to zero. \textcolor{black}{We
will show that the test statistic is $O_{\pr}(1)$, thus better capturing
the finite-sample behavior in the weak violation cases.} The local
alternative encompasses the null and fixed alternative as special
cases by considering different values of $\eta$. In particular, ${\rm H}_{0}$
corresponds to ${\rm H}_{a,n}$ with $\eta=0$. Also, ${\rm H}_{a}$
corresponds to ${\rm H}_{a,n}$ with $\eta=\pm\infty$; hence, considering
${\rm H}_{a}$ alone is not informative about the finite-sample behaviors
of the proposed test and estimator.

We detect biases in the RW data based on the following two key insights.
First, we obtain an initial estimator $\widehat{\psi}_{\rt}$ by solving
the estimating equation based solely on the RT data, $\sum_{i\in\mathcal{A}}\widehat{S}_{\rt,\psi}(V_{i})=0$.
It is important to emphasize that the propensity score in the RT $e_{1}(X)$
is known by design and, therefore, $\widehat{\psi}_{\rt}$ is always
consistent. Second, if Assumption \ref{Asump:rand-rwd} holds for
the RW data, $S_{\os,\psi_{0}}(V)$ is unbiased, but $S_{\os,\psi_{0}}(V)$
is no longer unbiased if it is violated. Therefore, large values of
$n^{-1/2}\sum_{i\in\mathcal{B}}\widehat{S}_{\rw,\widehat{\psi}_{\rt}}(V_{i})$
provide evidence of the violation of Assumption \ref{Asump:rand-rwd}.

To detect the violation of Assumption \ref{Asump:rand-rwd} for using
the RW data, we construct the test statistic 
\begin{equation}
T=\left\{ n^{-1/2}\sum_{i\in\mathcal{B}}\widehat{S}_{\os,\widehat{\psi}_{\rt}}(V_{i})\right\} ^{\T}\widehat{\Sigma}_{SS}^{-1}\left\{ n^{-1/2}\sum_{i\in\mathcal{B}}\widehat{S}_{\os,\widehat{\psi}_{\rt}}(V_{i})\right\} ,\label{eq:T}
\end{equation}
where $\Sigma_{SS}=\Gamma^{\T}{\cI}_{\rt}\Gamma+{\cI}_{\rw}$ is the
asymptotic variance of $n^{-1/2}\sum_{i\in\mathcal{B}}\widehat{S}_{\os,\widehat{\psi}_{\rt}}(V_{i})$,
$\Gamma={\cI}_{\rt}^{-1}{\cI}_{\rw}\rho^{-1/2}$, and $\widehat{\Sigma}_{SS}$
is a consistent estimator for $\Sigma_{SS}$. The test statistic $T$
measures the distance between $n^{-1/2}\sum_{i\in\mathcal{B}}S_{\os,\widehat{\psi}_{\rt}}(V_{i})$
and zero. If the idealistic assumption holds, we expect $T$ to be
small. By the standard asymptotic theory, we show in the supplementary
material that under ${\rm H}_{0},$ $T\stackrel{\cdot}{\sim}\chi_{p}^{2},$
a Chi-square distribution with degrees of freedom $p$, as $n\rightarrow\infty$.
This result serves to detect the violation of the assumption required
for the RW data.

\subsection{Elastic integration\label{subsec:Elastic} }

Let $c_{\gamma}=\chi_{p,\gamma}^{2}$ be the $100(1-\gamma)$th percentile
of $\chi_{p}^{2}$. For a small $\gamma,$ if $T\geq c_{\gamma}$,
there is strong evidence to reject ${\rm H}_{0}$ for the RW data;
i.e., there is a detectable bias for the RW data estimator. In this
case, we would only use the RT data for estimation. On the other hand,
if $T<c_{\gamma}$, there is no strong evidence that the RW data estimator
is biased; therefore, we would combine both the RT and RW data for
optimal estimation. Our strategy leads to the elastic integrative
estimator $\widehat{\psi}_{\elas}$ solving 
\begin{equation}
\sum_{i\in\mathcal{A}\cup\mathcal{B}}\left\{ \delta_{i}\widehat{S}_{\psi}(V_{i})+\bone(T<c_{\gamma})(1-\delta_{i})\widehat{S}_{\psi}(V_{i})\right\} =0.\label{eq:EE}
\end{equation}
The choice of $\gamma$ involves the bias-variance tradeoff. On the
one hand, under ${\rm H}_{0}$, the acceptance probability of integrating
the RW data is $\pr(T<c_{\gamma})=1-\gamma$. Therefore, for a relatively
large sample size, we will accept good-quality RW data with probability
$1-\gamma$ and reject good-quality RW data with type I error $\gamma$.
Hence, a small $\gamma$ is desirable; similarly, for ${\rm H}_{a,n}$
with small $\eta$. On the other hand, under ${\rm H}_{a,n}$ with
large $\eta$, the reverse is true, and hence a large $\gamma$ is
desirable.

To formally investigate the tradeoff, we characterize the asymptotic
distributions of the elastic integrative estimator $\widehat{\psi}_{\elas}$
under the null, fixed, and local alternatives. We do not discuss the
trivial cases when $\gamma=0$ and $1$, corresponding to $\widehat{\psi}_{\elas}=\widehat{\psi}_{\rt}$
or $\widehat{\psi}_{\eff}$. With $\gamma\in(0,1),$ $\widehat{\psi}_{\elas}$
mixes two distributions, namely, $\widehat{\psi}_{\rt}\mid(T\geq c_{\gamma})$
and $\widehat{\psi}_{\eff}\mid(T<c_{\gamma})$. \textcolor{black}{Each
distribution can be non-standard because the estimators and test are
constructed based on the same data and, therefore, may be asymptotically
dependent.}

To characterize those non-standard distributions, we decompose this
task into three steps. First, by the standard asymptotic theory, it
follows that $T\stackrel{\cdot}{\sim}\cZ_{1}^{\T}\cZ_{1}$, where
$\cZ_{1}$ is a standard $p$-variate normal random vector, $n^{1/2}(\widehat{\psi}_{\rt}-\psi_{0})\stackrel{\cdot}{\sim}\text{\ensuremath{\mathcal{N}_{\rt}}},$
and $n^{1/2}(\widehat{\psi}_{\eff}-\psi_{0})\stackrel{\cdot}{\sim}\mathcal{N}_{\rtrw},$
where $\mathcal{N}_{\rt}$ and $\mathcal{\mathcal{N}_{\eff}}$ are
some $p$-variate normal random vectors with variances $V_{\rt}=(\rho{\cI}_{\rt})^{-1}$
and $V_{\eff}=(\rho{\cI}_{\rt}+{\cI}_{\rw})^{-1}$, respectively.

\begin{figure}
\centering{}\includegraphics[scale=0.4]{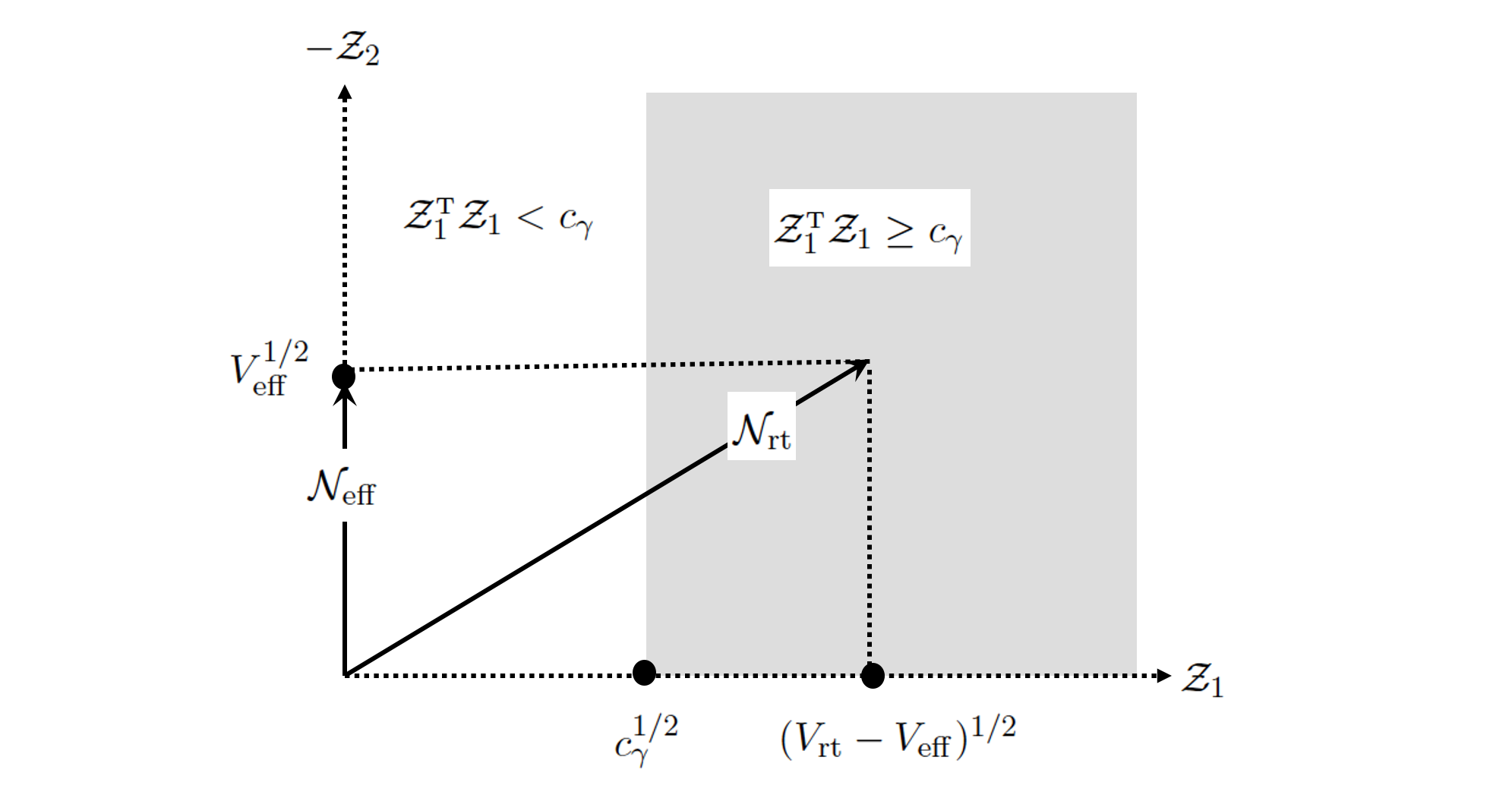} 
\begin{itemize}
\item {\scriptsize{}{}$\mathcal{N}_{\rt}=V_{\rteff}^{1/2}\cZ_{1}-V_{\eff}^{1/2}\cZ_{2}$
and $\mathcal{N}_{\rt}\mid(\cZ_{1}^{\T}\cZ_{1}\geq c_{\gamma})\sim V_{\rteff}^{1/2}\cZ_{1}\mid(\cZ_{1}^{\T}\cZ_{1}\geq c_{\gamma})-V_{\eff}^{1/2}\cZ_{2}$ }{\scriptsize\par}
\item {\scriptsize{}{}$\mathcal{N}_{\eff}=-V_{\eff}^{1/2}\cZ_{2}$ and
$\mathcal{N}_{\eff}\mid(\cZ_{1}^{\T}\cZ_{1}<c_{\gamma})\sim\mathcal{N}_{\eff}$}{\scriptsize\par}
\end{itemize}
\caption{\label{fig:representation}Representation of the normal distributions
$\mathcal{N}_{\rt}$ and $\mathcal{N}_{\eff}$ based on $\cZ_{1}$
and $\cZ_{2}$ with $p=1$.}
\end{figure}

Second, we find another standard $p$-variate normal random vector
$\cZ_{2}$ that is independent of $\cZ_{1}$, and decompose the normal
distributions $\text{\ensuremath{\mathcal{N}_{\rt}}}$ and $\mathcal{N}_{\rtrw}$
into two orthogonal components: i) one corresponds to $\cZ_{1}$ and
ii) the other one corresponds to $\cZ_{2}$. Importantly, component
i) would be affected by the test constraints induced by $\cZ_{1}^{\T}\cZ_{1}$,
but component ii) would not be affected. For $\mathcal{N}_{\rtrw},$
we show that it is fully represented by $\cZ_{2}$ as $\mathcal{N}_{\rtrw}=-V_{\eff}^{1/2}\cZ_{2}.$
Therefore, its distribution is not affected by $\cZ_{1}^{\T}\cZ_{1}<c_{\gamma}$;
that is, 
\[
\mathcal{N}_{\rtrw}\mid(\cZ_{1}^{\T}\cZ_{1}<c_{\gamma})\sim-V_{\eff}^{1/2}\cZ_{2}.
\]
For $\text{\ensuremath{\mathcal{N}_{\rt}}}$, we show that $\text{\ensuremath{\mathcal{N}_{\rt}}}=V_{\rteff}^{1/2}\cZ_{1}-V_{\eff}^{1/2}\cZ_{2}$
with $V_{\rteff}=V_{\rt}-V_{\eff}$. Due to the independence between
$\cZ_{1}$ and $\cZ_{2}$, $\text{\ensuremath{\mathcal{N}_{\rt}}}\mid(\cZ_{1}^{\T}\cZ_{1}\geq c_{\gamma})$
is a mixture distribution 
\[
\mathcal{N}_{\rt}\mid(\cZ_{1}^{\T}\cZ_{1}\geq c_{\gamma})\sim V_{\rteff}^{1/2}\cZ_{c_{\gamma}}^{\rmt}-V_{\eff}^{1/2}\cZ_{2},
\]
mixing a non-normal component, where $\cZ_{c}^{\rmt}$ represents
the truncated normal distribution $\cZ_{1}\mid(\cZ_{1}^{\T}\cZ_{1}\geq c)$,
and a normal component. For illustration, Figure \ref{fig:representation}
demonstrates the geometry of the decomposition of distributions with
scalar variables.

Third, we formally characterize the distribution of $\cZ_{c}^{\rmt}$,
a multivariate normal distribution with ellipsoid truncation \citep{tallis1963elliptical,li2018asymptotic}.
This step enables us to quantify the asymptotic bias and variance
of the proposed estimator; see Section \ref{subsec:Asymptotic-bias-and}.

Let $F_{p}(\cdot)$ be the cumulative distribution function (CDF)
of a $\chi_{p}^{2}$ random variable, and $F_{p}(\cdot;\lambda)$
be the CDF of a $\chi_{p}^{2}(\lambda)$ random variable, where $\chi_{p}^{2}$
and $\chi_{p}^{2}(\lambda)$ are the central Chi-square distribution
and the non-central Chi-square distribution with the non-centrality
parameter $\lambda$, respectively. Theorem \ref{Thm:elas} summarizes
the asymptotic distribution of $\widehat{\psi}_{\elas}$.

\begin{theorem}\label{Thm:elas} Suppose assumptions in Theorem \ref{Thm:Consistency-DML}
hold except that Assumption \ref{Asump:rand-rwd} may be violated.
Let $\cZ_{1}$ and $\cZ_{2}$ be independent normal random vectors
with mean $\mu_{1}=$$\Sigma_{SS}^{-1/2}\eta$ and $\mu_{2}=$$V_{\eff}^{1/2}\eta$,
respectively, and covariance $I_{p\times p}$. Let $\cZ_{c}^{\rmt}$
be the truncated normal distribution $\cZ_{1}\mid(\cZ_{1}^{\T}\cZ_{1}\geq c)$.
Let the elastic integrative estimator $\widehat{\psi}_{\elas}$ be
obtained by solving (\ref{eq:EE}). Then, $n^{1/2}(\widehat{\psi}_{\elas}-\psi_{0})$
has a limiting mixture distribution 
\begin{equation}
\mathcal{M}(\gamma;\eta)=\begin{cases}
\mathcal{M}_{1}(\gamma;\eta)=V_{\rteff}^{1/2}\cZ_{c_{\gamma}}^{\rmt}-V_{\eff}^{1/2}\cZ_{2}, & \text{w.p. }\xi,\\
\mathcal{M}_{2}(\eta)=-V_{\eff}^{1/2}\cZ_{2}, & \text{w.p. }1-\xi,
\end{cases}\label{eq:MixC1}
\end{equation}

\begin{enumerate}
\item Under ${\rm H}_{0},$ $\mu_{1}=\mu_{2}=0$ and $\xi=1-F_{p}(c_{\gamma})=\gamma$. 
\item Under ${\rm H}_{a},$ $\mu_{1}=\mu_{2}=\pm\infty$ and $\xi=1$; i.e.,
(\ref{eq:MixC1}) reduces to a normal distribution with mean $0$
and variance $V_{\rt}$. 
\item Under ${\rm H}_{a,n},$ $\mu_{1}=\Sigma_{SS}^{-1/2}\eta$ , $\mu_{2}=V_{\eff}^{1/2}\eta$
with $\eta\in\R^{p}$, and $\xi=1-F_{p}(c_{\gamma};\lambda)$, where
$\lambda=\eta^{\T}\Sigma_{SS}^{-1}\eta$. 
\end{enumerate}
\end{theorem}

In Theorem \ref{Thm:elas}, $\mathcal{M}(\gamma;\eta)$ in (\ref{eq:MixC1})
\textcolor{black}{is} a general characterization of the asymptotic
distribution of $n^{1/2}(\widehat{\psi}_{\elas}-\psi_{0})$. It implies
different asymptotic behaviors of $n^{1/2}(\widehat{\psi}_{\elas}-\psi_{0})$
depending on whether Assumption \ref{Asump:rand-rwd} is strongly,
weakly, or not violated. First, ${\rm H}_{a}$ corresponds to the
situation where Assumption \ref{Asump:rand-rwd} is strongly violated.
Under ${\rm H}_{a}$, $T$ rejects the RW data (i.e., $\cZ_{1}^{\T}\cZ_{1}\geq c_{\gamma}$
holds) with probability \textcolor{black}{converging} to one, $\cZ_{c_{\gamma}}^{\rmt}$
becomes $\cZ_{1}$, and $\mathcal{M}(\gamma;\eta=\pm\infty)$ becomes
$V_{\rteff}^{1/2}\cZ_{1}-V_{\eff}^{1/2}\cZ_{2}$, a normal distribution
with mean $0$ and variance $V_{\rt}$. As expected, under ${\rm H}_{a}$,
$n^{1/2}(\widehat{\psi}_{\elas}-\psi_{0})$ is asymptotically normal
and regular. Second, ${\rm H}_{0}$ and ${\rm H}_{a,n}$ correspond
to the situations when Assumption \ref{Asump:rand-rwd} is not and
weakly violated, respectively. Under ${\rm H}_{0}$ and ${\rm H}_{a,n}$,
$T$ has positive probabilities of accepting and rejecting the RW
data, $\widehat{\psi}_{\elas}$ switches between $\widehat{\psi}_{\eff}$
and $\widehat{\psi}_{\rt}$, and $n^{1/2}(\widehat{\psi}_{\elas}-\psi_{0})$
follows a limiting mixing distribution $\mathcal{M}(\gamma;\eta)$,
indexed by $\eta$. Although the exact form of $\mathcal{M}(\gamma;\eta)$
is complicated, the entire distribution and summary statistics such
as mean, variance, and quantiles can be simulated by rejective sampling.
Importantly, under ${\rm H}_{0}$ and ${\rm H}_{a,n}$, $n^{1/2}(\widehat{\psi}_{\elas}-\psi_{0})$
is non-normal and non-regular. The non-regularity is determined by
the local parameter $\eta$, which entails that the asymptotic distribution
of $n^{1/2}(\widehat{\psi}_{\elas}-\psi_{0})$ may change abruptly
when ${\rm H}_{0}$ is slightly violated. It is worth emphasizing
that the local asymptotics provides a better approach to demonstrate
the finite-sample properties of the test and estimators than the fixed
asymptotics does.

\subsection{Asymptotic bias and MSE \label{subsec:Asymptotic-bias-and}}

Based on Theorem \ref{Thm:elas}, it is essential to understand the
asymptotic behaviors of $\cZ_{c}^{\rmt}$ and the truncated multivariate
normal distribution in general. Toward that end, we derive the moment
generating functions (MGFs) of such distributions in the supplementary
material, which shed light on the moments of $n^{1/2}(\widehat{\psi}_{\elas}-\psi_{0})$.

\textcolor{black}{Corollary \ref{cor:var} provides the analytical
formula of the asymptotic bias and MSE of $n^{1/2}(\widehat{\psi}_{\elas}-\psi_{0})$.}

\begin{corollary}\label{cor:var}

Suppose assumptions in Theorem \ref{Thm:Consistency-DML} hold except
that Assumption \ref{Asump:rand-rwd} may be violated. 
\begin{enumerate}
\item Under ${\rm H}_{0},$ the bias and MSE of $n^{1/2}(\widehat{\psi}_{\elas}-\psi_{0})$
are $\bias=0$ and $\mse=V_{\eff}+V_{\rteff}\{1-F_{p+2}(c_{\gamma})\}.$ 
\item Under ${\rm H}_{a},$ the bias and MSE of $n^{1/2}(\widehat{\psi}_{\elas}-\psi_{0})$
are $\bias=0$ and $\mse=V_{\rt}.$ 
\item Under ${\rm H}_{a,n},$ the bias and MSE of $n^{1/2}(\widehat{\psi}_{\elas}-\psi_{0})$
are 
\begin{equation}
\bias(\gamma,\eta)=-V_{\eff}\eta F_{p+2}(c_{\gamma};\lambda),\label{eq:bias2}
\end{equation}
and 
\begin{eqnarray}
\mse(\gamma,\eta) & = & V_{\eff}+V_{\text{\rteff}}\{1-F_{p+2}(c_{\gamma};\lambda)\}\label{eq:mse2}\\
 &  & +(V_{\eff}\eta)^{\otimes2}\{2F_{p+2}(c_{\gamma};\lambda)-F_{p+4}(c_{\gamma};\lambda)\}\nonumber 
\end{eqnarray}
with $\lambda=\eta^{\T}\Sigma_{SS}^{-1}\eta$. 
\end{enumerate}
\end{corollary}

\begin{figure}
\centering 
\begin{centering}
\includegraphics[scale=0.7]{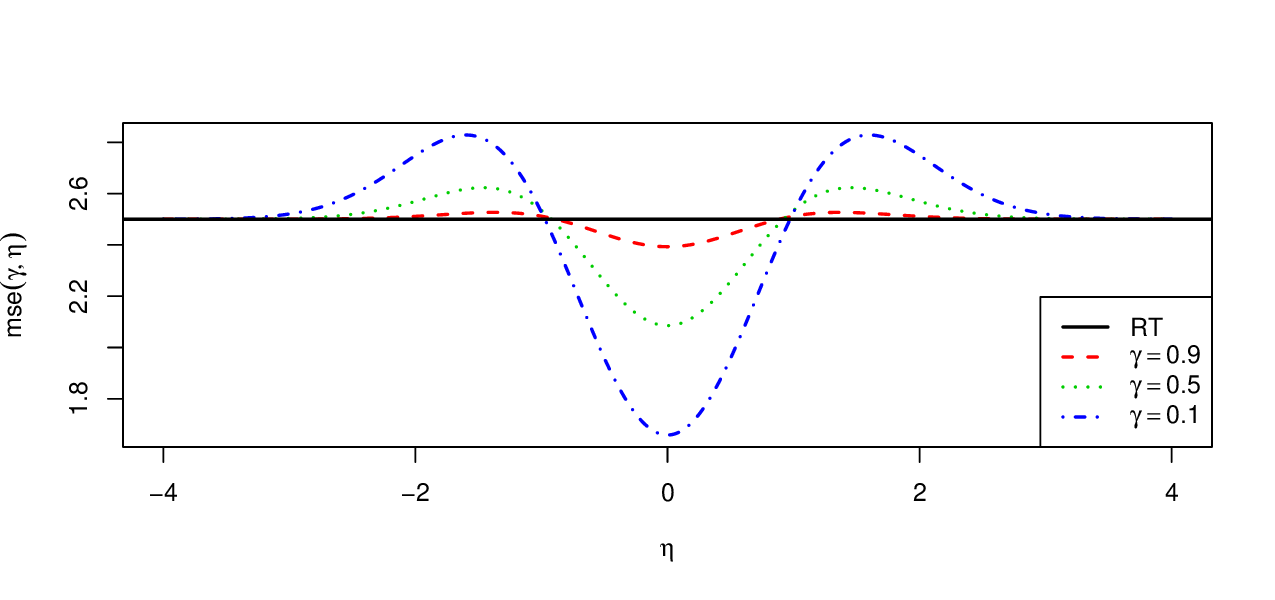} 
\par\end{centering}
\vspace{-0.25cm}

\caption{\label{fig:supp-eff}Illustration of the super-efficiency of $\widehat{\psi}_{\elas}$
in terms of mse$(\gamma,\eta)$ as a function of $\eta$ by varying
$\gamma\in\{0.9({\rm dashed}),0.5({\rm dotted}),0.1({\rm dotdash})\}$
compared to $\widehat{\psi}_{\rt}$.}
\end{figure}

\textcolor{black}{Corollary \ref{cor:var} enables us to demonstrate
the potential advantages and disadvantages of $\widehat{\psi}_{\elas}$
compared with $\widehat{\psi}_{\rt}$ and $\widehat{\psi}_{\eff}$
under different scenarios. To illustrate, we consider the case of
a scalar $\psi_{0}$, $V_{\eff}=1$, $V_{\rt}=2.5,$ and $\Sigma_{SS}=0.5.$
Figure \ref{fig:supp-eff} shows ${\rm mse}(\gamma,\eta)$ as a function
of $\eta$ by varying $\gamma\in\{0.9,0.5,0.1\}$ compared to $\widehat{\psi}_{\rt}$.
For a given $\gamma\in(0,1)$, when $\eta$ is small, $\widehat{\psi}_{\elas}$
is more efficient than $\widehat{\psi}_{\rt}$; and when $\eta$ increases,
the MSE of $\widehat{\psi}_{\elas}$ increases, exceeds, and gradually
returns to the MSE of $\widehat{\psi}_{\rt}$. This phenomenon reveals
the super-efficiency (related to the problem of non-regularity) of
$\widehat{\psi}_{\elas}$ at small values of $\eta$ at the cost of
the MSE inflation for some $\eta$ values. \citet{lecam1953some}
obtained an earlier result of super-efficiency for the famous Hodges
estimator. Also, $\widehat{\psi}_{\elas}$ with a smaller $\gamma$
achieves a larger deduction of the MSE at small values of $\eta$
but also more considerable inflation of the MSE at big values of $\eta$
compared to $\widehat{\psi}_{\rt}$, and vice versa. This observation
motivates our adaptive selection of $\gamma$ in Section \ref{subsec:Adaptive-selection}
to produce an elastic integrative estimator with small bias and mean
squared error for a possible value of $\eta$. Also, super-efficiency
and non-regularity are the root causes for the standard asymptotic
inference to fail, which motivates the proposed elastic confidence
intervals to provide uniformly valid confidence intervals (Section
\ref{subsec:Inference}); however, they can be conservative at certain
parameter values when the sample size is small (Section \ref{sec:Simulation}). }

\begin{remark}[Sample splitting and cross fitting]\label{rmk:splitting}

\textcolor{black}{Sample splitting and cross fitting are helpful tactics
to simplify asymptotic analyses by removing the dependence between
nuisance parameter estimation and primary parameter estimation \citep{chernozhukov2018double,kennedy2020optimal}.
To apply sample splitting to our context, one can divide the sample
into two parts for testing and estimation separately. While sample
splitting and cross fitting are beneficial in theoretical development,
they may come with expenses of heavier computation and fewer data
for estimating different components. Thus, we do not use sampling
splitting or cross fitting as a device to establish the theoretical
properties of the proposed pre-test estimator. Without sample splitting,
the test and estimators are intimately related, requiring careful
decompositions of the estimators into components that are asymptotically
dependent and independent of the test statistic, as shown in our three
steps toward Theorem \ref{Thm:elas}. Also, sample splitting can not
resolve the non-regularity issue of the pre-test estimator \citep{toyoda1979pre}.
This is because sample splitting cannot bypass additional randomness
due to pre-testing. Thus, the impact of pre-testing and superefficiency
remains an issue; see the simulation study in Section \ref{subsec:Regular-inference-fails}. }

\end{remark}

\begin{remark}[Soft thresholding to mitigate the non-regularity]

\label{rmk:soft}

\textcolor{black}{The proposed elastic integrative estimator involves
an indicator function to make a binary decision to include or exclude
the RW data from analysis. The indicator function serves as hard thresholding.
To alleviate the non-regularity issue and refine the proposed estimator,
one may use soft thresholding by imposing the smoothness of the indicator
function. For example, similar to \citet{yang2018trimming}, one can
use a smooth weight function $\Phi_{\epsilon}\left(c_{\gamma}-T\right)$
to replace $I(T<c_{\gamma})$, where $\Phi_{\epsilon}(z)$ is the
normal cumulative distribution with zero mean and variance $\epsilon^{2}$.
As $\epsilon\rightarrow0$, $\Phi_{\epsilon}(c_{\gamma}-T)$ becomes
closer to $I(T<c_{\gamma})$. Also, as suggested by a reviewer, one
can weigh the RW data based on the p-value from the test, i.e., $1-F_{p}(T)$.
A small p-value indicates a large bias in the RW data, and we should
give the RW data less weight. Conversely, a large p-value suggests
a small bias, and we should provide the RW data with more weight.
The third idea is to create bootstrap replications of the elastic
integrative estimator and obtain the average of the bootstrap replications
to impose smoothness. \citet{chakraborty2010inference} showed in
simulation that soft-thresholding reduces the non-regularity of Q-learner
in the dynamic treatment regime literature; however, they also provided
a caveat that soft-thresholding cannot eliminate the non-regularity.
Heuristically, the standard inference under the fixed alternative
still provides poor finite sample coverage properties. Therefore,
one still requires the local alternative asymptotics to derive inference
procedures with uniform validity as we did for the hard thresholding
estimator. We will leave this topic for future research. }

\end{remark}

\subsection{Inference\label{subsec:Inference}}

The nonparametric bootstrap method provides consistent inference in
many cases of \textcolor{black}{regular estimators}. However, this
feature prevents using the nonparametric bootstrap inference for $\widehat{\psi}_{\elas}$
because the indicator function of the preliminary test in (\ref{eq:EE})
renders $\widehat{\psi}_{\elas}$ a non-smooth and \textcolor{black}{non-regular}
estimator \citep{shao1994bootstrap}. We formally show in the supplementary
material the inconsistency of the nonparametric bootstrap inference
for $\widehat{\psi}_{\elas}$. Alternatively, \citet{laber2011adaptive}
proposed an adaptive confidence interval for the test error in classification,
a non-regular statistics, by bootstrapping the upper and lower bounds
of the test error. In this article, we propose an adaptive procedure
for robust inference of $\psi_{0}$ accommodating the strength of
violation of Assumption \ref{Asump:rand-rwd} in finite samples.

Let $e_{k}$ be a $p$-vector of zeros except that the $k$th component
is one, and let $e_{k}^{\T}\psi_{0}$ be the $k$th component of $\psi_{0}$,
for $k=1,\ldots,p$. Because the asymptotic distribution of $n^{1/2}e_{k}^{\T}(\widehat{\psi}_{\elas}-\psi_{0})$
is different under the local and fixed alternatives, we propose different
strategies for constructing CIs: under ${\rm H}_{a,n}$, the asymptotics
is non-standard, we construct a least favorable CI that guarantees
good coverage properties uniformly over possible values of the local
parameter; under ${\rm H}_{a}$, the asymptotics is standard, we construct
the usual Wald CI based on the normal limiting distribution.

First, under ${\rm H}_{a,n}$, we rewrite $\mathcal{M}(\gamma;\eta)$
in (\ref{eq:MixC1}) as $\cD^{{\rm NR}}(\mu_{1})+\cD^{{\rm R}},$
where $\cD^{{\rm NR}}(\mu_{1})=V_{\rteff}^{1/2}\cZ_{1}\bone(\cZ_{1}^{\T}\cZ_{1}$
$\geq c_{\gamma})$ is the non-regular component with $\cZ_{1}$ having
mean $\mu_{1}$, $\cD^{{\rm R}}=-V_{\eff}^{1/2}\cZ_{2}$ is the regular
component, and $\cD^{{\rm NR}}(\mu_{1})$ and $\cD^{{\rm R}}$ are
independent. For a fixed $\mu_{1},$ let $\widehat{Q}_{k,\alpha}(\mu_{1})$
be the approximated $100\alpha$th quantile of $\cD^{{\rm NR}}(\mu_{1})+\cD^{{\rm R}}$,
which can be obtained by rejective sampling. We can construct a $(1-\alpha)100\%$
confidence interval of $n^{1/2}e_{k}^{\T}(\widehat{\psi}_{\elas}-\psi_{0})$
as $[\widehat{Q}_{k,\alpha/2}(\mu_{1}),\widehat{Q}_{k,1-\alpha/2}(\mu_{1})]$.
Different CIs are required for different values of $\mu_{1}$. To
accommodate different possible values of $\mu_{1}$, one solution
is to construct the least favorable CI by taking the infimum of the
lower bound of the CI $\widehat{Q}_{k,\alpha/2}(\mu_{1})$ and the
supremum of the upper bound of the CI $\widehat{Q}_{k,1-\alpha/2}(\mu_{1})$
over all possible values of $\mu_{1}$. However, the range of $\mu_{1}$
can be vast, rendering the least favorable CI non-informative. We
identify the plausible values of $\mu_{1}$ following a multivariate
normal distribution with mean $n^{-1/2}\widehat{\Sigma}_{SS}^{-1/2}\sum_{i\in\mathcal{B}}\widehat{S}_{\rw,\widehat{\psi}_{\rt}}(V_{i})$
and variance $I_{p\times p}$. Let $\widetilde{\alpha}=1-(1-\alpha)^{1/2}$,
such that $(1-\widetilde{\alpha})^{2}=1-\alpha$ and let $\mathcal{B}_{1-\widetilde{\alpha}}^{{\rm N}}$
be a $1-\widetilde{\alpha}$ bounded region of a standard $p$-variate
normal distribution. Then, 
\[
\mathcal{B}_{1-\widetilde{\alpha}}=\left\{ \mu_{1}:\left\{ n^{-1/2}\widehat{\Sigma}_{SS}^{-1/2}\sum_{i\in\mathcal{B}}\widehat{S}_{\rw,\widehat{\psi}_{\rt}}(V_{i})-\mu_{1}\right\} \in\mathcal{B}_{1-\widetilde{\alpha}}^{{\rm N}}\right\} 
\]
is a bounded region of $\mu_{1}$ with asymptotic probability $1-\widetilde{\alpha}$.
We construct the $(1-\alpha)100\%$ least favorable CI for $n^{1/2}e_{k}^{\T}(\widehat{\psi}_{\elas}-\psi_{0})$
as $[\inf_{\mu_{1}\in\mathcal{B}_{1-\widetilde{\alpha}}}\widehat{Q}_{k,\widetilde{\alpha}/2}(\mu_{1}),\sup_{\mu_{1}\in\mathcal{B}_{1-\widetilde{\alpha}}}\widehat{Q}_{k,1-\widetilde{\alpha}/2}(\mu_{1})]$.
Here, using the wider $(1-\widetilde{\alpha})100\%$ quantile range
of $\widehat{Q}_{k}(\mu_{1})$ instead of the $(1-\alpha)$ quantile
range is necessary to guarantee the coverage of $(1-\alpha)$ due
to ignoring other possible values of $\mu_{1}$ outside $\cB_{1-\tilde{\alpha}}$.

Second, under ${\rm H}_{a},$ Assumption \ref{Asump:rand-rwd} is
strongly violated. As shown in Theorem \ref{Thm:elas}, $n^{1/2}e_{k}^{\T}(\widehat{\psi}_{\elas}-\psi_{0})$
is regular and asymptotically normal, denoted by $\mathcal{M}(\gamma;\pm\infty,\pm\infty)$.
Therefore, a $(1-\alpha)100\%$ confidence interval of $n^{1/2}e_{k}^{\T}(\widehat{\psi}_{\elas}-\psi_{0})$
can be constructed based on the $100\alpha/2$- and $100(1-\alpha/2)$-th
quantiles of the normal distribution $\mathcal{M}(\gamma;\pm\infty,\pm\infty)$,
denoted by $[\widehat{Q}_{k,\alpha/2}(\pm\infty),\widehat{Q}_{k,1-\alpha/2}(\pm\infty)]$.

Finally, because the least favorable CI may be unnecessarily wide
under ${\rm H}_{a}$, we require a strategy to distinguish between
${\rm H}_{a,n}$ corresponding to finite values of $\mu_{1}$ and
${\rm H}_{a}$ corresponding to $\mu_{1}=\pm\infty$. To do this,
we use the test statistic $T$. Under ${\rm H}_{a,n}$, $T=O_{\pr}(1)$;
while under ${\rm H}_{a},$ $T=\infty$. Therefore, we specify a sequence
of thresholds $\{\kappa_{n}:n\geq1\}$ that diverges to infinity as
$n\rightarrow\infty$ and compare $T$ to $\kappa_{n}$. Many choices
of $\kappa_{n}$ can be considered, e.g., $\kappa_{n}=(\log n){}^{1/2}$,
which is similar to the BIC criterion \citep{cheng2008robust,andrews2010inference}.
If $T\leq\kappa_{n}$, we choose the local alternative strategy to
construct the least favorable CI, and if $T>\kappa_{n}$, we choose
the fixed alternative strategy to construct a normal CI, leading to
an elastic CI 
\begin{equation}
{\rm ECI}_{k,1-\alpha}=\begin{cases}
[\inf_{\mu_{1}\in\mathcal{B}_{1-\widetilde{\alpha}}}\widehat{Q}_{k,\widetilde{\alpha}/2}(\mu_{1}),\sup_{\mu_{1}\in\mathcal{B}_{1-\widetilde{\alpha}}}\widehat{Q}_{k,1-\widetilde{\alpha}/2}(\mu_{1})], & \text{if }T\leq\kappa_{n},\\{}
[\widehat{Q}_{k,\alpha/2}(\pm\infty),\widehat{Q}_{k,1-\alpha/2}(\pm\infty)], & \text{if }T>\kappa_{n}.
\end{cases}\label{eq:elastCI}
\end{equation}

\begin{theorem}\label{Thm:elas-1} Suppose assumptions in Theorem
\ref{Thm:Consistency-DML} hold except that Assumption \ref{Asump:rand-rwd}
may be violated. The asymptotic coverage rate of the elastic CI of
$n^{1/2}e_{k}^{\T}(\widehat{\psi}_{\elas}-\psi_{0})$ in (\ref{eq:elastCI})
satisfies 
\begin{eqnarray*}
 &  & \lim_{n\rightarrow\infty}\pr\left\{ n^{1/2}e_{k}^{\T}(\widehat{\psi}_{\elas}-\psi_{0})\in{\rm ECI}_{k,1-\alpha}\right\} \geq1-\alpha,
\end{eqnarray*}
and the equality holds under ${\rm H}_{a}$.

\end{theorem}

\subsection{Adaptive selection of $\gamma$\label{subsec:Adaptive-selection}}

The selection of $\gamma$ involves the bas-variance tradeoff and
therefore is important to determine the MSE of $\widehat{\psi}_{\elas}$.
Corollary \ref{cor:var} indicates that under ${\rm H}_{a,n}$, the
MSE of $\widehat{\psi}_{\elas}$ in (\ref{eq:mse2}) involves two
terms: Term 1 is $V_{\eff}+V_{\rteff}\{1-F_{p+2}(c_{\gamma};\lambda)\}$,
and Term 2 involves $(V_{\eff}\eta)^{\otimes2}$. If $\eta$ is small,
the MSE is dominated by Term 1, which can be made small if we select
a small $\gamma$; while if $\eta$ is large, the MSE is dominated
by Term 2, which can be made small if we select a large $\gamma.$

The above observation motivates an adaptive selection of $\gamma.$
We propose to estimate $\eta$ by $\widehat{\eta}=n^{-1/2}\sum_{i\in\mathcal{B}}\widehat{S}_{\os,\widehat{\psi}_{\rt}}(V_{i})$
and select $\gamma$ that minimizes $\mse(\gamma;\widehat{\eta})$,
where $\mse(\gamma;\eta)$ is given by (\ref{eq:mse2}) or approximated
by rejective sampling. In practice, we can specify a grid of values
from $0$ to $1$ for $\gamma$, denoted by $\mathcal{G}$, simulate
the distribution of $\mathcal{M}(\gamma;\widehat{\eta})$ for all
$\gamma\in\mathcal{G}$, and finally choose $\gamma$ to be the one
in $\mathcal{G}$ that minimizes the MSE of $\mathcal{M}(\gamma;\widehat{\eta})$.
As corroborated by simulation, the selection strategy is effective
in the sense that when the signal of violation is weak, the selected
value of $\gamma$ is small and when the signal of violation is strong,
the selected value of $\gamma$ is large.

\section{Simulation study\label{sec:Simulation}}

We evaluate the finite sample performance of the proposed elastic
estimator via simulation for robustness against unmeasured confounding
and adaptive inference. Specifically, we compare the RT estimator,
the efficient combining estimator, and the elastic estimator under
settings that vary the strength of unmeasured confounding in the RW
data. \textcolor{black}{We also carry out simulation under a setting
when the transportability assumption is violated in the RW data; see
Section \ref{subsec:Importance-of-overlapping} in the supplementary
material.}

We first generate populations of size $10^{5}$. For each population,
we generate the covariate $X=(1,X_{1},X_{2},X_{3})^{\T}$, where $X_{j}\sim\rmN(1,1)$
for $j=1,2,3$, and the treatment effect modifier is $Z=(1,X_{1},X_{2})^{\T}$.
We generate $Y(a)$ by 
\begin{align}
Y(a)\mid X=\mu(X)+a\times\tau(Z)+\epsilon(a), & \ \ \ \ \epsilon(a)\sim\rmN(0,1),\nonumber \\
\mu(X)=X_{1}+X_{2}+X_{3}, & \ \ \ \ \tau(Z)=\psi_{0}+\psi_{1}X_{1}+\psi_{2}X_{2},\label{eq:mu}
\end{align}
for $a=0,1$. Throughout the simulation, we fix $\psi_{0}$ to be
zero and consider two cases for $(\psi_{1},\psi_{2})$: a) zero effect
modification $(\psi_{1},\psi_{2})=(0,0)$ and b) nonzero effect modification
$(\psi_{1},\psi_{2})=(1,1)$.

We then generate two samples from the target population. We generate
the RT selection indicator by $\delta\mid X\sim\mathrm{Bernoulli}\{\pi_{\delta}(X)\},$
where $\logit\{\pi_{\delta}(X)\}=-4.5-2X_{1}-2X_{2}.$ Under this
selection mechanism, the selection rate is around $0.6\%$, which
results in $m\approx620$ RT subjects. We also take a random sample
of size $n\in\{2000,5000\}$ from the population to form an RW sample.
In the RT sample, the treatment assignment is $A\mid X,\delta=1\sim\mathrm{Bernoulli}\{e_{1}(X)\},$
where $e_{1}(X)=0.5$. In the RW sample, $A\mid X,\delta=0\sim\mathrm{Bernoulli}\{e_{0}(X)\}$,
where logit$\{e_{0}(X)\}=\alpha-X_{1}-X_{2}-bX_{3}$ with adaptively
chosen $\alpha$ to ensure the mean of $e_{0}(X)$ to be around 0.5.
In addition, we vary $b$ to indicate the different strengths of unmeasured
confounding in the analysis (violation of Assumption \ref{Asump:rand-rwd}).
The observed outcome $Y$ in both samples is $Y=AY(1)+(1-A)Y(0)$.
\begin{figure}
\centering 
\begin{centering}
\includegraphics[scale=0.65]{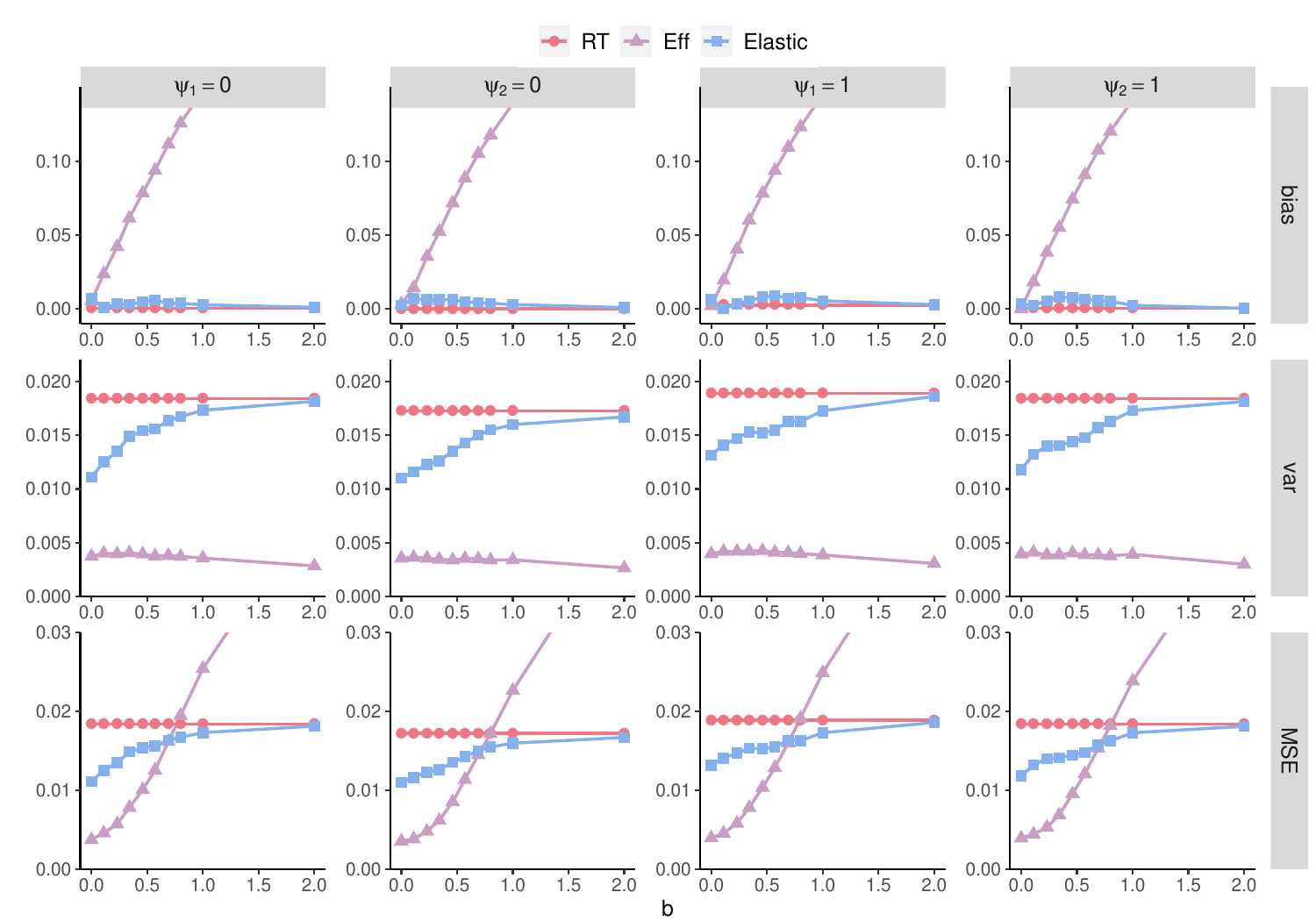} 
\par\end{centering}
\vspace{0.5cm}

\caption{\label{fig:violin}Summary statistics plots of estimators of $(\psi_{1},\psi_{2})$
with respect to the strength of unmeasured confounding labeled by
``b''. In each plot, the three estimators $\widehat{\psi}_{\rt}$,
$\widehat{\psi}_{\eff}$, and $\widehat{\psi}_{\elas}$ are labeled
by ``RT'', ``Eff'', and ``Elastic''. Each row of the plots corresponds
to a different metrics: ``bias'' for bias, ``var'' for variance,
``MSE'' for mean square error; each column of the plots corresponds
to one component of $(\psi_{1},\psi_{2})$ in the two cases: $\psi_{1}=0,\psi_{2}=0$,
$\psi_{1}=1$, and $\psi_{2}=1$ with $n=2000$.}
\end{figure}

\begin{table}[ht]
\caption{\label{table:cvg}Simulation results for coverage rates and widths
of 95\% confidence intervals for $\widehat{\psi}_{\rt},$ $\widehat{\psi}_{\eff},$
and $\widehat{\psi}_{\elas}$ (labeled as ``RT'', ``Eff'', and
``Elastic'') in the two cases: zero effect modification $\psi_{1}=\psi_{2}=0$
(left) and nonzero effect modification $\psi_{1}=\psi_{2}=1$ (right)
with $n=2000$; the slightly wider ECIs for $\widehat{\psi}_{\eff}$
(than CIs for $\widehat{\psi}_{\rt}$) are bolded }

\centering

\resizebox{\textwidth}{!}{

\begin{tabular}{ccccccccccccc}
\toprule 
 & \multicolumn{2}{c}{RT} & \multicolumn{2}{c}{Eff} & \multicolumn{2}{c}{Elastic} & \multicolumn{2}{c}{RT} & \multicolumn{2}{c}{Eff} & \multicolumn{2}{c}{Elastic}\tabularnewline
\midrule 
 & \multicolumn{6}{c}{Case 1: zero effect modification} & \multicolumn{6}{c}{Case 2: nonzero effect modification}\tabularnewline
$b$  & {\scriptsize{}{}$\psi_{1}=0$}  & {\scriptsize{}{}$\psi_{2}=0$}  & {\scriptsize{}{}$\psi_{1}=0$}  & {\scriptsize{}{}$\psi_{2}=0$}  & {\scriptsize{}{}$\psi_{1}=0$}  & {\scriptsize{}{}$\psi_{2}=0$}  & {\scriptsize{}{}$\psi_{1}=1$}  & {\scriptsize{}{}$\psi_{2}=1$}  & {\scriptsize{}{}$\psi_{1}=1$}  & {\scriptsize{}{}$\psi_{2}=1$}  & {\scriptsize{}{}$\psi_{1}=1$}  & {\scriptsize{}{}$\psi_{2}=1$} \tabularnewline
\midrule 
\multicolumn{13}{c}{Coverage Rate ($\%$)}\tabularnewline
\midrule 
0  & 94.1  & 94.1  & 93.8  & 93.7  & 92.7  & 92.5  & 94.3  & 93.8  & 95.0  & 94.2  & 92.7  & 92.5\tabularnewline
0.11  & 94.1  & 94.1  & 92.2  & 92.7  & 93.2  & 92.8  & 94.3  & 93.8  & 93.3  & 92.9  & 92.9  & 92.7\tabularnewline
0.23  & 94.1  & 94.0  & 88.5  & 89.8  & 92.8  & 92.8  & 94.3  & 93.8  & 89.8  & 89.0  & 93.3  & 92.7\tabularnewline
0.34  & 94.1  & 94.0  & 83.2  & 84.5  & 94.0  & 93.8  & 94.3  & 93.8  & 84.9  & 83.5  & 94.4  & 93.5\tabularnewline
0.46  & 94.1  & 94.0  & 74.7  & 76.3  & 94.5  & 94.5  & 94.3  & 93.8  & 76.8  & 75.8  & 94.5  & 94.4\tabularnewline
0.57  & 94.1  & 94.0  & 66.4  & 66.1  & 95.5  & 95.2  & 94.3  & 93.8  & 67.2  & 66.8  & 95.5  & 94.8\tabularnewline
0.69  & 94.1  & 94.1  & 56.1  & 56.3  & 95.5  & 95.8  & 94.3  & 93.8  & 56.8  & 55.9  & 95.3  & 94.6\tabularnewline
0.8  & 94.1  & 94.0  & 46.3  & 46.8  & 95.5  & 95.6  & 94.3  & 93.8  & 46.5  & 45.2  & 95.3  & 95.0\tabularnewline
1  & 94.1  & 94.0  & 31.5  & 31.1  & 95.5  & 95.0  & 94.3  & 93.8  & 30.9  & 29.4  & 95.5  & 94.9\tabularnewline
2  & 94.1  & 94.0  & 2.9  & 3.6  & 94.3  & 94.4  & 94.3  & 93.8  & 2.6  & 3.0  & 94.7  & 94.2\tabularnewline
\midrule 
\multicolumn{13}{c}{Width ($\times10^{-3}$)}\tabularnewline
\midrule 
0  & 528  & 528  & 243  & 242  & 472  & 473  & 529  & 530  & 243  & 243  & 472  & 474\tabularnewline
0.11  & 527  & 528  & 242  & 242  & 488  & 487  & 529  & 530  & 242  & 243  & 479  & 480\tabularnewline
0.23  & 527  & 528  & 241  & 242  & 496  & 497  & 529  & 530  & 241  & 242  & 498  & 500\tabularnewline
0.34  & 528  & 528  & 241  & 241  & 516  & 516  & 529  & 530  & 241  & 242  & 511  & 514\tabularnewline
0.46  & 528  & 528  & 239  & 240  & \textbf{530}  & \textbf{530}  & 529  & 530  & 240  & 240  & 524  & 526\tabularnewline
0.57  & 528  & 528  & 238  & 238  & \textbf{535}  & \textbf{535}  & 529  & 530  & 238  & 239  & \textbf{530}  & \textbf{532}\tabularnewline
0.69  & 528  & 528  & 235  & 236  & \textbf{534}  & \textbf{534}  & 529  & 530  & 236  & 236  & \textbf{529}  & \textbf{531}\tabularnewline
0.8  & 528  & 528  & 233  & 234  & \textbf{532}  & \textbf{532}  & 529  & 530  & 233  & 234  & \textbf{530}  & \textbf{532}\tabularnewline
1  & 528  & 528  & 229  & 230  & \textbf{529}  & \textbf{529}  & 529  & 530  & 229  & 230  & \textbf{530}  & \textbf{532}\tabularnewline
2  & 528  & 528  & 207  & 208  & 527  & 527  & 529  & 530  & 208  & 209  & 528  & 530\tabularnewline
\bottomrule
\end{tabular}} 
\end{table}

To assess the robustness of the elastic integrative estimator against
unmeasured confounding, we consider the omission of $X_{3}$ in all
estimators, resulting in unmeasured confounding in the RW data. The
strength of unmeasured confounding is indexed by $b$ in (\ref{eq:mu});
high values of $b$ indicate strong levels of unmeasured confounding
and vice versa. We specify the range of $b$ by $10$ values in an
irregular grid from $0$ to $2$ $\{0,0.11,0.23,0.34,0.46,0.57,0.69,0.80,1,2\}$,
which places more emphasis on the scenarios where Assumption \ref{Asump:rand-rwd}
is weakly violated. We compare the following estimators for the HTE
parameter $\psi$: 
\begin{enumerate}
\item RT $\widehat{\psi}_{\rt}$: the efficient estimator based only on
the RT data solving (\ref{eq:EE}) with $\bone(T<c_{\gamma})\equiv0$; 
\item Eff $\widehat{\psi}_{\eff}$: the efficient integrative estimator
solving (\ref{eq:EE}) with $\bone(T<c_{\gamma})\equiv1$; 
\item Elastic $\widehat{\psi}_{\elas}$: the proposed elastic integrative
estimator solving (\ref{eq:EE}) with adaptive selection of $\gamma$. 
\end{enumerate}
For all estimators, we estimate the propensity score function by a
logistic sieve model with the power series $X$, $X^{2}$ and their
two-way interactions (omitting $X_{3}$) and the outcome mean functions
by linear sieve models with the power series $X$, $X^{2}$ and their
two-way interactions (omitting $X_{3}$). \textcolor{black}{If higher-order
series is specified, it is necessary to select the series to balance
the bias and variance in estimating the nuisance functions, such as
using the penalized estimating equation approach \citep{lee2021improving}.
}The CIs are constructed for $\widehat{\psi}_{\aipw}$, $\widehat{\psi}_{\rt}$
and $\widehat{\psi}_{\eff}$ based on the perturbation-based resampling
with the replication size $100$ and for $\widehat{\psi}_{\elas}$
based on the elastic approach with $\kappa_{n}=(\log n){}^{1/2}$.
Sensitivity analysis shows that the coverage rates and widths of the
CIs stay close with $\kappa_{n}=0.5(\log n){}^{1/2}$ (Section \ref{subsec:Sensitivity-analysis-of-k}).

Figure~\ref{fig:violin} presents the plots of Monte Carlo biases,
variances, and MSEs of estimators based on 2000 simulated datasets
with numerical results reported in Table \ref{tab:SimNumericalResult}.
Table \ref{table:cvg} reports the coverage rates and widths of $95\%$
CIs. The RT estimator $\widehat{\psi}_{\rt}$ is unbiased across different
scenarios, and the coverage rates are close to the nominal level.
However, $\widehat{\psi}_{\rt}$ has larger variances than other integrative
estimators due to the small RT sample size. The efficient integrative
estimator $\widehat{\psi}_{\eff}$ gains efficiency over $\widehat{\psi}_{\rt}$
by leveraging the large sample size of the RW data. However, the bias
of $\widehat{\psi}_{\eff}$ increases as $b$ increases. Thus, $\widehat{\psi}_{\eff}$
has smaller MSEs than $\widehat{\psi}_{\rt}$ for small values of
$b$ but larger MSEs for large values of $b$. The coverage rates
of the CIs for $\widehat{\psi}_{\eff}$ deviate away from the nominal
level as $b$ increases. This can lead to an uncontrolled false discovery
of important treatment effect modifiers (see the case of zero effect
modification with $\psi_{1}=\psi_{2}=0$). The elastic integrative
estimator $\widehat{\psi}_{\elas}$ with the adaptive selection of
$\gamma$ reduces $\widehat{\psi}_{\eff}$'s biases across all scenarios
regardless of the strength of unmeasured confounding. The challenging
scenarios are indexed by $b$ around $0.44$ and $0.67$, where the
small biases of $\widehat{\psi}_{\elas}$ occur. In these scenarios,
the pre-testing (built in the elastic estimator) has difficulty in
detecting the RW sample's biases. However, $\widehat{\psi}_{\elas}$
with an adaptive selection of $\gamma$ achieves the smallest MSE
among all estimators across all scenarios (Figure \ref{fig:violin}
and Table \ref{tab:SimNumericalResult}).

To inspect the performance of the proposed data-adaptive selection
strategy, Table \ref{tab:Simulation-results-ofgamma} reports Monte
Carlo averages and standard deviations of the selected values for
the local parameter $\eta$, the threshold $c_{\gamma}$, and the
proportion of combining the RT and RW samples. As expected, $\widehat{\eta}$
increases as $b$ increases, indicating increased biases in the RW
sample. The selected $\gamma$ increases (as a result, the proportion
of combining the RT and RW samples decreases) as $b$ increases, which
shows the proposed adaptive selection strategy is effective. To compare
the adaptive selection strategy with the fixed threshold strategy,
a simulation study in Section \ref{subsec:fixing_gamma} shows that
the elastic integrative estimator $\widehat{\psi}_{\elas}$ with a
fixed threshold can have increased biases compared to a data-adaptive
selected threshold.

The coverage rates of the ECIs for $\widehat{\psi}_{\elas}$ are close
to the nominal level for all settings with different values of $b$.
The ECIs are narrower than the CIs for $\widehat{\psi}_{\rt}$ when
$b$ is small ($b\leq0.46$ for $\psi_{1}=\psi_{2}=0$ and $b\leq0.34$
for $\psi_{1}=\psi_{2}=1$), are wider than the CIs for $\widehat{\psi}_{\rt}$
when $b$ increases, and become close to the CIs for $\widehat{\psi}_{\rt}$
when $b$ reaches $1$ or larger. However, the conservativity of the
ECIs reduces as $n$ increases, and the ECIs can perform at least
as well as the CIs for $\widehat{\psi}_{\rt}$ for any $b$ (see Table
\ref{tab:SimNumericalResult-m5000-1} for $n=5000$).

\section{An application\label{sec:Application}}

We illustrate the potential benefit of the proposed elastic estimator
to evaluate the effect of adjuvant chemotherapy for early-stage resected
non-small cell lung cancer (NSCLC) using the CALGB 9633 data and a
large clinical oncology database, the NCDB. In CALGB 9633, we include
$319$ patients, with $163$ randomly assigned to observation ($A=0$)
and $156$ randomly assigned to chemotherapy ($A=1$). The NCDB cohort
is selected based on the same patient eligibility criteria as the
CALGB 9633 trial;\textcolor{black}{{} see Section \ref{sec:datasupp}
of the supplementary material.} The comparable NCDB sample includes
$15,166$ patients diagnosed with NSCLC between 2004 and 2016 in stage
IB disease, with $10,903$ on observation and $4316$\textcolor{red}{{}
}receiving chemotherapy after surgery. The numbers of treated and
controls are relatively balanced in the CALGB 9633 trial, while they
are unbalanced in the NCDB sample. We include five covariates in the
analysis: gender ($1=\textrm{male}$, $0=\textrm{female}$), age,
the indicator for histology ($1=\textrm{squamous}$, $0=\textrm{non-squamous}$),
race ($1=\text{white},0=\text{non-white}$), and tumor size in cm.
The outcome is the overall survival within three years after the surgery,
i.e., $Y=1$ if died due to all causes and $Y=0$ otherwise. We are
interested in \textcolor{black}{estimating} the HTE of adjuvant chemotherapy
over observation after resection for the patient population with the
same set of eligibility criteria as that of CALGB 9633. 
\begin{table}
\centering\caption{\label{tab:covbal}Covariate means with standard errors in parentheses
by sample and treatment group in the CALGB 9633 trial and NCDB samples}

\vspace{0cm}

\begin{tabular}{cccccccccc}
\toprule 
 & \multirow{1}{*}{$A$} & \multirow{1}{*}{$N$} & \multicolumn{2}{c}{age} & \multicolumn{2}{c}{tumor size} & male  & squamous  & white\tabularnewline
 &  &  & \multicolumn{2}{c}{(years)} & \multicolumn{2}{c}{(cm)} & ($\%$)  & ($\%$)  & ($\%$)\tabularnewline
\midrule 
RT:  & $0,1$  & 319  & \multicolumn{2}{c}{60.8 (9.62)} & \multicolumn{2}{c}{4.60 (2.08)} & 63.9  & 39.8  & 89.3\tabularnewline
CALGB 9633  & $1$  & 156  & \multicolumn{2}{c}{60.6 (10)} & \multicolumn{2}{c}{4.62 (2.09)} & 64.1  & 40.4  & 90.4\tabularnewline
 & $0$  & 163  & \multicolumn{2}{c}{61.1 (9.25)} & \multicolumn{2}{c}{4.57 (2.07)} & 63.8  & 39.3  & 88.3\tabularnewline
\midrule 
RW:  & $0,1$  & 15166  & \multicolumn{2}{c}{67.9 (10.2)} & \multicolumn{2}{c}{4.82 (1.71)} & 54.6  & 39.1  & 89.6\tabularnewline
NCDB  & $1$  & 4263  & \multicolumn{2}{c}{63.9 (9.23)} & \multicolumn{2}{c}{5.19 (1.79)} & 54.3  & 35.6  & 88.6\tabularnewline
 & $0$  & 10903  & \multicolumn{2}{c}{69.4 (10.1)} & \multicolumn{2}{c}{4.67 (1.65)} & 54.8  & 40.5  & 90.0\tabularnewline
\bottomrule
\end{tabular}
\end{table}

Table \ref{tab:covbal} reports the covariate means by sample and
treatment group. Due to treatment randomization, covariates are balanced
between the treated and the control in the CALGB 9633 trial sample.
While due to a lack of treatment randomization, covariates are relatively
unbalanced in the NCDB sample. Older patients with histology and smaller
tumors are likely to choose a conservative treatment on observation.
Moreover, we can not rule out the possibility of unmeasured confounders
in the NCDB sample. 
\begin{table}
\caption{\label{tab:real-results}Point estimate, standard error, and 95\%
Wald confidence interval of the causal risk difference between adjuvant
chemotherapy and observation based on the CALGB 9633 trial sample
and the NCDB sample: tumor size{*}$=($tumor size$-4.82)/1.72$}
\centering

\vspace{0cm}

\begin{tabular}{ccccccccc}
\toprule 
 & \multicolumn{4}{c}{Intercept ($\psi_{0,1}$)} & \multicolumn{4}{c}{$\text{tumor size}^{*}$ ($\psi_{0,2}$)}\tabularnewline
 & Est.  & S.E.  & \multicolumn{2}{c}{C.I.} & Est.  & S.E.  & \multicolumn{2}{c}{C.I.}\tabularnewline
\midrule 
RT  & -0.094  & 0.054  & (-0.202,  & 0.015)  & 0.002  & 0.055  & (-0.107,  & 0.111)\tabularnewline
{RW}  & {-0.076}  & {0.0085}  & {(-0.093,}  & {-0.059)}  & {-0.029}  & {0.009}  & {(-0.046,}  & {-0.011)}\tabularnewline
Eff  & -0.076  & 0.0083  & (-0.093,  & -0.059)  & -0.026  & 0.009  & (-0.043,  & -0.009)\tabularnewline
Elastic  & -0.076  & 0.0196  & (-0.115,  & -0.037)  & -0.026  & 0.029  & (-0.084,  & 0.032)\tabularnewline
\bottomrule
\end{tabular}
\end{table}

\begin{figure}
\centering 
\begin{centering}
\includegraphics[scale=0.8]{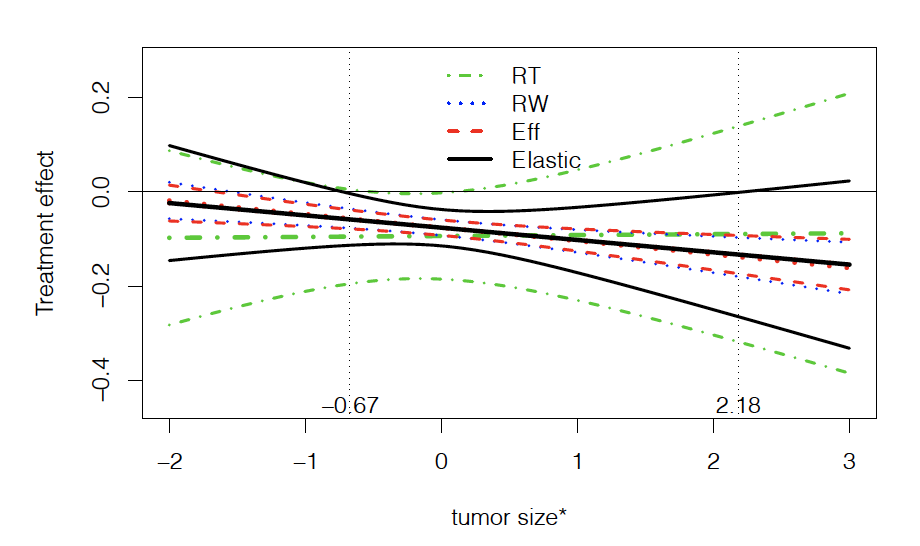} 
\par\end{centering}
\vspace{0cm}

\caption{\label{fig:tsize}Estimated treatment effect as a function of the
(standardized) tumor size along with the $95\%$ Wald confidence intervals:
tumor size{*}$=($tumor size$-4.82)/1.72$, RT, RW, and Eff are the
efficient estimator applied to the RT, RW, and combined sample, respectively,
and Elastic is the proposed elastic combining estimator.}
\end{figure}

We assume a linear HTE function with tumor size as the treatment effect
modifier. We compare the same set of estimators and variance estimators
considered in the simulation study and the efficient estimator applied
to the real-world NCDB cohort, denoted by $\widehat{\psi}_{\rw}$.
Table \ref{tab:real-results} reports the results. Figure \ref{fig:tsize}
shows the estimated treatment effect as a function of the standardized
tumor size. Due to the limited sample size of the trial sample, all
components in $\widehat{\psi}_{\rt}$ are not significant. Due to
the large sample size of the NCDB sample, $\widehat{\psi}_{\rw}$
and $\widehat{\psi}_{\eff}$ are close and reveal that adjuvant chemotherapy
significantly reduced cancer recurrence within three years after the
surgery. Patients with larger tumor sizes benefit more from adjuvant
chemotherapy. However, this finding may be subject to possible biases
of the NCDB sample. In the proposed elastic integrative analysis,
the test statistic is $T=1.9$; there is no strong evidence that the
NCDB presents hidden confounding in our analysis. As a result, the
elastic integrative estimator $\widehat{\psi}_{\elas}$ remains the
same as $\widehat{\psi}_{\eff}$. In reflection of the pre-testing
procedure, the estimated standard error of $\widehat{\psi}_{\elas}$
is larger than $\widehat{\psi}_{\eff}$'s. From Figure \ref{fig:tsize},
patients with tumor sizes in $[4.82+1.72\times(-0.67),4.82+1.72\times(2.18)]=[3.67,8.57]$
significantly benefit from adjuvant chemotherapy in improving overall
survival within three years after the surgery.

\section{Concluding remarks\label{sec:Concluding-remarks}}

The proposed elastic estimator integrates ``high-quality small data''
with ``big data'' to simultaneously leverage small but carefully
controlled unbiased experiments and massive but possibly biased RW
datasets for HTEs. Most causal inference methods require the no unmeasured
confounding assumption. However, this assumption may not hold for
the RW data due to the uncontrolled, real-world data collection mechanism
and is unverifiable based only on the RW data. Utilizing the design
advantage of RTs, we can gauge the reliability of the RW data and
decide whether or not to use RW data in an integrative analysis.

The key assumptions underpinning our framework are the structural
HTE model, i.e., Model (\ref{eq:SNMM}), HTE transportability, and
no unmeasured confounding. In practice, RTs usually consider much
narrower populations than seen in the real world. Improving the generalizability
or external validity of RT findings has been an important research
topic in the data integration literature \citep[e.g., ][]{cole2010generalizing,rudolph2017robust,lee2021improving}.
Besides Assumption \ref{Asump:rand-rct}(i), the positivity of trial
participation or the overlap of the covariate distribution between
the RT and RW samples is required in the problem of generalizability.
We emphasize that although, formally, we do not require the overlap
assumption between the RT and RW samples, its violation renders Model
(\ref{eq:SNMM}) and transportability vulnerable. When transporting
from the narrow RT sample to the broader RW sample, the reliable information
of treatment effects for the non-overlapping region essentially hinges
on the extrapolation from the RT sample. If there is no strong prior
knowledge, Model (\ref{eq:SNMM}) and transportability may not hold.
In this case, the RT estimate and the RW estimate of the HTE can be
inconsistent due to model misspecification even when there are no
unmeasured confounders. See a simulation study in Section \ref{subsec:Importance-of-overlapping}.
The inconsistency of the RW estimator with the RT estimator may reflect
violation of either transportability (e.g., due to model misspecification)
or unmeasured confounding.\textcolor{black}{{} Some practical strategies
(e.g., matching) can be implemented to select an RW sample with sufficient
overlap with the RT sample to improve their comparability and the
chance of successfully integrating the information from two separate
sources; see Section \ref{subsec:Strategies-for-selecting}.}

The elastic integrative estimator gains efficiency over the RT-only
estimator by integrating the reliable RW data and also automatically
detecting bias in the RW data and gears to the RT data. However, the
proposed estimator is non-regular and belongs to pre-test estimation
by construction \citep{giles1993pre}. To demonstrate the non-regularity
issue, we characterize the distribution of the elastic integrative
estimator under local alternatives, which better approximates the
finite-sample behaviors. Moreover, we provide a data-adaptive selection
of the threshold in the testing procedure, which guarantees small
MSEs of the estimator. Nonetheless, fixing the threshold may not control
bias well under ${\rm H}_{a,n}$; see a simulation study in Section
\ref{subsec:fixing_gamma}. If the investigator prefers small biases
in the elastic combining estimator, we recommend setting the lower
bounds of a grid for selecting $\gamma$. Although the elastic confidence
intervals demonstrate good coverage properties in our simulation under
all hypotheses ${\rm H}_{0}$, ${\rm H}_{a,n}$ and ${\rm H}_{a}$,
an open problem remains for the post-selection inference after a data-adaptive
selection of the threshold in the testing procedure, which will be
rigorously analyzed theoretically and empirically in the future study.

The proposed framework can also be extended to individualized treatment
regime learning \citep{chu2022targeted,wu2021transfer,wu2022integrativeR}
and the data integration problem of combining probability and non-probability
samples \citep{yang2019doubly,yang2018integration,yang2020statistical}.
However, an additional complication arises due to the mixed design-based
and super-population inference framework, which will be overcome in
future research.

\section*{Supplementary material}

\textcolor{black}{The R package ``ElasticIntegrative'' is available
at \href{https://github.com/Gaochenyin/ElasticIntegrative}{https://github.com/Gaochenyin/ElasticIntegrative}
for implementing the proposed method. Supplementary material online
includes technical details, proofs, additional simulation, the R package
\textquotedblleft ElasticIntegrative,\textquotedblright{} and a README
file providing instructions on downloading the R package and accessing
the codes for reproducing the simulation results.}

\bibliographystyle{dcu}
\bibliography{ci_all,ciwDuplicate}

\pagebreak{}

\pagenumbering{arabic} 
\renewcommand*{\thepage}{S\arabic{page}}

\global\long\def\theequation{S\arabic{equation}}%
\setcounter{equation}{0}

\global\long\def\thesection{S\arabic{section}}%
\setcounter{section}{0}

\global\long\def\thetable{S\arabic{table}}%
\setcounter{table}{0}

\global\long\def\thelemma{S\arabic{lemma}}%
\setcounter{lemma}{0}

\global\long\def\thetheorem{S\arabic{theorem}}%
\setcounter{theorem}{0}

\global\long\def\thecondition{S\arabic{condition}}%
\setcounter{condition}{0}

\global\long\def\theremark{S\arabic{remark}}%
\setcounter{remark}{0}

\global\long\def\thestep{S\arabic{step}}%
\setcounter{step}{0}

\global\long\def\theassumption{S\arabic{assumption}}%
\setcounter{assumption}{0}

\global\long\def\thefigure{S\arabic{figure}}%
\setcounter{figure}{0}

\global\long\def\theproposition{S\arabic{proposition}}%
\setcounter{proposition}{0} 
\begin{center}
\textbf{\huge{}{}{}{}{}{} Supplementary materials for \textquotedbl Elastic
integrative analysis of randomized trial and real-world data for treatment
heterogeneity estimation\textquotedbl}{\huge{}{}}\\
 {\huge{}{} }{\huge\par}
\par\end{center}

\begin{center}
 
\par\end{center}

\begin{center}
\textbf{\large{}{}{}by Yang, Gao, Zeng, and Wang}{\large\par}
\par\end{center}

\begin{center}
 
\par\end{center}

\bigskip{}

\setcounter{page}{1}

\noindent Section \ref{sec:DR} provides technical details and proofs
for (rate) doubly robust estimation.

\noindent Section \ref{sec:Test=00003D00003D000026Elastic} provides
technical details for the test and elastic estimator.

\noindent Section \ref{sec:Inference} provides technical details
for inference.

\noindent Section \ref{sec:Additional-simulation-results} provides
additional simulation results and studies.

\section{Technical details for rate doubly robust estimation \label{sec:DR}}

\subsection{Regularity conditions for rate double robustness}

Recall that $\mathbb{P}_{N}$ denotes the empirical measure over the
combined RT and RW data; i.e., $\mathbb{P}_{N}h(V)=N^{-1}\sum_{i\in\mathcal{A}\cup\mathcal{B}}h(V_{i})$.
Also, $\mathbb{P}\{h(V)\}=\int h(V)\de\mathbb{P}$ denotes the expectation
of $h(V)$ over the data generative distribution.

\begin{assumption}\label{asump:DML}The following regularity conditions
hold: 
\begin{enumerate}
\item $S_{\psi_{0}}(V)$ belongs to a Donsker class \citep{van1996weak}; 
\item $S_{\psi}(V)$ is differentiable in $\psi,$ and $\E\left\{ \partial S_{\psi_{0}}(V)/\partial\psi^{\T}\right\} $
exists and is invertible; 
\item there exists a constant $C$ such that $|\partial\tau_{\psi_{0}}(X)/\partial\psi|\leq C$
and $\large\vert\left\{ \sigma_{\delta}^{2}(X)\right\} ^{-1}\large\vert\leq C$
almost surely. 
\end{enumerate}
\end{assumption}

\subsection{Proof of Theorem \ref{Thm:Consistency-DML}}
\begin{proof}
Under Assumption \ref{asump:DML}, by the standard Taylor expansion,
we have 
\begin{eqnarray}
N^{1/2}\left(\widehat{\psi}_{\eff}-\psi_{0}\right) & = & \left[\E\left\{ \partial S_{\psi_{0}}(V)/\partial\psi^{\T}\right\} \right]^{-1}N^{1/2}\mathbb{P}_{N}\widehat{S}_{\psi_{0}}(V)+o_{\pp}(N^{-1/2}).\label{eq:taylor}
\end{eqnarray}
Moreover, we have 
\begin{align}
\mathbb{P}_{N}\widehat{S}_{\psi_{0}}(V) & =(\mathbb{P}_{N}-\mathbb{P})\widehat{S}_{\psi_{0}}(V)+\mathbb{P}\widehat{S}_{\psi_{0}}(V)\nonumber \\
 & =(\mathbb{P}_{N}-\mathbb{P})S_{\psi_{0}}(V)+\mathbb{P}\widehat{S}_{\psi_{0}}(V)+o_{\pp}(N^{-1/2})\nonumber \\
 & =\mathbb{P}_{N}S_{\psi_{0}}(V)+\mathbb{P}\widehat{S}_{\psi_{0}}(V)+o_{\pp}(N^{-1/2}),\label{eq:tau_tau0}
\end{align}
where the third equality follows because of $\mathbb{P}S_{\psi_{0}}(V)=0$
by Assumption \ref{assump:o1}.

We now show that the second term in (\ref{eq:tau_tau0}), $\mathbb{P}\widehat{S}_{\psi_{0}}(V)$,
is a small order term. We write 
\begin{equation}
\mathbb{P}\widehat{S}_{\psi_{0}}(V)=\mathbb{P}\left[\left\{ \frac{\partial\tau_{\psi_{0}}(X)}{\partial\psi}\right\} \left\{ \sigma_{\delta}^{2}(X)\right\} ^{-1}\left\{ \mu_{\delta}(X)-\widehat{\mu}_{\delta}(X)\right\} \{e_{\delta}(X)-\widehat{e}_{\delta}(X)\}\right].\label{eq:delta}
\end{equation}
By trial design, the propensity score of treatment is known; i.e.,
$\widehat{e}_{1}(X)=e_{1}(X)$. Therefore, we have 
\[
\mathbb{P}\left[\delta\left\{ \frac{\partial\tau_{\psi_{0}}(X)}{\partial\psi}\right\} \left\{ \sigma_{1}^{2}(X)\right\} ^{-1}\left\{ \mu_{1}(X)-\widehat{\mu}_{1}(X)\right\} \{e_{1}(X)-\widehat{e}_{1}(X)\}\right]=0.
\]
Then, (\ref{eq:delta}) becomes 
\[
\mathbb{P}\widehat{S}_{\psi_{0}}(V)=\mathbb{P}\left[(1-\delta)\left\{ \frac{\partial\tau_{\psi_{0}}(X)}{\partial\psi}\right\} \left\{ \sigma_{0}^{2}(X)\right\} ^{-1}\left\{ \mu_{0}(X)-\widehat{\mu}_{0}(X)\right\} \{e_{0}(X)-\widehat{e}_{0}(X)\}\right].
\]
By the Cauchy-Schwarz inequality, $\vert\mathbb{P}\widehat{S}_{\psi_{0}}(V)\vert$
is bounded by 
\begin{equation}
C^{2}\lVert\mu_{0}(X)-\widehat{\mu}_{0}(X)\rVert\times\rVert e_{0}(X)-\widehat{e}_{0}(X)\rVert.\label{eq:upperBound}
\end{equation}

Under the assumptions in Theorem \ref{Thm:Consistency-DML}, the product
\eqref{eq:upperBound} is $o_{\pp}(N^{-1/2})$. Therefore, $\mathbb{P}\widehat{S}_{\psi_{0}}(V)$
in \eqref{eq:tau_tau0} is asymptotically negligible. Therefore, (\ref{eq:tau_tau0})
becomes 
\begin{equation}
\mathbb{P}_{N}\widehat{S}_{\psi_{0}}(V)=o_{\pp}(N^{-1/2}).\label{eq:o1}
\end{equation}
Combining (\ref{eq:taylor}), (\ref{eq:tau_tau0}) and (\ref{eq:o1}),
the results in Theorem \ref{Thm:Consistency-DML} follow. 
\end{proof}

\subsection{Sieves estimation \label{subsec:Conditions}}

We illustrate Theorem \ref{Thm:Consistency-DML} by the method of
sieves. For simplicity, we consider the power series, although our
discussion extends to general sieve basis functions such as Fourier
series, splines, wavelets, and artificial neural networks (see, e.g.,
\citealp{chen2007large}). Let $d_{X}$ be the dimension of $X$.
For a $d_{X}$-vector of non-negative integers $\mathbb{\kappa}=(\kappa_{1},\ldots,\kappa_{d_{X}})$,
let $|\mathbf{\kappa}|=\sum_{l=1}^{d_{X}}\kappa_{l}$ and $X^{\mathbf{\kappa}}=\prod_{l=1}^{d_{X}}X_{l}^{\kappa_{l}}.$
Define a series $\{\mathbf{\kappa}(k):k=1,2,\ldots\}$ for all distinct
vectors of $\mathbf{\kappa}$ such that $|\mathbf{\kappa}(k)|\leq|\mathbf{\kappa}(k+1)|$.
Based on this series, we consider a $K$-vector $g(X)=\{g_{1}(X),\ldots,g_{K}(X)\}^{\top}=\{X^{\kappa(1)},\ldots,X^{\kappa(K)}\}^{\top}$.

To accommodate different type of variables, we approximate $e_{0}(X)$
and $\mu_{\delta}(X)$ by the generalized sieves functions 
\begin{equation}
\text{expit}\left\{ \alpha^{*\T}g(X)\right\} ,\qquad h\left\{ \eta_{\delta}^{*\top}g(X)\right\} ,\label{eq:sieve1}
\end{equation}
where expit$(\cdot)$ is the inverse of logit$(\cdot)$, $h(\cdot)$
is a certain link function, e.g., for a continuous outcome, $h(\cdot)$
is an identity function, and for a binary outcome, $h(\cdot)$ is
an expit function, and 
\begin{eqnarray*}
\alpha^{*} & = & \arg\min_{\alpha}\E[e_{0}(X)-\text{expit}\{\alpha^{\T}g(X)\}]^{2},\\
\eta_{\delta}^{*} & = & \arg\min_{\eta}\E\left[\mu_{\delta}(X)-h\{\eta_{\delta}^{\T}g(X)\}\right]^{2}.
\end{eqnarray*}

We provide the regularity conditions below, under which the sieves
estimators satisfy the conditions in Theorem \ref{Thm:Consistency-DML},
and therefore $\widehat{\psi}_{\eff}$ enjoys the properties in Theorem
\ref{Thm:Consistency-DML}.

\begin{assumption}\label{assump:sieveRegu}

The following regularity conditions hold: 
\begin{enumerate}
\item the density of $X$, $f(X)$, is bounded above and below away from
$0$ on $\mathcal{X}$; 
\item $\E[\{Y(a)\}^{2}]<\infty$, for $a=0,1$; 
\item $e_{0}(X)$ is $s_{1}$-times continuously differentiable, and $\mu_{\delta}(X)$
is $s_{2}$-times continuously differentiable; let $s_{0}=\min(s_{1},s_{2})$,
which satisfies that $s_{0}\geq3d_{X}$; 
\item there exist constant $l$ and $u$ such that $l\le\rho_{\mathrm{min}}\{g(X)^{\top}g(X)\}\le\rho_{\mathrm{max}}\{g(X)^{\top}g(X)\}\le u$
almost surely, where $\rho_{\mathrm{min}}(M)$ and $\rho_{\mathrm{\mathrm{max}}}(M)$
denote the minimum and maximum eigenvalues of a matrix; 
\item $K=O\left(n^{\frac{d_{X}}{s_{0}-d_{X}}}\right)$. 
\end{enumerate}
\end{assumption}

Under certain regularity conditions, we show that the method of sieves
allows flexible models for $e_{0}(X)$ and $\mu_{\delta}(X)$ and
also satisfies Assumption \ref{assump:o1}. Let $\mathcal{X}\subseteq\mathbb{R}^{d_{X}}$
be the support of $X$. Assume that $\mathcal{X}$ is a Cartesian
product of compact intervals, i.e. $\mathcal{X}=\prod_{j=1}^{p}\left[l_{j},u_{j}\right]$,
$l_{j},u_{j}\in\mathbb{R}$.

Following \citet{newey1997convergence}, under Assumption \ref{assump:sieveRegu},
the bounds for the deterministic differences between the true functions
and the sieves approximations are 
\begin{align}
\sup_{x\in\mathbb{\mathcal{X}}}|e_{0}(x)-\text{expit}\{\alpha^{*\T}g(x)\}| & =O\left\{ K^{1-s_{1}/(2d_{X})}\right\} ,\label{eq:true_psedoTrue}\\
\sup_{x\in\mathcal{X}}|\mu_{\delta}(x)-h\{\eta_{\delta}^{*\T}g(x)\}| & =O\left\{ K^{1-s_{2}/(2d_{X})}\right\} .\nonumber 
\end{align}
On the one hand, given (\ref{eq:true_psedoTrue}), the approximation
errors can be made sufficiently small if the number of basis functions
$K$ is large. On the other hand, in order to control the variance
of the sieves estimators, $K$ should increase slowly with the sample
size $n$. Concretely, under regularity conditions, $\widehat{\alpha}-\alpha^{*}=O_{\pp}(K/n)$
and $\widehat{\eta}_{\delta}-\eta_{\delta}^{*}=O_{\pp}(K/n)$. Therefore,
the bias of the sieves approximation for $e_{0}(X)$ is $O_{\pp}\left\{ K^{1-s_{1}/(2d_{X})}\right\} $
and the variance is $O_{\pp}\left(K/n\right)$. Similarly, the bias
of the sieves approximation for $\mu_{\delta}(X)$ is $O_{\pp}\left\{ K^{1-s_{2}/(2d_{X})}\right\} $
and the variance is $O_{\pp}\left(K/n\right)$. With $K=O\left\{ n^{d_{X}/(s_{0}-d_{X})}\right\} $
in Assumption \ref{assump:sieveRegu} (v), it balances the squared
bias and variance with both $O_{\pp}\left\{ n^{(2d_{X}-s_{0})/(s_{0}-d_{X})}\right\} $.
Then under Assumption \ref{assump:sieveRegu} (iii), Assumption \ref{assump:o1}
holds.

\section{Technical details for the test and elastic estimator\label{sec:Test=00003D00003D000026Elastic} }

\begin{table}[h]
\caption{\label{tab:Summary-of-notation}Notation and properties}

\centering

\vspace{0.5cm}

\begin{tabular}{lll}
\toprule 
Notation  & Definition  & Property\tabularnewline
\midrule 
${\cI}_{\rw}$  & $\E\{S_{\os,\psi_{0}}(V)^{\otimes2}\mid\delta=0\}$  & \tabularnewline
${\cI}_{\rt}$  & $\E\{S_{\rt,\psi_{0}}(V)^{\otimes2}\mid\delta=1\}$  & \tabularnewline
$\Gamma$  & ${\cI}_{\rt}^{-1}{\cI}_{\rw}\rho^{-1/2}$  & \tabularnewline
$\Sigma_{SS}$  & $\Gamma^{\T}{\cI}_{\rt}\Gamma+{\cI}_{\rw}$  & \tabularnewline
$V_{\rt}$  & $(\rho{\cI}_{\rt})^{-1}$  & \tabularnewline
$V_{\eff}$  & $(\rho{\cI}_{\rt}+{\cI}_{\rw})^{-1}$  & \tabularnewline
$V_{\rteff}$  & $V_{\rt}-V_{\eff}$  & $V_{\rteff}\Sigma_{SS}^{-1}=V_{\eff}^{2}$, $V_{\rteff}=V_{\eff}\Sigma_{SS}V_{\eff}$\tabularnewline
$\cZ_{\rw}$  & $\rmN(\mu_{\rw},{\cI}_{\rw})$  & \tabularnewline
$\cZ_{\rt}$  & $\rmN(0,{\cI}_{\rt})$  & \tabularnewline
\multicolumn{3}{l}{Under ${\rm H}_{0}$, $\mu_{\rw}=0$; under ${\rm H}_{a},$ $\mu_{\rw}=\infty$;
and under ${\rm H}_{a,n}$, $\mu_{\rw}=\eta$.}\tabularnewline
$\cZ_{1}$  & $\Sigma_{SS}^{-1/2}(\cZ_{\rw}-\Gamma^{\T}\cZ_{\rt})$  & $\sim\rmN(\Sigma_{SS}^{-1/2}\mu_{\rw},I_{p\times p})$\tabularnewline
$\cZ_{2}$  & $V_{\eff}^{1/2}(\cZ_{\rw}+\rho^{1/2}\cZ_{\rt})$  & $\sim\rmN(V_{\eff}^{1/2}\mu_{\rw},I_{p\times p})$\tabularnewline
 & $\mu_{1}=$$\Sigma_{SS}^{-1/2}\eta$ and $\mu_{2}=$$V_{\eff}^{1/2}\eta$  & \tabularnewline
\midrule 
\multirow{2}{*}{Definition} & Representation  & Orthogonal representation\tabularnewline
 & using $\cZ_{\rw}$ and $\cZ_{\rt}$  & using $\cZ_{1}$ and $\cZ_{2}$\tabularnewline
\midrule 
$T_{\infty}({\cZ}_{\rt},{\cZ}_{\rw})$  & $({\cZ}_{\rw}-\Gamma^{\T}{\cZ}_{\rt})^{\T}\Sigma_{SS}^{-1}({\cZ}_{\rw}-\Gamma^{\T}{\cZ}_{\rt})$  & $\cZ_{1}^{\T}\cZ_{1}$\tabularnewline
$\text{\ensuremath{\mathcal{N}_{\rt}}}({\cZ}_{\rt})$  & $-(\rho{\cI}_{\rt})^{-1}(\rho^{1/2}{\cZ}_{\rt})$  & $V_{\rteff}^{1/2}\cZ_{1}-V_{\eff}^{1/2}\cZ_{2}$\tabularnewline
bias  & $0$  & $(V_{\rteff}^{1/2}\Sigma_{SS}^{-1/2}-V_{\eff})\eta=0$\tabularnewline
var  & $(\rho{\cI}_{\rt})^{-1}$  & $V_{\rteff}+V_{\eff}=V_{\rt}$\tabularnewline
$\mathcal{N}_{\rtrw}({\cZ}_{\rt},{\cZ}_{\rw})$  & $-(\rho{\cI}_{\rt}+{\cI}_{\rw})^{-1}(\rho^{1/2}{\cZ}_{\rt}+{\cZ}_{\rw})$  & $-V_{\eff}^{1/2}\cZ_{2}$\tabularnewline
bias  & $-(\rho{\cI}_{\rt}+{\cI}_{\rw})^{-1}\mu_{\rw}$  & $-V_{\eff}\eta$\tabularnewline
var  & $(\rho{\cI}_{\rt}+{\cI}_{\rw})^{-1}$  & $V_{\eff}$\tabularnewline
\bottomrule
\end{tabular}
\end{table}

We introduce additional notation. For convenience, Table \ref{tab:Summary-of-notation}
summarizes additional notation and their properties for references.

To gauge the strength of the evidence, we characterize the asymptotic
distribution of $n^{-1/2}\sum_{i\in\mathcal{B}}$ $\widehat{S}_{\rw,\widehat{\psi}_{\rt}}(V_{i})$
under ${\rm H}_{0},$ ${\rm H}_{a}$ and ${\rm H}_{a,n}$. Toward
this end, we introduce two random variables 
\begin{align*}
{\cZ}_{\rt}\sim\rmN(0,{\cI}_{\rt}), & \quad{\cZ}_{\rw}\sim\rmN(\mu_{\rw},{\cI}_{\rw}),\\
{\cI}_{\rt}=\E\{S_{\rt,\psi_{0}}(V)^{\otimes2}\mid\delta=1\}, & \quad{\cI}_{\rw}=\E\{S_{\os,\psi_{0}}(V)^{\otimes2}\mid\delta=0\}.
\end{align*}
Under ${\rm H}_{0}$, $\mu_{\rw}=0$; under ${\rm H}_{a},$ $\mu_{\rw}=\infty$;
and under ${\rm H}_{a,n}$, $\mu_{\rw}=\eta$. By the generalized
information equality, we also have ${\cI}_{\rt}=\E\{\partial S_{\rt,\psi_{0}}(V)/\partial\psi\mid\delta=1\}$
and ${\cI}_{\rw}=\E\{\partial S_{\os,\psi_{0}}(V)/\partial\psi\mid\delta=0\}$.
Based on $\cZ_{\rt}$ and $\cZ_{\rw}$, by central limit theorem,
we have

\[
m^{-1/2}\sum_{i\in\mathcal{A}}S_{\rt,\psi_{0}}(V_{i})\stackrel{\cdot}{\sim}{\cZ}_{\rt},\ n^{-1/2}\sum_{i\in\mathcal{B}}S_{\rw,\psi_{0}}(V_{i})\stackrel{\cdot}{\sim}{\cZ}_{\rw}.
\]

Using $\cZ_{\rt}$ and $\cZ_{\rw}$ is also helpful to characterize
the asymptotic distributions of $\widehat{\psi}_{\rt}$, $\widehat{\psi}_{\eff}$,
and our proposed estimator in the later section. Building on the asymptotic
properties of the score functions, we have 
\begin{equation}
n^{1/2}(\widehat{\psi}_{\rt}-\psi_{0})\stackrel{\cdot}{\sim}\text{\ensuremath{\mathcal{N}_{\rt}}},\ \ n^{1/2}(\widehat{\psi}_{\eff}-\psi_{0})\stackrel{\cdot}{\sim}\mathcal{N}_{\rtrw},\label{eq:normal}
\end{equation}
where 
\begin{equation}
\text{\ensuremath{\mathcal{N}_{\rt}}}=-(\rho{\cI}_{\rt})^{-1}(\rho^{1/2}{\cZ}_{\rt}),\ \ \mathcal{N}_{\rtrw}=-(\rho{\cI}_{\rt}+{\cI}_{\rw})^{-1}(\rho^{1/2}{\cZ}_{\rt}+{\cZ}_{\rw}).\label{eq:normal2}
\end{equation}

The results in (\ref{eq:normal}) facilitate an easy comparison of
$\widehat{\psi}_{\rt}$ and $\widehat{\psi}_{\eff}$. Under the idealistic
assumption, $\V_{a}\{n^{1/2}(\widehat{\psi}_{\rt}-\psi_{0})\}=V_{\rt}=(\rho{\cI}_{\rt})^{-1}$,
and $\V_{a}\{n^{1/2}(\widehat{\psi}_{\eff}-\psi_{0})\}=V_{\eff}=(\rho{\cI}_{\rt}+{\cI}_{\rw})^{-1}$,
where $\rho$ can be viewed as the relative sample size of the RT
data compared with the RW data. It is clear that the difference between
$V_{\rt}$ and $V_{\eff}$ is $V_{\rteff}=(\rho{\cI}_{\rt})^{-1}-(\rho{\cI}_{\rt}+{\cI}_{\rw})^{-1}>0$
and $\widehat{\psi}_{\eff}$ gains efficiency by using the additional
information in the RW data.

We characterize the asymptotic distribution of $n^{-1/2}\sum_{i\in\mathcal{B}}\widehat{S}_{\rw,\widehat{\psi}_{\rt}}(V_{i})$
using $\cZ_{\rt}$ and $\cZ_{\rw}$, which is the building block to
constructing the test statistic.

\begin{proposition}\label{prop:scoreN}

Suppose assumptions in Theorem \ref{Thm:Consistency-DML} hold except
that Assumption \ref{Asump:rand-rwd} may be violated. Let $\Gamma={\cI}_{\rt}^{-1}{\cI}_{\rw}\rho^{-1/2}$.
Then, 
\begin{equation}
n^{-1/2}\sum_{i\in\mathcal{B}}\widehat{S}_{\rw,\widehat{\psi}_{\rt}}(V_{i})\stackrel{\cdot}{\sim}{\cZ}_{\rw}-\Gamma^{\T}{\cZ}_{\rt}.\label{eq:S-dist}
\end{equation}

\begin{enumerate}
\item Under ${\rm H}_{0},$ (\ref{eq:S-dist}) is a normal distribution
with mean $0$ and variance $\Sigma_{SS}$. 
\item Under ${\rm H}_{a},$ (\ref{eq:S-dist}) is $\pm\infty$. 
\item Under ${\rm H}_{a,n},$ (\ref{eq:S-dist}) is a normal distribution
with mean $\eta$ and variance $\Sigma_{SS}$. 
\end{enumerate}
\end{proposition}

In Proposition \ref{prop:scoreN}, because of the intrinsic connection
between ${\rm H}_{a,n}$ and the other two hypotheses, the result
in c) reduces to that in a) by considering $\eta=0$ and to that in
b) by considering $\eta=\pm\infty$.

\begin{proposition}\label{Thm:T}Suppose assumptions in Theorem \ref{Thm:Consistency-DML}
hold except that Assumption \ref{Asump:rand-rwd} may be violated. 
\begin{enumerate}
\item Under ${\rm H}_{0},$ we have $T\stackrel{\cdot}{\sim}\chi_{p}^{2},$
a Chi-square distribution with degrees of freedom $p$, as $n\rightarrow\infty.$ 
\item Under ${\rm H}_{a},$ we have $T\rightarrow\infty,$ almost surely,
as $n\rightarrow\infty.$ 
\item Under ${\rm H}_{a,n},$ we have $T\stackrel{\cdot}{\sim}\chi_{p}^{2}(\lambda),$
a non-central Chi-square distribution with degrees of freedom $p$
and non-centrality parameter $\lambda=\eta^{\T}\Sigma_{SS}^{-1}\eta$,
as $n\rightarrow\infty.$ 
\end{enumerate}
\end{proposition}

Not surprisingly, in Theorem \ref{Thm:T}, $\chi_{p}^{2}(\lambda)$
in c) becomes $\chi_{p}^{2}$ in a) by considering $\eta=0$ and to
$\infty$ in b) by considering $\eta=\pm\infty$.

Recall that the asymptotic distributions of $\widehat{\psi}_{\rt}$
and $\widehat{\psi}_{\eff}$ can be easily represented by $\text{\ensuremath{\mathcal{N}_{\rt}}}$
and $\mathcal{N}_{\rtrw}$ as in (\ref{eq:normal}). However, the
distributions of $\widehat{\psi}_{\rt}\mid(T\geq c_{\gamma})$ and
$\widehat{\psi}_{\eff}\mid(T<c_{\gamma})$ are those constrained to
the acceptance and rejection regions of the test. Below, we characterize
these asymptotic distributions, taking into account that the estimators
and the test may be correlated asymptotically.

Let the asymptotic distribution of $T$ be represented by 
\begin{equation}
T_{\infty}=({\cZ}_{\rw}-\Gamma^{\T}{\cZ}_{\rt})^{\T}\Sigma_{SS}^{-1}({\cZ}_{\rw}-\Gamma^{\T}{\cZ}_{\rt}).\label{eq:Tinfty}
\end{equation}
By (\ref{eq:T-dist-rep}), asymptotically, the event $T<c_{\gamma}$
corresponds to $T_{\infty}<c_{\gamma}$ and the event $T\geq c_{\gamma}$
corresponds to $T_{\infty}\geq c_{\gamma}$. We show in the supplementary
material that 
\begin{eqnarray}
\widehat{\psi}_{\rt}\mid(T\geq c_{\gamma}) & \stackrel{\cdot}{\sim} & \text{\ensuremath{\mathcal{N}_{\rt}}}\mid(T_{\infty}\geq c_{\gamma}),\label{eq:nt-rt}\\
\widehat{\psi}_{\eff}\mid(T<c_{\gamma}) & \stackrel{\cdot}{\sim} & \mathcal{N}_{\rtrw}\mid(T_{\infty}<c_{\gamma}).\label{eq:nt-rt+rw}
\end{eqnarray}
The limiting distributions in (\ref{eq:nt-rt}) and (\ref{eq:nt-rt+rw})
are multivariate normal distributions with ellipsoid truncation \citep{tallis1963elliptical}.
Specifically, (\ref{eq:nt-rt}) is the multivariate normal distribution
$\text{\ensuremath{\mathcal{N}_{\rt}}}$ outside the boundary of the
ellipsoid $T_{\infty}=c_{\gamma}$, and (\ref{eq:nt-rt+rw}) is the
multivariate normal distribution $\mathcal{N}_{\rtrw}$ inside the
boundary of the ellipsoid $T_{\infty}=c_{\gamma}$.

Based on (\ref{eq:Tinfty}), it is insightful to recognize that $T_{\infty}$
is fully characterized by $\cZ_{\rw}-\Gamma^{\T}\cZ_{\rt}$. We introduce
two normal random vectors 
\[
\cZ_{1}=\Sigma_{SS}^{-1/2}(\cZ_{\rw}-\Gamma^{\T}\cZ_{\rt}),\ \ \cZ_{2}=V_{\eff}^{1/2}(\cZ_{\rw}+\rho^{1/2}\cZ_{\rt}),
\]
which are multivariate normal distributions with means $\Sigma_{SS}^{-1/2}\mu_{\rw}$
and $V_{\eff}^{1/2}\mu_{\rw}$, respectively, and covariance $I_{p\times p}$.
Under ${\rm H}_{0}$, it is easy to verify that the covariance of
$\cZ_{1}$ and $\cZ_{2}$ is zero. Because uncorrelated normal random
vectors are\textbf{ }independent, $\cZ_{1}$ and $\cZ_{2}$ are independent.
Similarly, we can show that under ${\rm H}_{a}$ and ${\rm H}_{a,n}$,
$\cZ_{1}$ and $\cZ_{2}$ are independent. Translating the asymptotic
distributions (\ref{eq:nt-rt}) and (\ref{eq:nt-rt+rw}) into the
ones using $\cZ_{1}$ and $\cZ_{2}$ makes the characterization easier.
First, $T_{\infty}$ is equivalent to $\cZ_{1}^{\T}\cZ_{1}$.

\subsection{Proof of Proposition \ref{Thm:T}}
\begin{proof}
Because we construct the test statistic $T$ based on $\sum_{i\in\mathcal{B}}\widehat{S}_{\rw,\widehat{\psi}_{\rt}}(V_{i})$,
where $\widehat{\psi}_{\rt}$ satisfies $\sum_{i\in\mathcal{A}}\widehat{S}_{\rt,\widehat{\psi}_{\rt}}(V_{i})=0,$
we first investigate the statistical properties of $\sum_{i\in\mathcal{A}}\widehat{S}_{\rt,\widehat{\psi}_{\rt}}(V_{i})$
and $\sum_{i\in\mathcal{B}}\widehat{S}_{\rw,\widehat{\psi}_{\rt}}(V_{i})$,
both properly scaled.

By the Taylor expansion, we have 
\begin{eqnarray*}
m^{-1/2}\sum_{i\in\mathcal{A}}\widehat{S}_{\rt,\widehat{\psi}_{\rt}}(V_{i}) & = & m^{-1/2}\sum_{i\in\mathcal{A}}S_{\rt,\psi_{0}}(V_{i})\\
 &  & +m^{-1/2}\E\left\{ \sum_{i\in\mathcal{A}}\frac{\partial S_{\rt,\psi_{0}}(V_{i})}{\partial\psi^{\T}}\right\} (\widehat{\psi}_{\rt}-\psi_{0})+o_{\pp}(1)\\
 & = & m^{-1/2}\sum_{i\in\mathcal{A}}S_{\rt,\psi_{0}}(V_{i})+{\cI}_{\rt}\left\{ m^{1/2}(\widehat{\psi}_{\rt}-\psi_{0})\right\} +o_{\pp}(1),
\end{eqnarray*}
where $o_{\pp}(1)$ follows by a similar argument for (\ref{eq:o1}).
Then, we have 
\begin{equation}
m^{1/2}(\widehat{\psi}_{\rt}-\psi_{0})=-({\cI}_{\rt})^{-1}m^{-1/2}\sum_{i\in\mathcal{A}}S_{\rt,\psi_{0}}(V_{i})+o_{\pp}(1).\label{eq:p1}
\end{equation}

By the Taylor expansion, we have 
\begin{eqnarray}
n^{-1/2}\sum_{i\in\mathcal{B}}\widehat{S}_{\rw,\widehat{\psi}_{\rt}}(V_{i}) & = & n^{-1/2}\sum_{i\in\mathcal{B}}S_{\rw,\psi_{0}}(V_{i})\nonumber \\
 &  & +n^{-1/2}\E\left\{ \sum_{i\in\mathcal{B}}\frac{\partial S_{\rw,\psi_{0}}(V_{i})}{\partial\psi^{\T}}\right\} (\widehat{\psi}_{\rt}-\psi_{0})+o_{\pp}(1)\nonumber \\
 & = & n^{-1/2}\sum_{i\in\mathcal{B}}S_{\rw,\psi_{0}}(V_{i})+{\cI}_{\rw}\left\{ n^{1/2}(\widehat{\psi}_{\rt}-\psi_{0})\right\} +o_{\pp}(1),\label{eq:p2}
\end{eqnarray}
where $o_{\pp}(1)$ follows by a similar argument for (\ref{eq:o1}).

Combining (\ref{eq:p1}) and (\ref{eq:p2}) leads to 
\begin{multline}
n^{-1/2}\sum_{i\in\mathcal{B}}\widehat{S}_{\rw,\widehat{\psi}_{\rt}}(V_{i})=\left\{ n^{-1/2}\sum_{i\in\mathcal{B}}S_{\rw,\psi_{0}}(V_{i})\right\} \\
-{\cI}_{\rw}({\cI}_{\rt})^{-1}\left(\frac{n}{m}\right)^{1/2}\left\{ m^{-1/2}\sum_{i\in\mathcal{A}}S_{\rt,\psi_{0}}(V_{i})\right\} +o_{\pp}(1)\stackrel{\cdot}{\sim}{\cZ}_{\rw}-\Gamma^{\T}{\cZ}_{\rt}.\label{eq:T-dist}
\end{multline}
Specifically, ${\cZ}_{\rw}-\Gamma^{\T}{\cZ}_{\rt}$ follows $\rmN(0,\Sigma_{SS})$,
where $\Sigma_{SS}=\V({\cZ}_{\rw}-\Gamma^{\T}{\cZ}_{\rt})={\cI}_{\rw}+\Gamma^{\T}{\cI}_{\rt}\Gamma.$
Therefore, it follows that 
\begin{equation}
T\stackrel{\cdot}{\sim}({\cZ}_{\rw}-\Gamma^{\T}{\cZ}_{\rt})\Sigma_{SS}^{-1}({\cZ}_{\rw}-\Gamma^{\T}{\cZ}_{\rt})\sim\chi_{p}^{2}.\label{eq:T-dist-rep}
\end{equation}
\end{proof}

\subsection{Asymptotic distribution of an estimator given the test constraint }

We first provide a useful proposition.

\begin{proposition}\label{prop:weakCvg}Suppose the assumptions in
Theorem \ref{Thm:T} holds. For $p\times p$ matrices $A$ and $B$,
\begin{multline}
\left.A^{\T}\left\{ n^{-1/2}\sum_{i\in\mathcal{A}}S_{\psi_{0}}(V_{i})\right\} +B^{\T}\left\{ m^{-1/2}\sum_{i\in\mathcal{B}}S_{\psi_{0}}(V_{i})\right\} \right\vert (T<c_{\gamma})\\
\stackrel{\cdot}{\sim}A^{\T}{\cZ}_{\rt}+B^{\T}{\cZ}_{\rw}\mid(T_{\infty}<c_{\gamma}),\label{eq:prop1-1}
\end{multline}
and 
\begin{multline}
\left.A^{\T}\left\{ n^{-1/2}\sum_{i\in\mathcal{A}}S_{\psi_{0}}(V_{i})\right\} +B^{\T}\left\{ m^{-1/2}\sum_{i\in\mathcal{B}}S_{\psi_{0}}(V_{i})\right\} \right\vert (T\geq c_{\gamma})\\
\stackrel{\cdot}{\sim}A^{\T}{\cZ}_{\rt}+B^{\T}{\cZ}_{\rw}\mid(T_{\infty}\geq c_{\gamma}).\label{eq:prop1-2}
\end{multline}

\end{proposition} 
\begin{proof}
Denote $Q_{n}=A^{\T}\left\{ n^{-1/2}\sum_{i\in\mathcal{A}}S_{\psi_{0}}(V_{i})\right\} +B^{\T}\left\{ m^{-1/2}\sum_{i\in\mathcal{B}}S_{\psi_{0}}(V_{i})\right\} $
and $Q=A^{\T}{\cZ}_{\rt}+B^{\T}{\cZ}_{\rw}$. By Theorem \ref{Thm:T},
we have 
\[
\left(\begin{array}{c}
Q_{n}\\
T
\end{array}\right)\stackrel{\cdot}{\sim}\left(\begin{array}{c}
Q\\
T_{\infty}
\end{array}\right).
\]
By the continuous mapping theorem, we have 
\[
\left(\begin{array}{c}
Q_{n}\\
\bone(T<c_{\gamma})
\end{array}\right)\stackrel{\cdot}{\sim}\left(\begin{array}{c}
Q\\
\bone(T_{\infty}<c_{\gamma})
\end{array}\right).
\]
By the Portmanteau Theorem, for all bounded continuous functions $h:\R^{p}\mapsto\R$,
we have 
\[
\E\{h(Q_{n})\bone(T<c_{\gamma})\}\rightarrow\E\{h(Q)\bone(T_{\infty}<c_{\gamma})\}.
\]
For all bounded continuous functions $h:\R^{p}\mapsto\R$, we have
\begin{multline*}
\E\{h(Q_{n})\mid\bone(T<c_{\gamma})\}=\frac{\E\{h(Q_{n})\bone(T<c_{\gamma})\}}{\pr(T<c_{\gamma})}\\
\rightarrow\frac{\E\{h(Q)\bone(T_{\infty}<c_{\gamma})\}}{\pr\{\bone(T_{\infty}<c_{\gamma})\}}=\E\{h(Q)\mid\bone(T_{\infty}<c_{\gamma})\},
\end{multline*}
as $n\rightarrow\infty$. Applying the Portmanteau Theorem, $Q_{n}\mid\bone(T<c_{\gamma})\stackrel{\cdot}{\sim}Q\mid\bone(T_{\infty}<c_{\gamma})$.
The results (\ref{eq:prop1-1}) and (\ref{eq:prop1-2}) follow. 
\end{proof}

\paragraph*{Proof of Theorem \ref{Thm:elas}.}
\begin{proof}
Recall that $\widehat{\psi}_{\elas}$ satisfies 
\[
n^{-1/2}\sum_{i\in\mathcal{A}\cup\mathcal{B}}\{\delta_{i}S_{\widehat{\psi}_{\elas}}(V_{i};\widehat{\eta}_{1})+\bone(T<c_{\gamma})(1-\delta_{i})S_{\widehat{\psi}_{\elas}}(V_{i};\widehat{\alpha},\widehat{\eta}_{0})\}=0.
\]
We discuss two cases with $T<c_{\gamma}$ and $T\geq c_{\gamma}$
separately.

First, conditional on $T<c_{\gamma}$, $\widehat{\psi}_{\elas}$ satisfies
$0=n^{-1/2}\sum_{i\in\mathcal{A}\cup\mathcal{B}}\{\delta_{i}S_{\widehat{\psi}_{\elas}}(V_{i};\widehat{\eta}_{1})+(1-\delta_{i})S_{\widehat{\psi}_{\elas}}(V_{i};\widehat{\alpha},\widehat{\eta}_{0})\}.$
By the Taylor expansion, we then have 
\begin{eqnarray}
n^{1/2}(\widehat{\psi}_{\elas}-\psi_{0})\mid(T<c_{\gamma}) & = & \left.\left\{ -\cI_{\rtrw}^{-1}n^{-1/2}\sum_{i\in\mathcal{A}\cup\mathcal{B}}S_{\psi_{0}}(V_{i})+o_{\pp}(1)\right\} \right\vert (T<c_{\gamma})\nonumber \\
 & \stackrel{\cdot}{\sim} & \mathcal{N}_{\rtrw}\mid(T_{\infty}<c_{\gamma}),\label{eq:elas-1}
\end{eqnarray}
where $\cI_{\rtrw}=\rho{\cI}_{\rt}+{\cI}_{\rw},$ $o_{\pp}(1)$ follows
by a similar argument for (\ref{eq:o1}), and $\stackrel{\cdot}{\sim}$
follows by Proposition \ref{prop:weakCvg}.

Second, conditional on $T\geq c_{\gamma}$, $\widehat{\psi}_{\elas}$
satisfies $n^{-1/2}\sum_{i\in\mathcal{A}}\widehat{S}_{\widehat{\psi}_{\elas}}(V_{i})=0.$
By the Taylor expansion, we then have 
\begin{eqnarray}
n^{1/2}(\widehat{\psi}_{\elas}-\psi_{0})\mid(T\geq c_{\gamma}) & = & \left.\left\{ -{\cI}_{\rt}^{-1}\left(\frac{m}{n}\right)^{1/2}m^{-1/2}\sum_{i\in\mathcal{A}}S_{\psi_{0}}(V_{i})+o_{\pp}(1)\right\} \right\vert (T\geq c_{\gamma})\nonumber \\
 & \stackrel{\cdot}{\sim} & \mathcal{N}_{\rt}\mid(T_{\infty}\geq c_{\gamma}),\label{eq:elas-2}
\end{eqnarray}
where $o_{\pp}(1)$ follows by a similar argument for (\ref{eq:o1}),
and $\stackrel{\cdot}{\sim}$ follows by Proposition \ref{prop:weakCvg}. 
\end{proof}

\subsection{\textcolor{black}{Characterization of normal distributions with elliptical
truncations}}

We characterize the multivariate normal distributions with elliptical
truncations \citep{tallis1963elliptical} by MGFs. Let $F_{k}(\cdot)$
represent the cumulative density function of a chi-square distribution
with degrees of freedom $k$. Let $F_{k}(\cdot;\lambda)$ represent
the cumulative density function of a non-central chi-square distribution
with degrees of freedom $k$ and non-centrality parameter $\lambda$.
Proposition \ref{prop:truncN} is a general result.

\begin{proposition}\label{prop:truncN}Let $\cZ_{1}$ follows $\rmN(\mu_{1},I_{p\times p})$.
Then, the MGF of the truncated normal distribution $\cZ_{1}\mid\cZ_{1}^{\T}\cZ_{1}\geq a$
is 
\begin{equation}
m(t)=\frac{\exp(-\frac{1}{2}\mu_{1}^{\T}\mu_{1})\sum_{k=0}^{\infty}\{1-F_{p+2k}(a)\}\{(\mu_{1}+t)^{\T}(\mu_{1}+t)/2\}^{k}/k!}{1-F_{p}(a;\mu_{1}^{\T}\mu_{1})}.\label{eq:mgf-final}
\end{equation}
The first and second moments of $\cZ_{1}\mid\cZ_{1}^{\T}\cZ_{1}\geq a$
are 
\begin{eqnarray}
\E(\cZ_{1}\mid\cZ_{1}^{\T}\cZ_{1}\geq a) & = & \mu_{1}\cdot\frac{1-F_{p+2}(a;\mu_{1}^{\T}\mu_{1})}{1-F_{p}(a;\mu_{1}^{\T}\mu_{1})},\label{eq:mean}\\
\E(\cZ_{1}^{\otimes2}\mid\cZ_{1}^{\T}\cZ_{1}\geq a) & = & I_{p\times p}\frac{1-F_{p+2}(a;\mu_{1}^{\T}\mu_{1})}{1-F_{p}(a;\mu_{1}^{\T}\mu_{1})}+\mu_{1}\mu_{1}^{\T}\frac{1-F_{p+4}(a;\mu_{1}^{\T}\mu_{1})}{1-F_{p}(a;\mu_{1}^{\T}\mu_{1})}.\label{eq:variance}
\end{eqnarray}

\end{proposition}As a sanity check, we verify that for $a=0$, $m(t)$
is the MGF of a normal distribution with mean $\mu_{1}$ and variance
$I_{p\times p}$. When $a=0$, $F_{p+2k}(a)=0$ for all $k\geq0$.
Therefore, (\ref{eq:mgf-final}) reduces to 
\begin{eqnarray*}
m(t) & = & \exp(-\frac{1}{2}\mu_{1}^{\T}\mu_{1})\sum_{k=0}^{\infty}\{(\mu_{1}+t)^{\T}(\mu_{1}+t)/2\}^{k}/k!\\
 & = & \exp(-\frac{1}{2}\mu_{1}^{\T}\mu_{1})\{\exp(\mu_{1}+t)^{\T}(\mu_{1}+t)/2\}\\
 & = & \exp(\mu_{1}^{\T}t+\frac{1}{2}t^{\T}t),
\end{eqnarray*}
corresponding to the MGF of $\rmN(\mu_{1},I_{p\times p})$. Moreover,
by (\ref{eq:mean}) and (\ref{eq:variance}), the mean and variance
of $\cZ_{1}$ are 
\begin{eqnarray*}
\E(\cZ_{1}\mid\cZ_{1}^{\T}\cZ_{1}\geq0) & = & \mu_{1},\\
\E(\cZ_{1}^{2}\mid\cZ_{1}^{\T}\cZ_{1}\geq0) & = & I_{p\times p}+\mu_{1}\mu_{1}^{\T},
\end{eqnarray*}
respectively, corresponding to the mean and variance of $\rmN(\mu_{1},I_{p\times p})$. 
\begin{proof}
Define a set $\cC=\{\cZ_{1}\in\R^{p}:\cZ_{1}^{\T}\cZ_{1}\geq a\}$.
We derive the MGF $m(t)$ for $\cZ_{1}$ in the subspace $\cC$. By
definition, we have 
\begin{align}
m(t) & =\E\{\exp(t^{\intercal}Z_{1})\}\nonumber \\
 & \propto\int_{\mathbb{\cC}}\exp(t^{\T}z)\exp\left\{ -\frac{1}{2}(z-\mu_{1})^{\T}(z-\mu_{1})\right\} dz\nonumber \\
 & \propto\exp\left(\frac{1}{2}t^{\T}t+\mu_{1}^{\T}t\right)\int_{\mathbb{\cC}}\exp\left\{ -\frac{1}{2}(z-\mu_{1}-t)^{\T}(z-\mu_{1}-t)\right\} dz.\label{eq:mgf-2}
\end{align}
Let $\cZ^{\T}\cZ$ follow a non-central chi-square distribution with
parameters $p$ and $(\mu_{1}+t)^{\T}(\mu_{1}+t)$. The probability
of $\cZ^{\T}\cZ\geq a$ can be characterized by \citep{tallis1963elliptical}
\begin{align*}
P(\cZ^{\T}\cZ\geq a) & =\exp\{-(\mu_{1}+t)^{\T}(\mu_{1}+t)/2\}\sum_{k=0}^{\infty}\{1-F_{p+2k}(a)\}\{(\mu_{1}+t)^{\T}(\mu_{1}+t)/2\}^{k}/k!\\
 & =(2\pi)^{-p/2}\int_{\mathbb{\cC}}\exp\left\{ -\frac{1}{2}(z-\mu_{1}-t)^{\T}(z-\mu_{1}-t)\right\} dz.
\end{align*}
Continuing with (\ref{eq:mgf-2}), we have 
\[
m(t)\propto\exp(-\frac{1}{2}\mu_{1}^{\T}\mu_{1})\sum_{k=0}^{\infty}\{1-F_{p+2k}(a)\}\{(\mu_{1}+t)^{\T}(\mu_{1}+t)/2\}^{k}/k!.
\]
Because $m(0)=1$, (\ref{eq:mgf-final}) follows.

Let the normalizing constant be $C=\exp\left(-\frac{1}{2}\mu_{1}^{\T}\mu_{1}\right)\sum_{k=0}^{\infty}\{1-F_{p+2k}(a)\}(\mu_{1}^{\T}\mu_{1}/2)^{k}/k!=1-F_{p}(a;\mu_{1}^{\T}\mu_{1}).$
Taking the directive of $Cm(t)$ with respect to $t$ and evaluating
at $t=0$, we have $C\left.\de m(t)/\de t\right\vert _{t=0}$ becomes
\begin{eqnarray*}
 &  & \left.(\mu_{1}+t)\exp\left(-\frac{1}{2}\mu_{1}^{\T}\mu_{1}\right)\sum_{k=0}^{\infty}\{1-F_{p+2k+2}(a)\}\{(\mu_{1}+t)^{\T}(\mu_{1}+t)/2\}^{k}/k!\right\vert _{t=0}\\
 & = & \mu_{1}\exp\left(-\frac{1}{2}\mu_{1}^{\T}\mu_{1}\right)\sum_{k=0}^{\infty}\{1-F_{p+2k+2}(a)\}(\mu_{1}^{\T}\mu_{1}/2)^{k}/k!\\
 & = & \mu_{1}\left\{ 1-F_{p+2}(a;\mu_{1}^{\T}\mu_{1})\right\} .
\end{eqnarray*}
Thus, the first moment of $\cZ_{1}\mid\cZ_{1}^{\T}\cZ_{1}\geq a$
is 
\begin{eqnarray*}
\E(\cZ_{1}\mid\cZ_{1}^{\T}\cZ_{1}\geq a) & = & \mu_{1}\frac{\sum_{k=0}^{\infty}\{1-F_{p+2k+2}(a)\}(\mu_{1}^{\T}\mu_{1}/2)^{k}/k!}{\sum_{k=0}^{\infty}\{1-F_{p+2k}(a)\}(\mu_{1}^{\T}\mu_{1}/2)^{k}/k!},\\
 & = & \mu_{1}\frac{1-F_{p+2}(a;\mu_{1}^{\T}\mu_{1})}{1-F_{p}(a;\mu_{1}^{\T}\mu_{1})},
\end{eqnarray*}
as in (\ref{eq:mean}).

Taking the second directive of $Cm(t)$ with respect to $t$ and evaluating
at $t=0$, we have $C\left.\de^{2}m(t)/\de t^{2}\right\vert _{t=0}$
becomes 
\begin{eqnarray*}
 &  & \exp\left(-\frac{1}{2}\mu_{1}^{\T}\mu_{1}\right)\left[\sum_{k=0}^{\infty}\left.\{1-F_{p+2k+2}(a)\}\{(\mu_{1}+t)^{\T}(\mu_{1}+t)/2\}^{k}/k!\right\vert _{t=0}\right.\\
 &  & \left.+\left.(\mu_{1}+t)(\mu_{1}+t)^{\T}\sum_{k=0}^{\infty}\{1-F_{p+2k+4}(a)\}\{(\mu_{1}+t)^{\T}(\mu_{1}+t)/2\}^{k}/k!\right\vert _{t=0}\right]\\
 & = & \exp\left(-\frac{1}{2}\mu_{1}^{\T}\mu_{1}\right)\left[\sum_{k=0}^{\infty}\{1-F_{p+2k+2}(a)\}(\mu_{1}^{\T}\mu_{1}/2)^{k}/k!+\mu_{1}\mu_{1}^{\T}\sum_{k=0}^{\infty}\{1-F_{p+2k+4}(a)\}(\mu_{1}^{\T}\mu_{1}/2)^{k}/k!\right]\\
 & = & I_{p\times p}\{1-F_{p+2}(a;\mu_{1}^{\T}\mu_{1})\}+\mu_{1}\mu_{1}^{\T}\{1-F_{p+4}(a;\mu_{1}^{\T}\mu_{1})\},
\end{eqnarray*}
Thus, the second moment of $\cZ_{1}\mid\cZ_{1}^{\T}\cZ_{1}\geq a$
is 
\begin{eqnarray*}
\E(\cZ_{1}^{\otimes2}\mid\cZ_{1}^{\T}\cZ_{1}\geq a) & = & I_{p\times p}\frac{\sum_{k=0}^{\infty}\{1-F_{p+2k+2}(a)\}(\mu_{1}^{\T}\mu_{1}/2)^{k}/k!}{\sum_{k=0}^{\infty}\{1-F_{p+2k}(a)\}(\mu_{1}^{\T}\mu_{1}/2)^{k}/k!}\\
 &  & +\mu_{1}\mu_{1}^{\T}\frac{\sum_{k=0}^{\infty}\{1-F_{p+2k+4}(a)\}(\mu_{1}^{\T}\mu_{1}/2)^{k}/k!}{\sum_{k=0}^{\infty}\{1-F_{p+2k}(a)\}(\mu_{1}^{\T}\mu_{1}/2)^{k}/k!}.\\
 & = & I_{p\times p}\frac{1-F_{p+2}(a;\mu_{1}^{\T}\mu_{1})}{1-F_{p}(a;\mu_{1}^{\T}\mu_{1})}+\mu_{1}\mu_{1}^{\T}\frac{1-F_{p+4}(a;\mu_{1}^{\T}\mu_{1})}{1-F_{p}(a;\mu_{1}^{\T}\mu_{1})}.
\end{eqnarray*}
as in (\ref{eq:variance}). 
\end{proof}

\paragraph{Proof of the decomposition $\text{\ensuremath{\mathcal{N}_{\rt}}}=V_{\rteff}^{1/2}\cZ_{1}-V_{\eff}^{1/2}\cZ_{2}$.}
\begin{proof}
First, we show 
\begin{eqnarray*}
({\cI}_{\rw}+\rho{\cI}_{\rt})^{-1}(\rho^{1/2}+\Gamma^{\T}) & = & ({\cI}_{\rw}+\rho{\cI}_{\rt})^{-1}(\rho^{1/2}+{\cI}_{\rt}^{-1}{\cI}_{\rw}\rho^{-1/2})\\
 & = & ({\cI}_{\rw}+\rho{\cI}_{\rt})^{-1}(\rho\cI_{\rt}+{\cI}_{\rw})\cI_{\rt}^{-1}\rho^{-1/2}\\
 & = & (\rho{\cI}_{\rt})^{-1}\rho^{1/2}.
\end{eqnarray*}
Then, we have 
\begin{eqnarray*}
\text{\ensuremath{\mathcal{N}_{\rt}}}({\cZ}_{\rt}) & = & -(\rho{\cI}_{\rt})^{-1}(\rho^{1/2}{\cZ}_{\rt})\\
 & = & -({\cI}_{\rw}+\rho{\cI}_{\rt})^{-1}(\rho^{1/2}+\Gamma^{\T}){\cZ}_{\rt}\\
 & = & -(\rho{\cI}_{\rt}+{\cI}_{\rw})^{-1}(\cZ_{\rw}+\rho^{1/2}\cZ_{\rt})+(\rho{\cI}_{\rt}+{\cI}_{\rw})^{-1}(\cZ_{\rw}-\Gamma^{\T}\cZ_{\rt})\\
 & = & -(\rho{\cI}_{\rt}+{\cI}_{\rw})^{-1/2}\left\{ V_{\eff}^{1/2}(\cZ_{\rw}+\rho^{1/2}\cZ_{\rt})\right\} +\left\{ (\rho{\cI}_{\rt}+{\cI}_{\rw})^{-1}\Sigma_{SS}^{1/2}\right\} \left\{ \Sigma_{SS}^{-1/2}(\cZ_{\rw}-\Gamma^{\T}\cZ_{\rt})\right\} \\
 & = & -(\rho{\cI}_{\rt}+{\cI}_{\rw})^{-1/2}\cZ_{2}+\left\{ (\rho{\cI}_{\rt}+{\cI}_{\rw})^{-1}\Sigma_{SS}^{1/2}\right\} \cZ_{1}\\
 & = & -(\rho{\cI}_{\rt}+{\cI}_{\rw})^{-1/2}\cZ_{2}+\left\{ (\rho{\cI}_{\rt})^{-1}-(\rho{\cI}_{\rt}+{\cI}_{\rw})^{-1}\right\} ^{1/2}\cZ_{1}\\
 & = & -V_{\eff}^{1/2}\cZ_{2}+V_{\rteff}^{1/2}\cZ_{1}.
\end{eqnarray*}
\end{proof}

\section{Technical details for inference\label{sec:Inference} }

\subsection{Inconsistency of the nonparametric bootstrap}

\begin{theorem}\label{Thm:nonparboot}

Let $\widehat{\psi}_{\elas}^{*}$ be the nonparametric bootstrap replicate
of $\widehat{\psi}_{\elas}$. Under assumptions in Theorem \ref{Thm:Consistency-DML},
the bootstrap distribution of $n^{1/2}(\widehat{\psi}_{\elas}^{*}-\widehat{\psi}_{\elas})$
give the observed data is inconsistent for the distribution of $n^{1/2}(\widehat{\psi}_{\elas}-\psi_{0})$.

\end{theorem}

The reason for the inconsistency of the bootstrap estimator is exactly
that $\bone(T<c_{\gamma})=0$ does not imply that Assumption \ref{Asump:rand-rwd}
is violated. When this happens, $\bone(T^{*}<c_{\gamma})$ is asymptotically
non-degenerate, making the bootstrap estimator inconsistent.

\paragraph*{Proof of Theorem \ref{Thm:nonparboot}.}
\begin{proof}
We now show that the nonparametric bootstrap inference for $\widehat{\psi}_{\elas}$
is inconsistent.

Let the bootstrap RT and RW resamples be indexed by $\mathcal{A}^{*}$
and $\mathcal{B}^{*}$, respectively. Denote the bootstrap replicates
of $\widehat{\psi}_{\rt}$, $\widehat{\alpha},$ and $\widehat{\eta}^{\T}=(\widehat{\eta}_{0}^{\T},\widehat{\eta}_{1}^{\T})$
as $\widehat{\psi}_{\rt}^{*}$, $\widehat{\alpha}^{*},$ and $\widehat{\eta}^{*\T}=(\widehat{\eta}_{0}^{*\T},\widehat{\eta}_{1}^{*\T})$,
respectively. Then the bootstrap replicate of $T$ is

\begin{equation}
T^{*}=\left\{ \sum_{i\in\mathcal{B^{*}}}S_{\rw,\widehat{\psi}_{\rt}^{*}}(V_{i};\widehat{\alpha}^{*},\widehat{\eta}_{0}^{*})\right\} ^{\T}\widehat{\Sigma}_{SS}^{*-1}\left\{ \sum_{i\in\mathcal{B}^{*}}S_{\rw,\widehat{\psi}_{\rt}^{*}}(V_{i};\widehat{\alpha}^{*},\widehat{\eta}_{0}^{*})\right\} .\label{eq:T-1}
\end{equation}
The bootstrap replicate $\widehat{\psi}_{\elas}^{*}$ solves 
\begin{equation}
\sum_{i\in\mathcal{A}^{*}\cup\mathcal{B}^{*}}\left\{ \delta_{i}S_{\psi}(V_{i};\widehat{\eta}_{1}^{*})+\bone(T^{*}<c_{\gamma})(1-\delta_{i})S_{\psi}(V_{i};\widehat{\alpha}^{*},\widehat{\eta}_{0}^{*})\right\} =0.\label{eq:EE-2}
\end{equation}
By the Taylor expansion, we have 
\begin{eqnarray}
 &  & n^{1/2}(\widehat{\psi}_{\elas}^{*}-\widehat{\psi}_{\elas})\nonumber \\
 & = & -(\Gamma_{\elas}^{\T})^{-1}n^{-1/2}\sum_{i\in\mathcal{A}^{*}\cup\mathcal{B}^{*}}\left\{ \delta_{i}S_{\widehat{\psi}_{\elas}}(V_{i})+\bone(T^{*}<c_{\gamma})(1-\delta_{i})S_{\widehat{\psi}_{\elas}}(V_{i})\right\} +o_{\pp}(1)\\
 & = & -(\Gamma_{\elas}^{\T})^{-1}n^{-1/2}\sum_{i\in\mathcal{A}^{*}\cup\mathcal{B}^{*}}\left\{ \delta_{i}S_{\psi_{0}}(V_{i})+\bone(T^{*}<c_{\gamma})(1-\delta_{i})S_{\psi_{0}}(V_{i})\right\} \nonumber \\
 &  & -(\Gamma_{\elas}^{\T})^{-1}n^{-1}\sum_{i\in\mathcal{A}^{*}\cup\mathcal{B}^{*}}\left\{ \delta_{i}\frac{\partial S_{\psi_{0}}(V_{i})}{\partial\psi^{\T}}+\bone(T^{*}<c_{\gamma})(1-\delta_{i})\frac{\partial S_{\psi_{0}}(V_{i})}{\partial\psi^{\T}}\right\} \nonumber \\
 &  & \times n^{1/2}(\widehat{\psi}_{\elas}-\psi_{0})+o_{\pp}(1).\label{eq:elasBoot}
\end{eqnarray}
Suppose that the observed $T$ is $T\geq c_{\gamma}$. Then by (\ref{eq:elas-1}),
\begin{eqnarray*}
n^{1/2}(\widehat{\psi}_{\elas}-\psi_{0}) & = & -(\Gamma_{\elas}^{\T})^{-1}n^{-1/2}\sum_{i\in\mathcal{A}\cup\mathcal{B}}\delta_{i}S_{\psi_{0}}(V_{i})+o_{\pp}(1)\\
 & \stackrel{\cdot}{\sim} & -(\Gamma_{\elas}^{\T}\rho^{1/2})^{-1}{\cZ}_{\rt}.
\end{eqnarray*}

Let ${\cZ}_{\rt}^{*}\sim\N({\cZ}_{\rt},{\cI}_{\rt})$ and ${\cZ}_{\rw}^{*}\sim\N({\cZ}_{\rw},{\cI}_{\rw})$.
Then, (\ref{eq:elasBoot}) becomes 
\begin{eqnarray}
 & \stackrel{\cdot}{\sim} & -(\Gamma_{\elas}^{\T}\rho^{1/2})^{-1}\left\{ {\cZ}_{\rt}^{*}+\bone(T^{*}<c_{\gamma})\rho^{1/2}{\cZ}_{\rw}^{*}\right\} +(\Gamma_{\elas}^{\T}\rho^{1/2})^{-1}{\cZ}_{\rt}\nonumber \\
 & = & -(\Gamma_{\elas}^{\T}\rho^{1/2})^{-1}\left\{ {\cZ}_{\rt}^{*}-{\cZ}_{\rt}+\bone(T^{*}<c_{\gamma})\rho^{1/2}{\cZ}_{\rw}^{*}\right\} \nonumber \\
 & = & -(\Gamma_{\elas}^{\T}\rho^{1/2})^{-1}\left\{ {\cZ}_{\rt}^{*}-{\cZ}_{\rt}+\bone(T^{*}<c_{\gamma})\rho^{1/2}({\cZ}_{\rw}^{*}-{\cZ}_{\rw})\right\} \label{eq:d1}\\
 &  & -(\Gamma_{\elas}^{\T}\rho^{1/2})^{-1}\bone(T^{*}<c_{\gamma})\rho^{1/2}{\cZ}_{\rw}\label{eq:d2}
\end{eqnarray}
Note, (\ref{eq:d1}) has the same limiting distribution as $n^{1/2}(\widehat{\psi}_{\elas}-\psi_{0})$,
while (\ref{eq:d2}) is asymptotically non-degenerate. Therefore,
the bootstrap inference is inconsistent. 
\end{proof}

\subsection{Proof of Theorem \ref{Thm:elas-1}}
\begin{proof}
Under the local alternatives, we have 
\begin{eqnarray*}
 &  & \lim_{n\rightarrow\infty}\pr\left\{ n^{1/2}e_{k}^{\T}(\widehat{\psi}_{\elas}-\psi_{0})\in{\rm ECI}_{k,1-\alpha}\right\} \\
 & \geq & \lim_{n\rightarrow\infty}\pr\left\{ n^{1/2}e_{k}^{\T}(\widehat{\psi}_{\elas}-\psi_{0})\in[\inf_{\mu_{1}\in\mathcal{B}_{1-\widetilde{\alpha}}}\widehat{Q}_{k,\widetilde{\alpha}/2}(\mu_{1}),\sup_{\mu_{1}\in\mathcal{B}_{1-\widetilde{\alpha}}}\widehat{Q}_{k,1-\widetilde{\alpha}/2}(\mu_{1})]\mid\mu_{1}\in\mathcal{B}_{1-\widetilde{\alpha}}\right\} \times(1-\widetilde{\alpha})\\
 & \geq & (1-\widetilde{\alpha})^{2}=1-\alpha.
\end{eqnarray*}
Under the fixed alternative, we have 
\begin{eqnarray*}
 &  & \lim_{n\rightarrow\infty}\pr\left\{ n^{1/2}e_{k}^{\T}(\widehat{\psi}_{\elas}-\psi_{0})\in{\rm ECI}_{k,1-\alpha}\right\} \\
 & = & \lim_{n\rightarrow\infty}\pr\left\{ n^{1/2}e_{k}^{\T}(\widehat{\psi}_{\elas}-\psi_{0})\in[\widehat{Q}_{k,\alpha/2}(\pm\infty),\widehat{Q}_{k,1-\alpha/2}(\pm\infty)]\right\} =1-\alpha.
\end{eqnarray*}
\end{proof}

\section{Additional simulation results and studies\label{sec:Additional-simulation-results}}

\subsection{Comparing AIPW and SES\label{subsec:Comparing-AIPW-and-SES} }

\begin{figure}
\centering\includegraphics[scale=0.65]{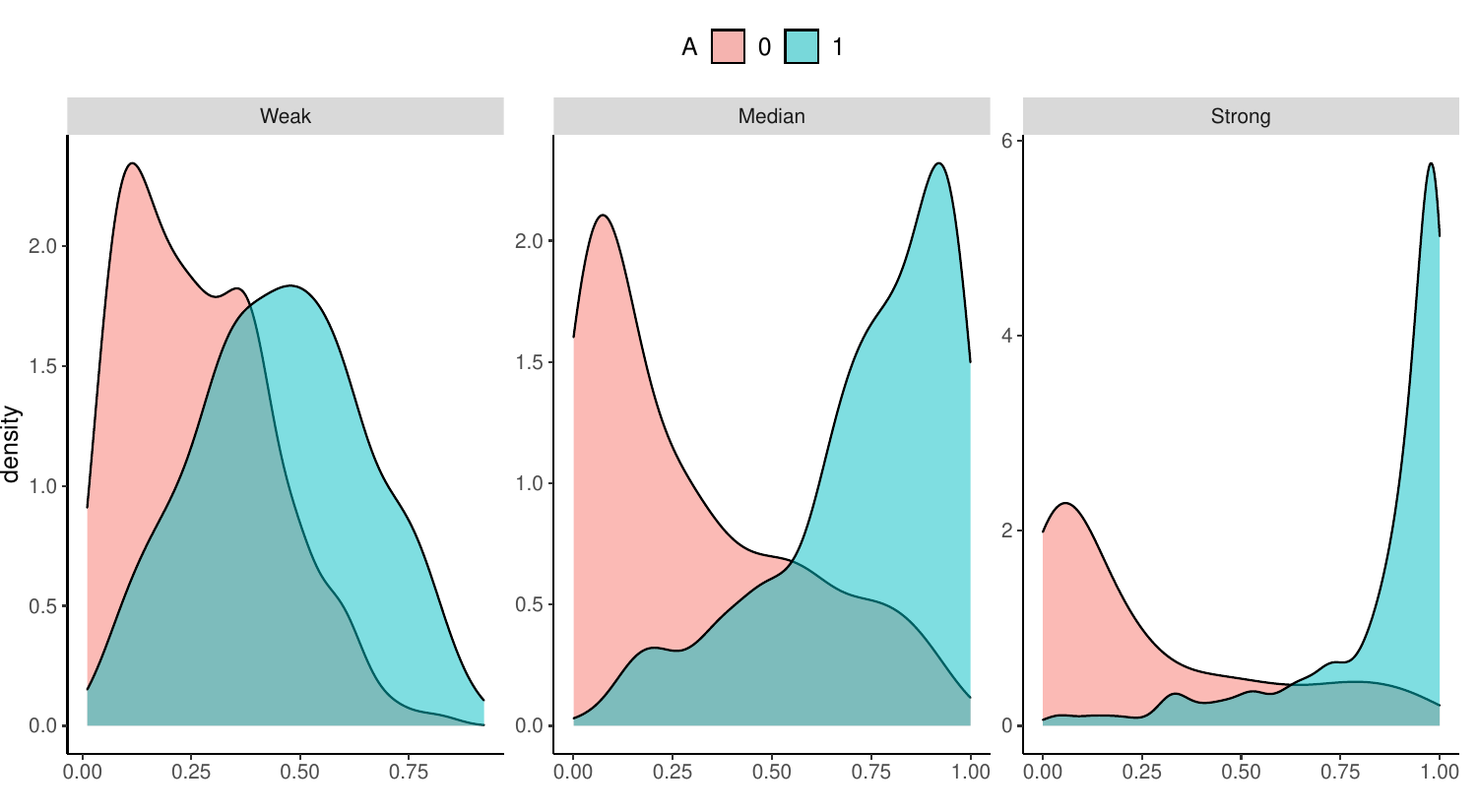}

\caption{\label{fig:separation}Illustration of the propensity score distributions
by treatment group under \textcolor{black}{weak, median and strong
separations (Section \ref{subsec:Comparing-AIPW-and-SES}).}}
\end{figure}

In this simulation study, we compare the performances of the AIPW-adjusted
approach and the SES approach based on RT data. The data generating
mechanism is the same as in Section \ref{sec:Simulation} except that
we now consider different propensity score distributions. Specifically,
consider $A\mid X,\delta=1\sim\mathrm{Bernoulli}\{e_{1}(X)\},$ where
${\rm logit}\ e_{1}(X)=\alpha_{0}+\alpha_{1}X_{1}+\alpha_{2}X_{2}$
and 
\begin{itemize}
\item[{i)}] (weak separation of propensity score distributions by treatment group)
$\alpha=(-2,-1,-1)$, 
\item[{ii)}] (median separation of propensity score distributions by treatment
group) $\alpha=(-2,-2,-2)$, 
\item[{iii)}] (strong separation of propensity score distributions by treatment
group) $\alpha=(-2,-3,-3)$. 
\end{itemize}
Figure \ref{fig:separation} shows the propensity score distributions
by treatment group and demonstrates the degrees of separation in the
three scenarios.

The estimators for comparison are the following: 
\begin{enumerate}
\item RT.AIPW: the AIPW-adjustment outcome approach of \citet{kennedy2020optimal}
that fits 
\[
\frac{A_{i}\{Y_{i}-\widehat{\mu}_{\delta,1}(X_{i})\}}{\widehat{e}_{\delta}(X_{i})}-\frac{(1-A_{i})\{Y_{i}-\widehat{\mu}_{\delta}(X_{i})\}}{1-\widehat{e}_{\delta}(X_{i})}+\widehat{\mu}_{\delta,1}(X_{i})-\widehat{\mu}_{\delta}(X_{i})
\]
against $Z_{i}$ based only on the RT data, where $\mu_{\delta,1}(X_{i})=\E\{Y(1)\mid X,\delta=1\}$,
and 
\item RT.SES: the efficient estimator based only on the RT data solving
the SES (\ref{eq:EE}) with the combining indicator $\bone(T<c_{\gamma})\equiv0$. 
\end{enumerate}
Table \ref{tab:SimNumericalResult-AIPW-EE} reports the simulation
results for comparing RT.AIPW and RT.SES. Across the three cases,
both estimators have small biases. RT.AIPW has larger variances and
MSEs than RT.SES. By construction, RT.AIPW takes the inverse of the
estimated propensity scores the to remove confounding biases, which
can be unstable when some propensity scores are close to zero or one.
Differently, the SES in (\ref{eq:eff psi}) uses the mean independence
of $H_{\psi_{0}}-\mu_{\delta}(X)$ and $A-e_{\delta}(X)$ to construct
unbiased estimating equations, thus avoiding taking the inverse of
the estimated propensity score.

\begin{table}[h]
\caption{\label{tab:SimNumericalResult-AIPW-EE}Simulation results of Monte
Carlo biases, standard deviations and root-mean-square errors of estimators
in the three cases: \textcolor{black}{weak separation $\alpha=(-2,-1,-1)$,
median separation $\alpha=(-2,-2,-2)$, and strong separation $\alpha=(-2,-3,-3)$
with $n=2000$ regarding the performances of the AIPW-adjusted outcome
approach (labeled as RT.AIPW) and the SES approach (labeled as RT.SES)
(Section \ref{subsec:Comparing-AIPW-and-SES})}}

\vspace{0.15cm}

\centering

\resizebox{\textwidth}{!}{

\begin{tabular}{ccccccccccccc}
\hline 
 & \multicolumn{2}{c}{RT.AIPW} & \multicolumn{2}{c}{RT.SES} & \multicolumn{2}{c}{RT.AIPW} & \multicolumn{2}{c}{RT.SES} & \multicolumn{2}{c}{RT.AIPW} & \multicolumn{2}{c}{RT.SES}\tabularnewline
\hline 
 & \multicolumn{4}{c}{Case 1: weak separation} & \multicolumn{4}{c}{Case 2: median separation} & \multicolumn{4}{c}{Case 3: strong separation}\tabularnewline
 & \multicolumn{1}{l}{{\scriptsize{}{}{}$\psi_{1}=1$}} & \multicolumn{1}{l}{{\scriptsize{}{}{}$\psi_{2}=1$}} & \multicolumn{1}{l}{{\scriptsize{}{}{}$\psi_{1}=1$}} & \multicolumn{1}{l}{{\scriptsize{}{}{}$\psi_{2}=1$}} & \multicolumn{1}{l}{{\scriptsize{}{}{}$\psi_{1}=1$}} & \multicolumn{1}{l}{{\scriptsize{}{}{}$\psi_{2}=1$}} & \multicolumn{1}{l}{{\scriptsize{}{}{}$\psi_{1}=1$}} & \multicolumn{1}{l}{{\scriptsize{}{}{}$\psi_{2}=1$}} & \multicolumn{1}{l}{{\scriptsize{}{}{}$\psi_{1}=1$}} & \multicolumn{1}{l}{{\scriptsize{}{}{}$\psi_{2}=1$}} & \multicolumn{1}{l}{{\scriptsize{}{}{}$\psi_{1}=1$}} & \multicolumn{1}{l}{{\scriptsize{}{}{}$\psi_{2}=1$}}\tabularnewline
\hline 
Bias ($\times10^{-2}$)  & -1  & -2  & -1  & -1  & 1  & -1  & 0  & -1  & 3  & 1  & -2  & -2\tabularnewline
S.D. ($\times10^{-3}$)  & 242  & 248  & 180  & 186  & 365  & 381  & 260  & 249  & 622  & 555  & 394  & 397\tabularnewline
root-MSE ($\times10^{-3}$)  & 242  & 249  & 180  & 186  & 365  & 381  & 260  & 250  & 623  & 555  & 394  & 397\tabularnewline
Coverage rate ($\%$)  & 94.4  & 91.8  & 95.4  & 93.2  & 92.8  & 93.2  & 94.6  & 94.6  & 91.0  & 92.0  & 93.2  & 91.6\tabularnewline
Width ($\times10^{-3}$)  & 955  & 942  & 709  & 710  & 1390  & 1420  & 974  & 971  & 14569  & 5077  & 1448  & 1448\tabularnewline
\hline 
\end{tabular}} 
\end{table}

\subsection{Additional simulation results in Section \ref{sec:Simulation}}

Table \ref{tab:SimNumericalResult} reports the detailed numerical
results of the estimators in Section \ref{sec:Simulation}, including
absolute biases, standard deviations, and MSEs. In addition, Tables
\ref{tab:SimNumericalResult-m5000} and \ref{tab:SimNumericalResult-m5000-1}
report the numerical results of the estimators, including absolute
biases, standard deviations, and MSEs, and the confidence intervals,
including coverage rates and widths when $m=5000$. Table \ref{tab:Simulation-results-ofgamma}
reports Monte Carlo averages and standard deviations of the estimators
for the local parameter $\eta$, the threshold $c_{\gamma}$, and
the proportion of combining the RT and RW samples.

\begin{table}[h]
\caption{\label{tab:SimNumericalResult}Simulation results of Monte Carlo biases,
standard deviations, and root-mean-square errors of estimators $\widehat{\psi}_{\rt},$
$\widehat{\psi}_{\eff},$ and $\widehat{\psi}_{\elas}$ (labeled as
``RT'', ``Eff'', and ``Elastic'') in the two cases: zero effect
modification $\psi_{1}=\psi_{2}=0$ (left) and nonzero effect modification
$\psi_{1}=\psi_{2}=1$ (right) with $n=2000$ (Section \ref{sec:Simulation})}

\centering

\vspace{0.35cm}

\begin{tabular}{ccccccccccccc}
\toprule 
 & \multicolumn{2}{c}{RT} & \multicolumn{2}{c}{Eff} & \multicolumn{2}{c}{Elastic} & \multicolumn{2}{c}{RT} & \multicolumn{2}{c}{Eff} & \multicolumn{2}{c}{Elastic}\tabularnewline
\midrule 
 & \multicolumn{6}{c}{Case 1: zero effect modification} & \multicolumn{6}{c}{Case 2: nonzero effect modification}\tabularnewline
$b$  & {\scriptsize{}{}{}$\psi_{1}=0$}  & {\scriptsize{}{}{}$\psi_{2}=0$}  & {\scriptsize{}{}{}$\psi_{1}=0$}  & {\scriptsize{}{}{}$\psi_{2}=0$}  & {\scriptsize{}{}{}$\psi_{1}=0$}  & {\scriptsize{}{}{}$\psi_{2}=0$}  & {\scriptsize{}{}{}$\psi_{1}=1$}  & {\scriptsize{}{}{}$\psi_{2}=1$}  & {\scriptsize{}{}{}$\psi_{1}=1$}  & {\scriptsize{}{}{}$\psi_{2}=1$}  & {\scriptsize{}{}{}$\psi_{1}=1$}  & {\scriptsize{}{}{}$\psi_{2}=1$}\tabularnewline
\midrule 
\multicolumn{13}{c}{Bias ($\times10^{-2}$)}\tabularnewline
\midrule 
0  & 0  & 0  & 0  & 0  & 0  & 0  & 0  & 0  & 0  & 0  & 0  & 0 \tabularnewline
0.11  & 0  & 0  & -2  & -2  & 0  & 0  & 0  & 0  & -2  & -2  & 0  & 0 \tabularnewline
0.23  & 0  & 0  & -4  & -4  & 0  & 0  & 0  & 0  & -4  & -4  & 0  & 0 \tabularnewline
0.34  & 0  & 0  & -6  & -6  & 0  & 1  & 0  & 0  & -6  & -6  & 0  & 0 \tabularnewline
0.46  & 0  & 0  & -8  & -8  & 1  & 1  & 0  & 0  & -8  & -8  & 0  & 0 \tabularnewline
0.57  & 0  & 0  & -9  & -9  & 1  & 1  & 0  & 0  & -9  & -9  & 0  & 0 \tabularnewline
0.69  & 0  & 0  & -11  & -11  & 1  & 1  & 0  & 0  & -11  & -11  & 0  & 0 \tabularnewline
0.8  & 0  & 0  & -12  & -12  & 1  & 0  & 0  & 0  & -12  & -12  & 0  & 0 \tabularnewline
1  & 0  & 0  & -14  & -14  & 0  & 0  & 0  & 0  & -14  & -15  & 0  & 0 \tabularnewline
2  & 0  & 0  & -21  & -21  & 0  & 0  & 0  & 0  & -21  & -21  & 0  & 0 \tabularnewline
\midrule 
\multicolumn{13}{c}{S.D. ($\times10^{-3}$)}\tabularnewline
\midrule 
0  & 136  & 137  & 64  & 63  & 113  & 112  & 135  & 138  & 61  & 63  & 110  & 111 \tabularnewline
0.11  & 136  & 137  & 64  & 63  & 114  & 114  & 135  & 138  & 61  & 63  & 111  & 113 \tabularnewline
0.23  & 136  & 137  & 65  & 62  & 117  & 116  & 135  & 138  & 61  & 63  & 113  & 115 \tabularnewline
0.34  & 136  & 137  & 64  & 62  & 120  & 119  & 135  & 138  & 61  & 62  & 117  & 118 \tabularnewline
0.46  & 136  & 137  & 63  & 62  & 122  & 121  & 135  & 138  & 61  & 61  & 121  & 122 \tabularnewline
0.57  & 136  & 137  & 62  & 61  & 124  & 125  & 135  & 138  & 59  & 60  & 123  & 126 \tabularnewline
0.69  & 136  & 137  & 62  & 60  & 127  & 128  & 135  & 138  & 59  & 60  & 127  & 130 \tabularnewline
0.8  & 136  & 137  & 61  & 60  & 128  & 131  & 135  & 138  & 59  & 59  & 128  & 131 \tabularnewline
1  & 136  & 137  & 60  & 60  & 131  & 133  & 135  & 138  & 58  & 58  & 130  & 134 \tabularnewline
2  & 136  & 137  & 53  & 53  & 135  & 136  & 135  & 138  & 52  & 52  & 133  & 137 \tabularnewline
\midrule 
\multicolumn{13}{c}{root-MSE ($\times10^{-3}$)}\tabularnewline
\midrule 
0  & 136  & 137  & 64  & 63  & 113  & 112  & 135  & 138  & 61  & 63  & 110  & 111 \tabularnewline
0.11  & 136  & 137  & 67  & 66  & 115  & 114  & 135  & 138  & 64  & 66  & 111  & 113 \tabularnewline
0.23  & 136  & 137  & 76  & 73  & 117  & 116  & 135  & 138  & 72  & 75  & 113  & 115 \tabularnewline
0.34  & 136  & 137  & 86  & 84  & 120  & 119  & 135  & 138  & 83  & 85  & 118  & 118 \tabularnewline
0.46  & 136  & 137  & 100  & 97  & 122  & 121  & 135  & 138  & 97  & 98  & 121  & 122 \tabularnewline
0.57  & 136  & 137  & 112  & 110  & 124  & 125  & 135  & 138  & 109  & 110  & 124  & 126 \tabularnewline
0.69  & 136  & 137  & 125  & 124  & 127  & 128  & 135  & 138  & 124  & 124  & 127  & 130 \tabularnewline
0.8  & 136  & 137  & 136  & 136  & 128  & 131  & 135  & 138  & 135  & 137  & 128  & 131 \tabularnewline
1  & 136  & 137  & 155  & 156  & 131  & 133  & 135  & 138  & 155  & 156  & 130  & 134 \tabularnewline
2  & 136  & 137  & 214  & 212  & 135  & 136  & 135  & 138  & 212  & 213  & 133  & 137 \tabularnewline
\bottomrule
\end{tabular}
\end{table}

\begin{table}[h]
\caption{\label{tab:SimNumericalResult-m5000}Simulation results of Monte Carlo
biases, standard deviations and root-mean-square errors of estimators
$\widehat{\psi}_{\rt},$ $\widehat{\psi}_{\eff},$ and $\widehat{\psi}_{\elas}$
(labeled as ``RT'', ``Eff'', and ``Elastic'') in the two cases:
zero effect modification $\psi_{1}=\psi_{2}=0$ (left) and nonzero
effect modification $\psi_{1}=\psi_{2}=1$ (right) with $n=5000$
(Section \ref{sec:Simulation})}

\centering

\begin{longtable}[c]{ccccccccccccc}
\toprule 
 & \multicolumn{2}{c}{RT} & \multicolumn{2}{c}{Eff} & \multicolumn{2}{c}{Elastic} & \multicolumn{2}{c}{RT} & \multicolumn{2}{c}{Eff} & \multicolumn{2}{c}{Elastic}\tabularnewline
\midrule 
 & \multicolumn{6}{c}{Case 1: zero effect modification} & \multicolumn{6}{c}{Case 2: nonzero effect modification}\tabularnewline
$b$  & {\scriptsize{}{}{}$\psi_{1}=0$}  & {\scriptsize{}{}{}$\psi_{2}=0$}  & {\scriptsize{}{}{}$\psi_{1}=0$}  & {\scriptsize{}{}{}$\psi_{2}=0$}  & {\scriptsize{}{}{}$\psi_{1}=0$}  & {\scriptsize{}{}{}$\psi_{2}=0$}  & {\scriptsize{}{}{}$\psi_{1}=1$}  & {\scriptsize{}{}{}$\psi_{2}=1$}  & {\scriptsize{}{}{}$\psi_{1}=1$}  & {\scriptsize{}{}{}$\psi_{2}=1$}  & {\scriptsize{}{}{}$\psi_{1}=1$}  & {\scriptsize{}{}{}$\psi_{2}=1$}\tabularnewline
\midrule 
\multicolumn{13}{c}{Bias ($\times10^{-2}$)}\tabularnewline
\midrule 
0  & 0  & 0  & 0  & 0  & 0  & 0  & 0  & 0  & 0  & 0  & 0  & 0 \tabularnewline
0.11  & 0  & 0  & -2  & -1  & 1  & 1  & 0  & 0  & -1  & -1  & 1  & 0 \tabularnewline
0.23  & 0  & 0  & -3  & -3  & 1  & 1  & 0  & 0  & -3  & -3  & 1  & 1 \tabularnewline
0.34  & 0  & 0  & -4  & -4  & 1  & 1  & 0  & 0  & -4  & -4  & 1  & 1 \tabularnewline
0.46  & 0  & 0  & -6  & -6  & 1  & 1  & 0  & 0  & -6  & -6  & 1  & 1 \tabularnewline
0.57  & 0  & 0  & -7  & -7  & 1  & 1  & 0  & 0  & -7  & -7  & 1  & 1 \tabularnewline
0.69  & 0  & 0  & -8  & -8  & 1  & 1  & 0  & 0  & -8  & -8  & 0  & 0 \tabularnewline
0.8  & 0  & 0  & -9  & -9  & 0  & 0  & 0  & 0  & -9  & -9  & 0  & 0 \tabularnewline
1  & 0  & 0  & -10  & -10  & 0  & 0  & 0  & 0  & -10  & -10  & 0  & 0 \tabularnewline
2  & 0  & 0  & -15  & -14  & 0  & 0  & 0  & 0  & -14  & -15  & 0  & 0 \tabularnewline
\midrule 
\multicolumn{13}{c}{S.D. ($\times10^{-3}$)}\tabularnewline
\midrule 
0  & 136  & 137  & 47  & 45  & 104  & 104  & 135  & 138  & 46  & 45  & 104  & 104 \tabularnewline
0.11  & 136  & 137  & 47  & 45  & 105  & 106  & 135  & 138  & 46  & 46  & 103  & 106 \tabularnewline
0.23  & 136  & 137  & 47  & 45  & 108  & 110  & 135  & 138  & 45  & 46  & 106  & 108 \tabularnewline
0.34  & 136  & 137  & 47  & 44  & 113  & 113  & 135  & 138  & 45  & 46  & 111  & 113 \tabularnewline
0.46  & 136  & 137  & 46  & 44  & 118  & 118  & 135  & 138  & 44  & 45  & 117  & 120 \tabularnewline
0.57  & 136  & 137  & 46  & 44  & 123  & 123  & 135  & 138  & 44  & 45  & 122  & 125 \tabularnewline
0.69  & 136  & 137  & 45  & 44  & 128  & 128  & 135  & 138  & 43  & 44  & 127  & 129 \tabularnewline
0.8  & 136  & 137  & 45  & 44  & 131  & 132  & 135  & 138  & 43  & 44  & 130  & 132 \tabularnewline
1  & 136  & 137  & 43  & 43  & 134  & 135  & 135  & 138  & 42  & 42  & 132  & 135 \tabularnewline
2  & 136  & 137  & 37  & 38  & 135  & 136  & 135  & 138  & 37  & 38  & 133  & 137 \tabularnewline
\midrule 
\multicolumn{13}{c}{root-MSE ($\times10^{-3}$)}\tabularnewline
\midrule 
0  & 136  & 137  & 47  & 45  & 104  & 104  & 135  & 138  & 46  & 45  & 104  & 104 \tabularnewline
0.11  & 136  & 137  & 49  & 48  & 106  & 106  & 135  & 138  & 48  & 48  & 103  & 106 \tabularnewline
0.23  & 136  & 137  & 55  & 54  & 109  & 110  & 135  & 138  & 54  & 54  & 106  & 109 \tabularnewline
0.34  & 136  & 137  & 63  & 62  & 114  & 114  & 135  & 138  & 61  & 63  & 111  & 113 \tabularnewline
0.46  & 136  & 137  & 72  & 72  & 119  & 119  & 135  & 138  & 71  & 72  & 117  & 121 \tabularnewline
0.57  & 136  & 137  & 81  & 82  & 123  & 124  & 135  & 138  & 80  & 81  & 122  & 125 \tabularnewline
0.69  & 136  & 137  & 90  & 91  & 128  & 129  & 135  & 138  & 90  & 91  & 127  & 129 \tabularnewline
0.8  & 136  & 137  & 98  & 100  & 131  & 132  & 135  & 138  & 98  & 99  & 130  & 132 \tabularnewline
1  & 136  & 137  & 112  & 113  & 134  & 135  & 135  & 138  & 111  & 113  & 132  & 135 \tabularnewline
2  & 136  & 137  & 150  & 150  & 135  & 136  & 135  & 138  & 149  & 151  & 133  & 137 \tabularnewline
\bottomrule
\end{longtable}
\end{table}

\begin{table}[h]
\caption{\label{tab:SimNumericalResult-m5000-1}Simulation results of coverage
rates and widths of confidence intervals for $\widehat{\psi}_{\rt},$
$\widehat{\psi}_{\eff},$ and $\widehat{\psi}_{\elas}$ (labeled as
``RT'', ``Eff'', and ``Elastic'') in the two cases: zero effect
modification $\psi_{1}=\psi_{2}=0$ (left) and nonzero effect modification
$\psi_{1}=\psi_{2}=1$ (right) with $n=5000$ (Section \ref{sec:Simulation})}

\centering

\begin{longtable}[c]{ccccccccccccc}
\toprule 
 & \multicolumn{2}{c}{RT} & \multicolumn{2}{c}{Eff} & \multicolumn{2}{c}{Elastic} & \multicolumn{2}{c}{RT} & \multicolumn{2}{c}{Eff} & \multicolumn{2}{c}{Elastic}\tabularnewline
\midrule 
 & \multicolumn{6}{c}{Case 1: zero effect modification} & \multicolumn{6}{c}{Case 2: nonzero effect modification}\tabularnewline
$b$  & {\scriptsize{}{}{}$\psi_{1}=0$}  & {\scriptsize{}{}{}$\psi_{2}=0$}  & {\scriptsize{}{}{}$\psi_{1}=0$}  & {\scriptsize{}{}{}$\psi_{2}=0$}  & {\scriptsize{}{}{}$\psi_{1}=0$}  & {\scriptsize{}{}{}$\psi_{2}=0$}  & {\scriptsize{}{}{}$\psi_{1}=1$}  & {\scriptsize{}{}{}$\psi_{2}=1$}  & {\scriptsize{}{}{}$\psi_{1}=1$}  & {\scriptsize{}{}{}$\psi_{2}=1$}  & {\scriptsize{}{}{}$\psi_{1}=1$}  & {\scriptsize{}{}{}$\psi_{2}=1$}\tabularnewline
\midrule 
\multicolumn{13}{c}{Coverage Rate ($\%$)}\tabularnewline
\midrule 
0  & 94.3  & 94.2  & 94.4  & 95.8  & 94.1  & 93.0  & 94.5  & 93.5  & 95.3  & 95.9  & 94.2  & 93.2 \tabularnewline
0.11  & 94.4  & 94.2  & 93.1  & 94.2  & 94.5  & 93.4  & 94.6  & 93.5  & 93.7  & 94.0  & 94.3  & 93.5 \tabularnewline
0.23  & 94.4  & 94.2  & 90.5  & 90.1  & 95.2  & 94.5  & 94.6  & 93.5  & 90.6  & 90.1  & 95.1  & 94.8 \tabularnewline
0.34  & 94.4  & 94.2  & 84.0  & 84.2  & 95.7  & 95.3  & 94.6  & 93.5  & 85.2  & 84.5  & 95.7  & 94.8 \tabularnewline
0.46  & 94.4  & 94.2  & 76.8  & 75.8  & 96.0  & 95.5  & 94.6  & 93.5  & 76.2  & 75.7  & 96.2  & 95.2 \tabularnewline
0.57  & 94.4  & 94.2  & 67.2  & 65.8  & 96.0  & 95.4  & 94.6  & 93.5  & 67.0  & 65.7  & 96.2  & 95.0 \tabularnewline
0.69  & 94.4  & 94.2  & 56.6  & 56.6  & 95.5  & 94.9  & 94.6  & 93.5  & 56.4  & 56.3  & 96.0  & 94.5 \tabularnewline
0.8  & 94.4  & 94.2  & 48.4  & 47.4  & 95.0  & 94.6  & 94.6  & 93.5  & 47.2  & 47.9  & 95.5  & 94.2 \tabularnewline
1  & 94.4  & 94.2  & 33.8  & 32.3  & 95.0  & 94.1  & 94.6  & 93.5  & 33.8  & 31.4  & 95.1  & 94.0 \tabularnewline
2  & 94.4  & 94.2  & 2.8  & 2.8  & 94.5  & 94.0  & 94.6  & 93.5  & 3.5  & 2.5  & 94.8  & 93.7 \tabularnewline
\midrule 
\multicolumn{13}{c}{Width ($\times10^{-3}$)}\tabularnewline
\midrule 
0  & 529  & 528  & 181  & 181  & 468  & 468  & 531  & 529  & 181  & 181  & 469  & 467 \tabularnewline
0.11  & 529  & 528  & 181  & 180  & 479  & 479  & 531  & 529  & 181  & 181  & 479  & 477 \tabularnewline
0.23  & 529  & 528  & 181  & 180  & 501  & 500  & 531  & 529  & 180  & 180  & 500  & 498 \tabularnewline
0.34  & 529  & 528  & 179  & 178  & 516  & 515  & 531  & 529  & 180  & 179  & 517  & 514 \tabularnewline
0.46  & 529  & 528  & 178  & 177  & 524  & 523  & 531  & 529  & 178  & 177  & 526  & 524 \tabularnewline
0.57  & 529  & 528  & 177  & 176  & 527  & 526  & 531  & 529  & 176  & 176  & 528  & 526 \tabularnewline
0.69  & 529  & 528  & 175  & 174  & 528  & 528  & 531  & 529  & 174  & 174  & 529  & 527 \tabularnewline
0.8  & 529  & 528  & 172  & 172  & 529  & 528  & 531  & 529  & 172  & 172  & 530  & 527 \tabularnewline
1  & 529  & 528  & 168  & 168  & 529  & 528  & 531  & 529  & 168  & 168  & 530  & 527 \tabularnewline
2  & 529  & 528  & 148  & 148  & 529  & 528  & 531  & 529  & 148  & 149  & 530  & 528 \tabularnewline
\bottomrule
\end{longtable}
\end{table}

\begin{table}[h]
\caption{\label{tab:Simulation-results-ofgamma}Simulation results of Monte
Carlo averages and standard deviations of the estimators for the local
parameter $\eta=(\eta_{0},\eta_{1},\eta_{2})$, the thresholds $\gamma$
and $c_{\gamma}$,\textcolor{red}{{} }and the proportion of combining
the RT and RW samples $\pr({\rm comb})$ for the elastic combining
estimator in the two cases: zero effect modification $\psi_{1}=\psi_{2}=0$
(top) and nonzero effect modification $\psi_{1}=\psi_{2}=1$ (bottom)
with $n=2000$ (Section \ref{sec:Simulation})}

\vspace{0.15cm}

\centering %
\begin{tabular}{ccccccccccccccc}
\toprule 
\multicolumn{15}{c}{Case 1: zero effect modification}\tabularnewline
\midrule 
 &  & \multicolumn{2}{c}{$\eta_{0}$ } & \multicolumn{2}{c}{$\eta_{1}$ } & \multicolumn{2}{c}{$\eta_{2}$ } & \multicolumn{2}{c}{$\text{\ensuremath{\gamma}}$} & \multicolumn{2}{c}{$c_{\gamma}$} &  & \multicolumn{2}{c}{$\pr({\rm comb})$}\tabularnewline
$b$  &  & Est  & S.D.  & Est  & S.D.  & Est  & S.D.  & Est  & S.D.  & Est  & S.D.  &  & Est  & S.D.\tabularnewline
\midrule 
0  &  & 0.00  & 1.54  & -0.01  & 1.98  & 0.01  & 1.99  & 0.30  & 0.44  & 34.04  & 22.87  &  & 0.68  & 0.47 \tabularnewline
0.11  &  & -0.45  & 1.54  & -0.46  & 1.98  & -0.44  & 1.98  & 0.35  & 0.46  & 31.51  & 23.74  &  & 0.63  & 0.48 \tabularnewline
0.23  &  & -0.94  & 1.55  & -0.95  & 2.00  & -0.94  & 2.00  & 0.43  & 0.48  & 27.09  & 24.52  &  & 0.54  & 0.50 \tabularnewline
0.34  &  & -1.39  & 1.57  & -1.40  & 2.02  & -1.39  & 2.03  & 0.54  & 0.48  & 21.58  & 24.42  &  & 0.43  & 0.49 \tabularnewline
0.46  &  & -1.88  & 1.59  & -1.89  & 2.05  & -1.88  & 2.06  & 0.67  & 0.45  & 15.26  & 22.70  &  & 0.30  & 0.46 \tabularnewline
0.57  &  & -2.32  & 1.62  & -2.32  & 2.09  & -2.31  & 2.10  & 0.75  & 0.41  & 10.62  & 20.09  &  & 0.21  & 0.41 \tabularnewline
0.69  &  & -2.80  & 1.66  & -2.80  & 2.14  & -2.79  & 2.15  & 0.83  & 0.35  & 6.82  & 16.79  &  & 0.13  & 0.34 \tabularnewline
0.8  &  & -3.23  & 1.70  & -3.23  & 2.20  & -3.23  & 2.20  & 0.89  & 0.29  & 4.06  & 13.30  &  & 0.08  & 0.27 \tabularnewline
1  &  & -3.98  & 1.77  & -3.98  & 2.29  & -3.98  & 2.29  & 0.94  & 0.21  & 1.80  & 8.92  &  & 0.03  & 0.18 \tabularnewline
2  &  & -7.18  & 2.19  & -7.19  & 2.84  & -7.17  & 2.82  & 1.00  & 0.03  & 0.01  & 0.17  &  & 0.00  & 0.00 \tabularnewline
\bottomrule
\end{tabular}

\vspace{0.25cm}

\begin{tabular}{ccccccccccccccc}
\toprule 
\multicolumn{15}{c}{Case 2: nonzero effect modification}\tabularnewline
\midrule 
 &  & \multicolumn{2}{c}{$\eta_{0}$ } & \multicolumn{2}{c}{$\eta_{1}$ } & \multicolumn{2}{c}{$\eta_{2}$ } & \multicolumn{2}{c}{$\gamma$} & \multicolumn{2}{c}{$c_{\gamma}$} &  & \multicolumn{2}{c}{$\pr({\rm comb})$}\tabularnewline
$b$  &  & Est  & S.D.  & Est  & S.D.  & Est  & S.D.  & Est  & S.D.  & Est  & S.D.  &  & Est  & S.D.\tabularnewline
\midrule 
0  &  & 0.03  & 1.54  & 0.04  & 1.98  & 0.04  & 1.99  & 0.30  & 0.44  & 34.07  & 22.85  &  & 0.68  & 0.46 \tabularnewline
0.11  &  & -0.42  & 1.54  & -0.41  & 1.99  & -0.41  & 2.00  & 0.33  & 0.46  & 32.56  & 23.41  &  & 0.65  & 0.48 \tabularnewline
0.23  &  & -0.91  & 1.55  & -0.90  & 2.00  & -0.90  & 2.01  & 0.41  & 0.48  & 27.94  & 24.43  &  & 0.56  & 0.50 \tabularnewline
0.34  &  & -1.36  & 1.57  & -1.35  & 2.03  & -1.35  & 2.04  & 0.52  & 0.48  & 22.58  & 24.52  &  & 0.45  & 0.50 \tabularnewline
0.46  &  & -1.84  & 1.59  & -1.83  & 2.06  & -1.83  & 2.07  & 0.64  & 0.46  & 16.24  & 23.07  &  & 0.32  & 0.47 \tabularnewline
0.57  &  & -2.27  & 1.62  & -2.27  & 2.10  & -2.27  & 2.11  & 0.74  & 0.41  & 11.13  & 20.45  &  & 0.22  & 0.41 \tabularnewline
0.69  &  & -2.76  & 1.66  & -2.75  & 2.14  & -2.75  & 2.16  & 0.83  & 0.35  & 6.98  & 16.98  &  & 0.13  & 0.34 \tabularnewline
0.8  &  & -3.18  & 1.70  & -3.17  & 2.20  & -3.18  & 2.21  & 0.88  & 0.30  & 4.31  & 13.64  &  & 0.08  & 0.27 \tabularnewline
1  &  & -3.94  & 1.78  & -3.93  & 2.29  & -3.94  & 2.30  & 0.94  & 0.22  & 2.00  & 9.41  &  & 0.04  & 0.19 \tabularnewline
2  &  & -7.12  & 2.20  & -7.11  & 2.84  & -7.11  & 2.84  & 1.00  & 0.04  & 0.02  & 0.24  &  & 0.00  & 0.00 \tabularnewline
\bottomrule
\end{tabular}
\end{table}

\subsection{Importance of overlapping between the RT and RW samples\label{subsec:Importance-of-overlapping}}

This simulation study creates an artificial scenario to illustrate
the danger of extrapolation when transporting from the narrow RT sample
to the broader RW sample with the HTE transportability and overlap
assumption violated. Instead of the data generation distributions
in Section \ref{sec:Simulation}, we consider a new data distribution,
in which $X_{1},X_{2}\sim\mathcal{N}(0,1)$ for both RT and RW finite
population. For the HTE, we now consider $\tau(Z)=\psi_{0}+\psi_{1}X_{1}+\psi_{2}|X_{2}|$
with $\psi_{0}=0,\psi_{1}=\psi_{2}=1$. To generate the RT sample,
we first generate the RT selection indicator by $\delta\mid X\sim\mathrm{Bernoulli}\{\pi_{\delta}(X)\},$
where $\logit\{\pi_{\delta}(X)\}=-4.5-2X_{1}-2X_{2}$, and then set
$\delta=0$ if $X_{2}<0$. We also select a random sample of size
$m=2000$ from the population to form an RW sample. In this case,
the RT sample is a narrower sample where $X_{2}$ can only be positive.
The HTE for the RT sample (and the overlap RW sample) is $\tau(Z)=\psi_{0}+\psi_{1}X_{1}+\psi_{2}X_{2}$,
but the HTE for the non-overlap region of the RW sample is not. In
the RW sample, we consider $b=0$ to represent a scenario without
unmeasured confounding for treatment assignment.

Table \ref{tab:Sim-misspecification} reports the simulation results.
The RT estimator $\widehat{\psi}_{\rt}$ has small biases. Even without
unmeasured confounding, the efficient combining estimator $\widehat{\psi}_{\eff}$
is biased with low coverage rates for all parameters, due to model
misspecification of the HTE for the RW population and violation of
Assumption \ref{Asump:rand-rwd}(i) and (iii). The elastic combining
approach rejects the RW sample for combining by $97\%$ times over
simulation, and thus $\widehat{\psi}_{\elas}$ stays close to $\widehat{\psi}_{\rt}$
and has small biases and satisfactory coverage properties.

\begin{table}[h]
\caption{\label{tab:Sim-misspecification}\textcolor{black}{Simulation results
for Monte Carlo biases, standard deviations, root-mean-square errors,
coverage rates and widths of $95\%$ confidence intervals for }$\widehat{\psi}_{\rt},$
$\widehat{\psi}_{\eff},$ and $\widehat{\psi}_{\elas}$ (labeled as
``RT'', ``Eff'', and ``Elastic'')\textcolor{black}{{} }when
the transportability of the HTE does not hold in the RW data with
$n=2000$\textcolor{black}{{} (Section \ref{subsec:Importance-of-overlapping})}}

\centering

\vspace{0.15cm}

\begin{tabular}{ccccccccc}
\toprule 
 & \multicolumn{2}{c}{RT} &  & \multicolumn{2}{c}{Eff} &  & \multicolumn{2}{c}{Elastic}\tabularnewline
 & $\psi_{1}$  & $\psi_{2}$  &  & $\psi_{1}$  & $\psi_{2}$  &  & $\psi_{1}$  & $\psi_{2}$\tabularnewline
\midrule 
Bias ($\times10^{-2}$)  & 0.6  & 1.3  &  & 7.5  & -95.6  &  & 0.6  & 1.3 \tabularnewline
S.D. ($\times10^{-3}$)  & 79  & 188  &  & 39  & 62  &  & 79  & 185\tabularnewline
root-MSE ($\times10^{-3}$)  & 79.5  & 188  &  & 84  & 958  &  & 79  & 185\tabularnewline
Coverage rate ($\%$)  & 95.0  & 94.2  &  & 48.6  & 0.0  &  & 94.6  & 94.4\tabularnewline
Width ($\times10^{-3}$)  & 311  & 730  &  & 150  & 233  &  & 311  & 728\tabularnewline
\bottomrule
\end{tabular}
\end{table}

\pagebreak{}

\subsection{Sensitivity analysis of ECIs to the choice of the cut-off value $\kappa_{n}$\label{subsec:Sensitivity-analysis-of-k}}

{In this simulation study, we conduct a sensitivity analysis to assess
the performance of the ECIs to the choice of the cut-off value $\kappa_{n}.$
All data generating distributions are the same as in Section \ref{sec:Simulation}.
Table \ref{tab:Sensitivity-analysis-ofECI} reports the sensitivity
results of coverage rates and widths of the ECIs to the choice of
the cut-off value $\kappa_{n}\in\{0.5(\log n){}^{1/2},(\log n){}^{1/2}\}$.
When $b$ is small, the width of the ECIs decreases with $\kappa_{n}$,
while when $b$ is large, the coverage rate and width of the ECIs
increase with $\kappa_{n}$. To help explain the results, denote ${\rm ECI}_{k,1-\alpha}^{[1]}=[\inf_{\eta\in\mathcal{B}_{1-\widetilde{\alpha}}}\widehat{Q}_{k,\widetilde{\alpha}/2}(\mu_{1}),\sup_{\eta\in\mathcal{B}_{1-\widetilde{\alpha}}}\widehat{Q}_{k,1-\widetilde{\alpha}/2}(\mu_{1})]$,
${\rm ECI}_{k,1-\alpha}^{[2]}=[\widehat{Q}_{k,\alpha/2}(\pm\infty),\widehat{Q}_{k,1-\alpha/2}(\pm\infty)],$
and ${\rm ECI}_{k,1-\alpha}={\rm ECI}_{k,1-\alpha}^{[1]}\bone(T\leq\kappa_{n})+{\rm ECI}_{k,1-\alpha}^{[2]}\bone(T>\kappa_{n})$.
When $\kappa_{n}$ increases from $0.5(\log n){}^{1/2}$ to $(\log n){}^{1/2}$,
the frequency of choosing ${\rm ECI}_{k,1-\alpha}^{[2]}$ over ${\rm ECI}_{k,1-\alpha}^{[1]}$
decreases; see $\pr(T>\kappa_{n})$ reported in Table \ref{tab:Sensitivity-analysis-ofECI-1}.
When $b$ is small, ${\rm ECI}_{k,1-\alpha}^{[2]}$ can be more conservative
than ${\rm ECI}_{k,1-\alpha}^{[1]}$, and when $b$ is large, ${\rm ECI}_{k,1-\alpha}^{[1]}$
can be more conservative than ${\rm ECI}_{k,1-\alpha}^{[2]}$. This
is because when $b$ is small, $\eta$ is small, the likelihood of
integrating the RW sample is large, and $[\widehat{Q}_{k,\widetilde{\alpha}/2}(\mu_{1}),\widehat{Q}_{k,1-\widetilde{\alpha}/2}(\mu_{1})]$
for each small $\eta$ (thus small $\mu_{1}$) is narrower than ${\rm ECI}_{k,1-\alpha}^{[2]}$,
and thus ${\rm ECI}_{k,1-\alpha}^{[2]}$ can be more conservative
than ${\rm ECI}_{k,1-\alpha}^{[1]}$, and vice versa. Overall, the
coverage rates and widths of ECIs are close over the choice of $\kappa_{n}.$
}

\begin{table}[h]
\caption{\label{tab:Sensitivity-analysis-ofECI}Sensitivity analysis of coverage
rates and widths of the elastic confidence intervals to the choice
of the cut-off value $\kappa_{n}$ in the two cases: zero effect modification
$\psi_{1}=\psi_{2}=0$ (left) and nonzero effect modification $\psi_{1}=\psi_{2}=1$
(right) with $n=2000$ (Section \ref{sec:Simulation}); the narrower
ECIs are bolded }

\vspace{0.15cm}

\centering

\begin{tabular}{cccccccccccccc}
\toprule 
 & \multicolumn{6}{c}{Case 1: zero effect modification} & \multicolumn{7}{c}{Case 2: nonzero effect modification}\tabularnewline
 &  &  & \multicolumn{2}{c}{{\small{}{}{}$0.5(\log n)^{1/2}$}} & \multicolumn{2}{c}{{\small{}{}{}$(\log n)^{1/2}$}} &  &  & \multicolumn{2}{c}{{\small{}{}{}$0.5(\log n)^{1/2}$}} & \multicolumn{2}{c}{{\small{}{}{}$(\log n)^{1/2}$}} & \tabularnewline
\midrule 
$b$  &  &  & {\scriptsize{}{}{}$\psi_{1}=0$}  & {\scriptsize{}{}{}$\psi_{2}=0$}  & {\scriptsize{}{}{}$\psi_{1}=0$}  & {\scriptsize{}{}{}$\psi_{2}=0$}  &  &  & {\scriptsize{}{}{}$\psi_{1}=1$}  & {\scriptsize{}{}{}$\psi_{2}=1$}  & {\scriptsize{}{}{}$\psi_{1}=1$}  & {\scriptsize{}{}{}$\psi_{2}=1$}  & \tabularnewline
\midrule 
\multicolumn{13}{c}{Coverage Rate ($\%$)} & \tabularnewline
\midrule 
0  &  &  & 94.2  & 93.5  & 92.7  & 92.5  &  &  & 94.3  & 94.2  & 92.7  & 92.5  & \tabularnewline
0.11  &  &  & 93.8  & 94.2  & 93.2  & 92.8  &  &  & 94.8  & 94.5  & 92.9  & 92.7  & \tabularnewline
0.23  &  &  & 94.5  & 94.6  & 92.8  & 92.8  &  &  & 95.2  & 94.8  & 93.3  & 92.7  & \tabularnewline
0.34  &  &  & 94.5  & 94.8  & 94.0  & 93.8  &  &  & 95.5  & 95.0  & 94.4  & 93.5  & \tabularnewline
0.46  &  &  & 95.0  & 95.5  & 94.5  & 94.5  &  &  & 95.8  & 95.3  & 94.5  & 94.4  & \tabularnewline
0.57  &  &  & 95.7  & 96.0  & 95.5  & 95.2  &  &  & 96.2  & 95.5  & 95.5  & 94.8  & \tabularnewline
0.69  &  &  & 95.9  & 96.2  & 95.5  & 95.8  &  &  & 95.9  & 95.3  & 95.3  & 94.6  & \tabularnewline
0.8  &  &  & 95.8  & 96.0  & 95.5  & 95.6  &  &  & 95.7  & 95.2  & 95.3  & 95.0  & \tabularnewline
1  &  &  & 95.5  & 95.0  & 95.5  & 95.0  &  &  & 95.5  & 95.0  & 95.5  & 94.9  & \tabularnewline
2  &  &  & 94.3  & 94.5  & 94.3  & 94.4  &  &  & 94.7  & 94.2  & 94.7  & 94.2  & \tabularnewline
\midrule 
\multicolumn{13}{c}{Width ($\times10^{-3}$)} & \tabularnewline
\midrule 
0  &  &  & 480  & 481  & \textbf{472 }  & \textbf{473 }  &  &  & 479  & 480  & \textbf{472 }  & \textbf{474 }  & \tabularnewline
0.11  &  &  & 488  & 489  & \textbf{488 }  & \textbf{487 }  &  &  & 486  & 488  & \textbf{479 }  & \textbf{480 }  & \tabularnewline
0.23  &  &  & 498  & 499  & \textbf{496 }  & \textbf{497 }  &  &  & \textbf{497 }  & \textbf{499 }  & 498  & 500  & \tabularnewline
0.34  &  &  & \textbf{509 }  & \textbf{510 }  & 516  & 516  &  &  & \textbf{507 }  & \textbf{509 }  & 511  & 514  & \tabularnewline
0.46  &  &  & \textbf{518 }  & \textbf{519 }  & 530  & 530  &  &  & \textbf{518 }  & \textbf{519 }  & 524  & 526  & \tabularnewline
0.57  &  &  & \textbf{524 }  & \textbf{524 }  & 535  & 535  &  &  & \textbf{524 }  & \textbf{526 }  & 530  & 532  & \tabularnewline
0.69  &  &  & \textbf{528 }  & \textbf{528 }  & 534  & 534  &  &  & \textbf{526 }  & \textbf{528 }  & 529  & 531  & \tabularnewline
0.8  &  &  & \textbf{528 }  & \textbf{528 }  & 532  & 532  &  &  & \textbf{527 }  & \textbf{529 }  & 530  & 532  & \tabularnewline
1  &  &  & \textbf{528 }  & \textbf{528 }  & 529  & 529  &  &  & \textbf{528 }  & \textbf{530 }  & 530  & 532  & \tabularnewline
2  &  &  & 527  & 528  & 527  & \textbf{527}  &  &  & 528  & 530  & 528  & 530  & \tabularnewline
\bottomrule
\end{tabular}
\end{table}

\begin{table}[h]
\caption{\label{tab:Sensitivity-analysis-ofECI-1}Sensitivity analysis of $\pr(T>\kappa_{n})$
to the choice of the cut-off value $\kappa_{n}$ in the two cases:
zero effect modification $\psi_{1}=\psi_{2}=0$ (left) and nonzero
effect modification $\psi_{1}=\psi_{2}=1$ (right) with $n=2000$
(Section \ref{sec:Simulation})}

\vspace{0.15cm}

\centering %
\begin{tabular}{cccccccccccccc}
\toprule 
 & \multicolumn{6}{c}{Case 1: zero effect modification} & \multicolumn{7}{c}{Case 2: nonzero effect modification}\tabularnewline
\midrule 
$\kappa_{n}$  &  &  & \multicolumn{2}{c}{{\small{}{}{}$0.5(\log n)^{1/2}$}} & \multicolumn{2}{c}{{\small{}{}{}$(\log n)^{1/2}$}} &  &  & \multicolumn{2}{c}{{\small{}{}{}$0.5(\log n)^{1/2}$}} & \multicolumn{2}{c}{{\small{}{}{}$(\log n)^{1/2}$}} & \tabularnewline
$b$  &  &  & Est  & S.D.  & Est  & S.D.  &  &  & Est  & S.D.  & Est  & S.D.  & \tabularnewline
\midrule 
0  &  &  & 0.76  & 0.43  & 0.54  & 0.50  &  &  & 0.76  & 0.43  & 0.55  & 0.50  & \tabularnewline
0.11  &  &  & 0.80  & 0.40  & 0.61  & 0.49  &  &  & 0.79  & 0.41  & 0.59  & 0.49  & \tabularnewline
0.23  &  &  & 0.86  & 0.35  & 0.67  & 0.47  &  &  & 0.85  & 0.36  & 0.67  & 0.47  & \tabularnewline
0.34  &  &  & 0.91  & 0.29  & 0.77  & 0.42  &  &  & 0.90  & 0.30  & 0.77  & 0.42  & \tabularnewline
0.46  &  &  & 0.96  & 0.21  & 0.87  & 0.34  &  &  & 0.95  & 0.22  & 0.86  & 0.34  & \tabularnewline
0.57  &  &  & 0.98  & 0.13  & 0.92  & 0.26  &  &  & 0.98  & 0.16  & 0.92  & 0.27  & \tabularnewline
0.69  &  &  & 0.99  & 0.09  & 0.97  & 0.18  &  &  & 0.99  & 0.10  & 0.96  & 0.20  & \tabularnewline
0.8  &  &  & 1.00  & 0.04  & 0.98  & 0.12  &  &  & 1.00  & 0.07  & 0.98  & 0.15  & \tabularnewline
1  &  &  & 1.00  & 0.00  & 1.00  & 0.06  &  &  & 1.00  & 0.00  & 1.00  & 0.07  & \tabularnewline
2  &  &  & 1.00  & 0.00  & 1.00  & 0.00  &  &  & 1.00  & 0.00  & 1.00  & 0.00  & \tabularnewline
\bottomrule
\end{tabular}
\end{table}

\subsection{The elastic combining estimator with a fixed threshold $c_{\gamma}$\label{subsec:fixing_gamma}}

In this simulation study, we assess the performance of the elastic
combining estimator with a fixed threshold $c_{\gamma}$. All data
generating distributions are the same as in Section \ref{sec:Simulation}.
For the elastic combining estimator, we fix $c_{\gamma}$ to be the
$95$th quantile of a $\chi_{3}^{2}$ distribution (i.e., $7.81$).
Figure~\ref{fig:violin} presents the plots of Monte Carlo biases,
variances, and MSEs of estimators based on 2000 simulated datasets.
The elastic integrative estimator $\widehat{\psi}_{\elas}$ with a
fixed threshold can increase biases compared to using a data-adaptive
selected threshold.

\begin{figure}[h]
\centering 
\begin{centering}
\includegraphics[scale=0.65]{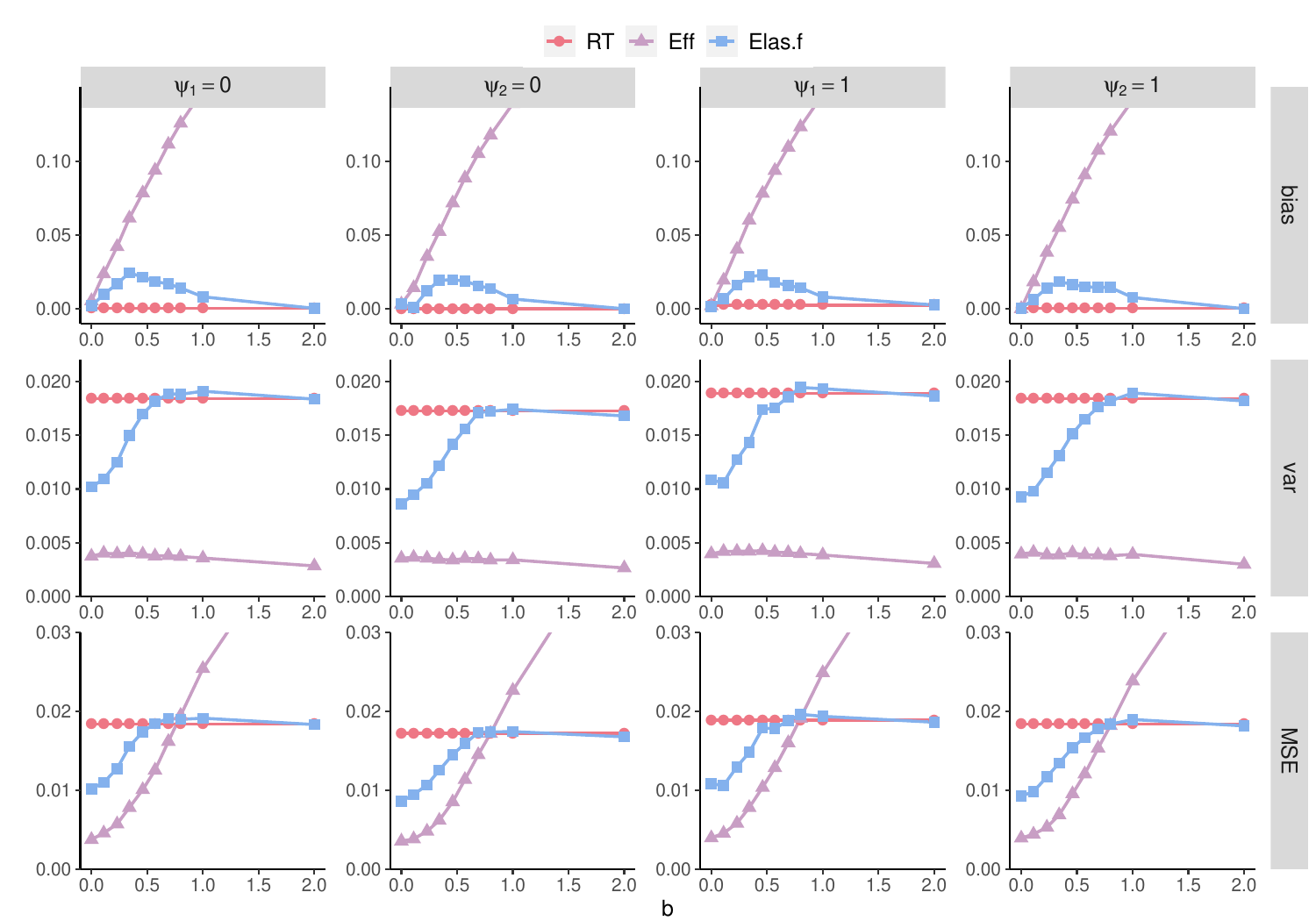} 
\par\end{centering}
\vspace{0.5cm}

\caption{\label{fig:violin-1}Summary statistics plots of estimators of $(\psi_{1},\psi_{2})$
with respect to the strength of unmeasured confounding labeled by
``b''. In each plot, the three estimators are labeled by ``RT'',
``Eff'', and ``Elas.f'' with $c_{\gamma}\equiv7.81$ (i.e., the
$95$th quantile of a $\chi_{3}^{2}$ distribution). Each row of the
plots corresponds to a different metrics: ``bias'' for bias, ``var''
for variance, ``MSE'' for mean square error; each column of the
plots corresponds to one component of $(\psi_{1},\psi_{2})$ in the
two cases: $\psi_{1}=0,\psi_{2}=0$, $\psi_{1}=1$, and $\psi_{2}=1$
with $n=2000$ (Section \ref{subsec:fixing_gamma}).}
\end{figure}

\subsection{Regular inference fails for the pre-test estimator with sample splitting\label{subsec:Regular-inference-fails}}

In this simulation study, we illustrate the failure of regular inference
for the pre-test estimator in the context of data integration with
sampling splitting. The data generating distribution is the same as
in Section \ref{sec:Simulation} except that we double the sample
sizes for the RT and RW samples. The RT sample is randomly split into
two folds, and similarly, the RW sample is randomly split into two
folds. We couple one fold from the RT sample and one fold from the
RW sample to form a pre-testing dataset and the rest to form an estimation
dataset. In the pre-testing dataset, we conduct the pre-test to decide
whether to combine the RT sample and the RW sample. In the estimation
dataset, we conduct the analysis based on the decision from the pre-testing,
including the point and variance estimation and regular Wald inference.
For comparison, we also compute the RT and Eff estimator and their
regular Wald inferences.

Table \ref{table:cvg-ss} reports the results for the $95\%$ confidence
intervals. Similar conclusions in Section \ref{sec:Simulation} can
be drawn for the point estimators in this simulation study (not shown).
The coverage rates for the RT estimator are close to the nominal level;
while the coverage rates for the Eff estimator are more off as $b$
increases. Importantly, the pre-test estimator has poor coverage rates,
especially when $b$ is around $0.57$, demonstrating the failure
of regular inference for the pre-test estimator with sample splitting.

\begin{table}[h]
\caption{\label{table:cvg-ss}Simulation results for coverage rates and widths
of $95\%$ confidence intervals in the case: nonzero effect modification
$\psi_{1}=\psi_{2}=1$ with $n=2000$ when using sample splitting
(Section \ref{subsec:Regular-inference-fails})}

\centering

\vspace{0.15cm}

\resizebox{\textwidth}{!}{

\begin{tabular}{ccccccccccccccc}
\toprule 
 &  & \multicolumn{2}{c}{RT} & \multicolumn{2}{c}{Eff} & \multicolumn{2}{c}{Pre-testing} &  & \multicolumn{2}{c}{RT} & \multicolumn{2}{c}{Eff} & \multicolumn{2}{c}{Pre-testing}\tabularnewline
$b$  &  & {\scriptsize{}{}{}$\psi_{1}=1$}  & {\scriptsize{}{}{}$\psi_{2}=1$}  & {\scriptsize{}{}{}$\psi_{1}=1$}  & {\scriptsize{}{}{}$\psi_{2}=1$}  & {\scriptsize{}{}{}$\psi_{1}=1$}  & {\scriptsize{}{}{}$\psi_{2}=1$}  &  & {\scriptsize{}{}{}$\psi_{1}=1$}  & {\scriptsize{}{}{}$\psi_{2}=1$}  & {\scriptsize{}{}{}$\psi_{1}=1$}  & {\scriptsize{}{}{}$\psi_{2}=1$}  & {\scriptsize{}{}{}$\psi_{1}=1$}  & {\scriptsize{}{}{}$\psi_{2}=1$}\tabularnewline
\midrule 
\multicolumn{8}{c}{Coverage Rate ($\%$)} &  & \multicolumn{6}{c}{Width ($\times10^{-2}$)}\tabularnewline
\midrule 
0  &  & 94.4  & 94.8  & 94.4  & 94.4  & 94.0  & 95.2  &  & 542  & 546  & 245  & 246  & 357  & 356\tabularnewline
0.11  &  & 94.4  & 94.8  & 93.4  & 94.0  & 93.0  & 95.2  &  & 542  & 546  & 245  & 246  & 361  & 365\tabularnewline
0.23  &  & 94.4  & 94.8  & 89.8  & 88.4  & 91.6  & 92.2  &  & 542  & 545  & 244  & 244  & 397  & 399\tabularnewline
0.34  &  & 94.4  & 94.8  & 84.2  & 83.6  & 90.6  & 88.8  &  & 542  & 545  & 243  & 243  & 430  & 433\tabularnewline
0.46  &  & 94.4  & 94.8  & 74.4  & 77.2  & 89.4  & 90.0  &  & 542  & 546  & 242  & 243  & 471  & 475\tabularnewline
0.57  &  & 94.4  & 94.8  & 63.6  & 69.6  & 88.6  & 89.6  &  & 542  & 546  & 240  & 241  & 500  & 503\tabularnewline
0.69  &  & 94.6  & 94.8  & 52.8  & 59.0  & 90.8  & 91.2  &  & 542  & 545  & 237  & 238  & 519  & 518\tabularnewline
0.8  &  & 94.4  & 94.8  & 42.6  & 47.8  & 92.8  & 92.8  &  & 542  & 546  & 236  & 235  & 533  & 534\tabularnewline
1  &  & 94.4  & 94.8  & 28.6  & 32.2  & 93.6  & 94.2  &  & 542  & 545  & 231  & 230  & 541  & 543\tabularnewline
2  &  & 94.4  & 94.8  & 1.8  & 2.4  & 94.4  & 94.6  &  & 542  & 545  & 210  & 210  & 542  & 545\tabularnewline
\bottomrule
\end{tabular}} 
\end{table}

\section{\textcolor{black}{Details of the real-data application} \label{sec:datasupp}}

\subsection{\textcolor{black}{Eligibility criteria }}

\textcolor{black}{The CALGB 9633 trial was designed to determine the
efficacy of adjuvant chemotherapy compared with observation with the
following eligibility criteria \citep{strauss2008adjuvant}:} 
\begin{itemize}
\item \textcolor{black}{Histologically documented non-small cell lung cancer
(NSCLC)} 
\item \textcolor{black}{Complete surgical resection (lobectomy or pneumonectomy)} 
\item \textcolor{black}{Stage IB disease (T2N0M0) with tumor size $\geq3$cm} 
\item \textcolor{black}{Randomization occurs within 4-8 weeks of surgery} 
\item \textcolor{black}{No prior chemotherapy and radiotherapy for NSCLC} 
\item \textcolor{black}{Age $\geq18$} 
\item \textcolor{black}{Performance status $0-1$} 
\item \textcolor{black}{No concomitant malignancy} 
\end{itemize}
\textcolor{black}{The NCDB is a clinical oncology registry database
that captures the information from approximately $75\%$ of all newly
diagnosed cancer patients in the US. Thus, the NCDB involves a more
diverse patient population than the RT sample. To make the two RT
and RW samples represent the patients from comparable populations,
we have used the same eligibility criteria as those of the RT sample
to identify eligible patients from the NCDB. }

\subsection{\textcolor{black}{Strategies for selecting an RW sample with sufficient
overlap with the RT sample \label{subsec:Strategies-for-selecting}}}

\textcolor{black}{We can employ the following strategies to select
an RW sample with sufficient overlap with the RT sample. First, consider
the RW sample to be extracted from a large population-based database
or electronic health records or claim data. The RT tends to use restrictive
criteria in the phase of new treatment evaluation to ensure a more
homogeneous patient sample with less severe baseline functional status
or comorbidity for safety consideration when testing a new treatment.
The large population-based database allows the investigators to find
comparable populations or samples by applying the same eligibility
criteria of the RT sample to identify comparable RW patients. Second,
suppose comparability still fails to achieve by applying the same
eligibility criteria. In this case, one can apply an additional matching
procedure to improve the comparability between the RT and RW samples
and the chance of successfully integrating the information from two
separate sources. In our motivating application, we may subsample
the cohort of NCDB patients who met the RT eligible criteria. The
distribution of baseline covariates, e.g., race, sex, age, histology,
and tumor size, can be further matched. Toward this end, one can use
various matching algorithms, such as $K$-nearest neighbor matching,
based on the based covariates. } 
\end{document}